\documentclass[10pt,tightenlines,showpacs,letterpaper,aps,prd,longbibliography,twocolumn,nofootinbib,superscriptaddress]{revtex4-2}
\usepackage{rotating}
\usepackage{floatpag}
\rotfloatpagestyle{empty}

\usepackage{dsfont}
\usepackage{amsmath,amsthm,amssymb}
\usepackage{xspace,enumerate,epsfig}
\usepackage{graphicx}
\graphicspath{{.}{./figures/}}
\usepackage{marvosym}

\usepackage{tikzfig}
\usepackage{stmaryrd}
\usepackage{docmute}
\usepackage{keycommand}

\newcommand{\Bob}{$\mathcal{B}$}
\newcommand{\Alice}{$\mathcal{A}$}

\newkeycommand{\pointmap}[style=point][1]{\,\tikz{\node[style=\commandkey{style}] (x) at (0,-0.3) {$#1$};\node [style=none] (3) at (0, 1.0) {};\node [style=none] (3) at (0, -1.0) {};\draw (x)--(0,0.7);}\,}

\newkeycommand{\typedkpoint}[style=kpoint,edgestyle=][2]{%
\,\begin{tikzpicture}
  \begin{pgfonlayer}{nodelayer}
    \node [style=none] (0) at (0, 1) {};
    \node [style=\commandkey{style}] (1) at (0, -0.75) {$#2$};
    \node [style=right label] (2) at (0.25, 0.5) {$#1$};
  \end{pgfonlayer}
  \begin{pgfonlayer}{edgelayer}
    \draw [\commandkey{edgestyle}] (1) to (0.center);
  \end{pgfonlayer}
\end{tikzpicture}\,}

\newkeycommand{\typedkpointdag}[style=kpoint adjoint,edgestyle=][2]{%
\,\begin{tikzpicture}
  \begin{pgfonlayer}{nodelayer}
    \node [style=none] (0) at (0, -1) {};
    \node [style=\commandkey{style}] (1) at (0, 0.75) {$#2$};
    \node [style=right label] (2) at (0.25, -0.5) {$#1$};
  \end{pgfonlayer}
  \begin{pgfonlayer}{edgelayer}
    \draw [\commandkey{edgestyle}] (0.center) to (1);
  \end{pgfonlayer}
\end{tikzpicture}\,}

\newcommand{\bistateadj}[1]{\,\begin{tikzpicture}[yshift=-3mm]
  \begin{pgfonlayer}{nodelayer}
    \node [style=kpointadj, minimum width=1 cm, inner sep=2pt] (0) at (0, 0.5) {$\psi$};
    \node [style=none] (1) at (-0.75, 0.25) {};
    \node [style=none] (2) at (0.75, 0.25) {};
    \node [style=none] (3) at (-0.75, -0.5) {};
    \node [style=none] (4) at (0.75, -0.5) {};
  \end{pgfonlayer}
  \begin{pgfonlayer}{edgelayer}
    \draw (1.center) to (3.center);
    \draw (2.center) to (4.center);
  \end{pgfonlayer}
\end{tikzpicture}\,}

\newkeycommand{\pointketbra}[style1=copoint,style2=point][2]{\,%
\begin{tikzpicture}
  \begin{pgfonlayer}{nodelayer}
    \node [style=none] (0) at (0, -1.5) {};
    \node [style=\commandkey{style1}] (1) at (0, -0.75) {$#1$};
    \node [style=\commandkey{style2}] (2) at (0, 0.75) {$#2$};
    \node [style=none] (3) at (0, 1.5) {};
  \end{pgfonlayer}
  \begin{pgfonlayer}{edgelayer}
    \draw (0.center) to (1);
    \draw (2) to (3.center);
  \end{pgfonlayer}
\end{tikzpicture}\,}

\newkeycommand{\twocopointketbra}[style1=copoint,style2=copoint,style3=point][3]{\,%
\begin{tikzpicture}
  \begin{pgfonlayer}{nodelayer}
    \node [style=none] (0) at (-0.75, -1.5) {};
    \node [style=none] (1) at (0.75, -1.5) {};
    \node [style=\commandkey{style1}] (2) at (-0.75, -0.75) {$#1$};
    \node [style=\commandkey{style2}] (3) at (0.75, -0.75) {$#2$};
    \node [style=\commandkey{style3}] (4) at (0, 0.75) {$#3$};
    \node [style=none] (5) at (0, 1.5) {};
  \end{pgfonlayer}
  \begin{pgfonlayer}{edgelayer}
    \draw (0.center) to (2);
    \draw (1.center) to (3);
    \draw (4) to (5.center);
  \end{pgfonlayer}
\end{tikzpicture}\,}

\newkeycommand{\twopointketbra}[style1=copoint,style2=point,style3=point][3]{\,%
\begin{tikzpicture}
  \begin{pgfonlayer}{nodelayer}
    \node [style=none] (0) at (0, -1.5) {};
    \node [style=none] (1) at (-0.75, 1.5) {};
    \node [style=\commandkey{style1}] (2) at (0, -0.75) {$#1$};
    \node [style=\commandkey{style2}] (3) at (-0.75, 0.75) {$#2$};
    \node [style=\commandkey{style3}] (4) at (0.75, 0.75) {$#3$};
    \node [style=none] (5) at (0.75, 1.5) {};
  \end{pgfonlayer}
  \begin{pgfonlayer}{edgelayer}
    \draw (0.center) to (2);
    \draw (1.center) to (3);
    \draw (4) to (5.center);
  \end{pgfonlayer}
\end{tikzpicture}\,}

\newkeycommand{\pointbraket}[style1=point,style2=copoint][2]{\,%
\begin{tikzpicture}
  \begin{pgfonlayer}{nodelayer}
    \node [style=\commandkey{style1}] (0) at (0, -0.5) {$#1$};
    \node [style=\commandkey{style2}] (1) at (0, 0.5) {$#2$};
  \end{pgfonlayer}
  \begin{pgfonlayer}{edgelayer}
    \draw (0) to (1);
  \end{pgfonlayer}
\end{tikzpicture}\,}

\newkeycommand{\kpointbraket}[style1=kpoint,style2=kpoint adjoint][2]{\,%
\begin{tikzpicture}
  \begin{pgfonlayer}{nodelayer}
    \node [style=\commandkey{style1}] (0) at (0, -0.75) {$#1$};
    \node [style=\commandkey{style2}] (1) at (0, 0.75) {$#2$};
  \end{pgfonlayer}
  \begin{pgfonlayer}{edgelayer}
    \draw (0) to (1);
  \end{pgfonlayer}
\end{tikzpicture}\,}

\newkeycommand{\kpointmap}[style1=kpoint,style2=map][2]{\,%
\begin{tikzpicture}
  \begin{pgfonlayer}{nodelayer}
    \node [style=\commandkey{style1}] (0) at (0, -1.1) {$#1$};
    \node [style=\commandkey{style2}] (1) at (0, 0.2) {$#2$};
    \node [style=none] (2) at (0, 1.5) {};
  \end{pgfonlayer}
  \begin{pgfonlayer}{edgelayer}
    \draw (0) to (1);
    \draw (1) to (2.center);
  \end{pgfonlayer}
\end{tikzpicture}%
\,}

\newkeycommand{\sandwichtwo}[style1=point,style2=map,style3=map,style4=copoint][4]{\,%
\begin{tikzpicture}
  \begin{pgfonlayer}{nodelayer}
    \node [style=\commandkey{style2}] (0) at (0, -0.75) {$#2$};
    \node [style=\commandkey{style3}] (1) at (0, 0.75) {$#3$};
    \node [style=\commandkey{style1}] (2) at (0, -2) {$#1$};
    \node [style=\commandkey{style4}] (3) at (0, 2) {$#4$};
  \end{pgfonlayer}
  \begin{pgfonlayer}{edgelayer}
    \draw (2) to (0);
    \draw (0) to (1);
    \draw (1) to (3);
  \end{pgfonlayer}
\end{tikzpicture}\,}

\newkeycommand{\longbraket}[style1=point,style2=copoint][2]{\,%
\begin{tikzpicture}
  \begin{pgfonlayer}{nodelayer}
    \node [style=\commandkey{style1}] (0) at (0, -2) {$#1$};
    \node [style=\commandkey{style2}] (1) at (0, 2) {$#2$};
  \end{pgfonlayer}
  \begin{pgfonlayer}{edgelayer}
    \draw (0) to (1);
  \end{pgfonlayer}
\end{tikzpicture}\,}

\newkeycommand{\onb}[style=point]{\ensuremath{\left\{\tikz{\node[style=\commandkey{style}] (x) at (0,-0.3) {$j$};\draw (x)--(0,0.7);}\right\}}\xspace}

\newkeycommand{\copointmap}[style=copoint][1]{\,\tikz{\node[style=\commandkey{style}] (x) at (0,0.3) {$#1$};\draw (x)--(0,-0.7);}\,}

\tikzstyle{cdiag}=[matrix of math nodes, row sep=3em, column sep=3em, text height=1.5ex, text depth=0.25ex,inner sep=0.5em]
\tikzstyle{arrow above}=[transform canvas={yshift=0.5ex}]
\tikzstyle{arrow below}=[transform canvas={yshift=-0.5ex}]

\usetikzlibrary{shapes.multipart}

\tikzstyle{bwSpider}=[
       rectangle split,
       rectangle split parts=2,
       rectangle split part fill={black,white},
 minimum size=3.6 mm, inner sep=-2mm, draw=black,scale=0.5,rounded corners=0.8 mm
       ]
 \tikzstyle{wbSpider}=[
       rectangle split,
       rectangle split parts=2,
       rectangle split part fill={white,black},
 minimum size=3.6 mm, inner sep=-2mm, draw=black,scale=0.5,rounded corners=0.8 mm
       ]
\tikzstyle{qWire}=[line width = .5pt, color=black, double distance=1pt]
\tikzstyle{hyperbit}=[line width = 2pt, color=black]
\tikzstyle{pWire}=[line width = 1pt, color=red!60!black]
\tikzstyle{gWire}=[line width = 1.5pt, color=green!60!black]
\tikzstyle{cWire}=[color=blue,thin]
\tikzstyle{env}=[copoint,regular polygon rotate=0,minimum width=0.2cm, fill=black]

\tikzstyle{probs}=[shape=semicircle,fill=white,draw=black,shape border rotate=180,minimum width=1.2cm]

\tikzstyle{every picture}=[baseline=-0.25em,scale=0.5]
\tikzstyle{dotpic}=[] 
\tikzstyle{diredges}=[every to/.style={diredge}]
\tikzstyle{math matrix}=[matrix of math nodes,left delimiter=(,right delimiter=),inner sep=2pt,column sep=1em,row sep=0.5em,nodes={inner sep=0pt},text height=1.5ex, text depth=0.25ex]

\tikzstyle{inline text}=[text height=1.5ex, text depth=0.25ex,yshift=0.5mm]
\tikzstyle{label}=[font=\footnotesize,text height=1.5ex, text depth=0.25ex,yshift=0.5mm]
\tikzstyle{left label}=[label,anchor=east,xshift=1.5mm]
\tikzstyle{right label}=[label,anchor=west,xshift=-1mm]

\tikzstyle{braceedge}=[decorate,decoration={brace,amplitude=2mm,raise=-1mm}]
\tikzstyle{small braceedge}=[decorate,decoration={brace,amplitude=1mm,raise=-1mm}]

\tikzstyle{doubled}=[line width=1.6pt] 
\tikzstyle{boldedge}=[doubled,shorten <=-0.17mm,shorten >=-0.17mm]
\tikzstyle{boldedgegray}=[doubled,gray,shorten <=-0.17mm,shorten >=-0.17mm]
\tikzstyle{singleedgegray}=[gray]

\tikzstyle{semidoubled}=[line width=1.4pt] 
\tikzstyle{semiboldedgegray}=[semidoubled,gray,shorten <=-0.17mm,shorten >=-0.17mm]

\tikzstyle{boxedge}=[semiboldedgegray]

\tikzstyle{boldedgedashed}=[very thick,dashed,shorten <=-0.17mm,shorten >=-0.17mm]
\tikzstyle{vboldedgedashed}=[doubled,dashed,shorten <=-0.17mm,shorten >=-0.17mm]
\tikzstyle{left hook arrow}=[left hook-latex]
\tikzstyle{right hook arrow}=[right hook-latex]
\tikzstyle{sembracket}=[line width=0.5pt,shorten <=-0.07mm,shorten >=-0.07mm]

\tikzstyle{causal edge}=[->,thick,gray]
\tikzstyle{causal nondir}=[thick,gray]
\tikzstyle{timeline}=[thick,gray, dashed]

\tikzstyle{cedge}=[<->,thick,gray!70!white]

\tikzstyle{empty diagram}=[draw=gray!40!white,dashed,shape=rectangle,minimum width=1cm,minimum height=1cm]
\tikzstyle{empty diagram small}=[draw=gray!50!white,dashed,shape=rectangle,minimum width=0.6cm,minimum height=0.5cm]

\tikzstyle{dot}=[inner sep=0mm,minimum width=2mm,minimum height=2mm,draw,shape=circle]
\tikzstyle{leak}=[white dot, shape=regular polygon, minimum size=3.3 mm, regular polygon sides=3, outer sep=-0.2mm, regular polygon rotate=270]
\tikzstyle{proj}=[draw,fill=white,chamfered rectangle,chamfered rectangle angle=30, minimum width=2mm, minimum height= 1mm,scale=0.5, outer sep=-0.2mm]
\tikzstyle{wide proj}=[draw,fill=white,chamfered rectangle,chamfered rectangle angle=30, minimum width=15mm,minimum height=1mm,scale=0.5, outer sep=-0.2mm]
\tikzstyle{very wide proj}=[draw,fill=white,chamfered rectangle,chamfered rectangle angle=30, minimum width=25mm,minimum height=1mm,scale=0.5, outer sep=-0.2mm]
\tikzstyle{very very wide proj}=[draw,fill=white,chamfered rectangle,chamfered rectangle angle=30, minimum width=35mm,minimum height=1mm,scale=0.5, outer sep=-0.2mm]
\tikzstyle{preleak}=[proj]

\tikzstyle{split proj out}=[regular polygon,regular polygon sides=3,draw,scale=0.75,inner sep=-0.5pt,minimum width=3.3mm,fill=white,regular polygon rotate=180]
\tikzstyle{split proj in}=[regular polygon,regular polygon sides=3,draw,scale=0.75,inner sep=-0.5pt,minimum width=3.3mm,fill=white]
\tikzstyle{Vleak}=[white dot, shape=regular polygon, minimum size=3.3 mm, regular polygon sides=3, outer sep=-0.2mm, regular polygon rotate=90]
\tikzstyle{dleak}=[white dot, line width=1.6pt, shape=regular polygon, minimum size=3.3 mm, regular polygon sides=3, outer sep=-0.2mm, regular polygon rotate=270]

\tikzstyle{Wsquare}=[white dot, shape=regular polygon, rounded corners=0.8 mm, minimum size=3.3 mm, regular polygon sides=3, outer sep=-0.2mm]
\tikzstyle{Wsquareadj}=[white dot, shape=regular polygon, rounded corners=0.8 mm, minimum size=3.3 mm, regular polygon sides=3, outer sep=-0.2mm, regular polygon rotate=180]
\tikzstyle{ddot}=[inner sep=0mm, doubled, minimum width=2.5mm,minimum height=2.5mm,draw,shape=circle]

\tikzstyle{black dot}=[dot,fill=black]
\tikzstyle{white dot}=[dot,fill=white,,text depth=-0.2mm]
\tikzstyle{white Wsquare}=[Wsquare,fill=gray,,text depth=-0.2mm]
\tikzstyle{white Wsquareadj}=[Wsquareadj,fill=white,,text depth=-0.2mm]
\tikzstyle{green dot}=[white dot] 
\tikzstyle{gray dot}=[dot,fill=gray!40!white,,text depth=-0.2mm]
\tikzstyle{red dot}=[gray dot] 

\tikzstyle{black ddot}=[ddot,fill=black]
\tikzstyle{white ddot}=[ddot,fill=white]
\tikzstyle{gray ddot}=[ddot,fill=gray!40!white]

\tikzstyle{gray edge}=[gray!60!white]

\tikzstyle{small dot}=[inner sep=0.5mm,minimum width=0pt,minimum height=0pt,draw,shape=circle]

\tikzstyle{small black dot}=[small dot,fill=black]
\tikzstyle{small white dot}=[small dot,fill=white]
\tikzstyle{small gray dot}=[small dot,fill=gray!40!white]

\tikzstyle{causal dot}=[inner sep=0.4mm,minimum width=0pt,minimum height=0pt,draw=white,shape=circle,fill=gray!40!white]

\tikzstyle{phase dimensions}=[minimum size=5mm,font=\footnotesize,rectangle,rounded corners=2.5mm,inner sep=0.2mm,outer sep=-2mm]
\tikzstyle{dphase dimensions}=[minimum size=5mm,font=\footnotesize,rectangle,rounded corners=2.5mm,inner sep=0.2mm,outer sep=-2mm]

\tikzstyle{white phase dot}=[dot,fill=white,phase dimensions]
\tikzstyle{white phase ddot}=[ddot,fill=white,dphase dimensions]

\tikzstyle{white rect ddot}=[draw=black,fill=white,doubled,minimum size=5mm,font=\footnotesize,rectangle,rounded corners=2.5mm,inner sep=0.2mm]
\tikzstyle{gray rect ddot}=[draw=black,fill=gray!40!white,doubled,minimum size=6mm,font=\footnotesize,rectangle,rounded corners=3mm]

\tikzstyle{gray phase dot}=[dot,fill=gray!40!white,phase dimensions]
\tikzstyle{gray phase ddot}=[ddot,fill=gray!40!white,dphase dimensions]
\tikzstyle{grey phase dot}=[gray phase dot]
\tikzstyle{grey phase ddot}=[gray phase ddot]

\tikzstyle{small phase dimensions}=[minimum size=4mm,font=\tiny,rectangle,rounded corners=2mm,inner sep=0.2mm,outer sep=-2mm]
\tikzstyle{small dphase dimensions}=[minimum size=4mm,font=\tiny,rectangle,rounded corners=2mm,inner sep=0.2mm,outer sep=-2mm]

\tikzstyle{small gray phase dot}=[dot,fill=gray!40!white,small phase dimensions]
\tikzstyle{small gray phase ddot}=[ddot,fill=gray!40!white,small dphase dimensions]

\tikzstyle{small map}=[draw,shape=rectangle,minimum height=4mm,minimum width=4mm,fill=white]

\tikzstyle{cnot}=[fill=white,shape=circle,inner sep=-1.4pt]

\tikzstyle{asym hadamard}=[fill=white,draw,shape=NEbox,inner sep=0.6mm,font=\footnotesize,minimum height=4mm]
\tikzstyle{asym hadamard conj}=[fill=white,draw,shape=NWbox,inner sep=0.6mm,font=\footnotesize,minimum height=4mm]
\tikzstyle{asym hadamard dag}=[fill=white,draw,shape=SEbox,inner sep=0.6mm,font=\footnotesize,minimum height=4mm]

\tikzstyle{hadamard}=[fill=white,draw,inner sep=0.6mm,font=\footnotesize,minimum height=4mm,minimum width=4mm]
\tikzstyle{small hadamard}=[fill=white,draw,inner sep=0.6mm,minimum height=1.5mm,minimum width=1.5mm]
\tikzstyle{small hadamard rotate}=[small hadamard,rotate=45]
\tikzstyle{dhadamard}=[hadamard,doubled]
\tikzstyle{small dhadamard}=[small hadamard,doubled]
\tikzstyle{small dhadamard rotate}=[small hadamard rotate,doubled]
\tikzstyle{antipode}=[white dot,inner sep=0.3mm,font=\footnotesize]

\tikzstyle{scalar}=[diamond,draw,inner sep=0.5pt,font=\small]
\tikzstyle{dscalar}=[diamond,doubled, draw,inner sep=0.5pt,font=\small]

\tikzstyle{small box}=[rectangle,inline text,fill=white,draw,minimum height=5mm,yshift=-0.5mm,minimum width=5mm,font=\small]
\tikzstyle{small gray box}=[small box,fill=gray!30]
\tikzstyle{medium box}=[rectangle,inline text,fill=white,draw,minimum height=5mm,yshift=-0.5mm,minimum width=10mm,font=\small]
\tikzstyle{square box}=[small box] 
\tikzstyle{medium gray box}=[small box,fill=gray!30]
\tikzstyle{semilarge box}=[rectangle,inline text,fill=white,draw,minimum height=5mm,yshift=-0.5mm,minimum width=12.5mm,font=\small]
\tikzstyle{large box}=[rectangle,inline text,fill=white,draw,minimum height=5mm,yshift=-0.5mm,minimum width=15mm,font=\small]
\tikzstyle{large gray box}=[small box,fill=gray!30]

\tikzstyle{Bayes box}=[rectangle,fill=black,draw, minimum height=3mm, minimum width=3mm]

\tikzstyle{gray square point}=[small box,fill=gray!50]

\tikzstyle{dphase box white}=[dhadamard]
\tikzstyle{dphase box gray}=[dhadamard,fill=gray!50!white]
\tikzstyle{phase box white}=[hadamard]
\tikzstyle{phase box gray}=[hadamard,fill=gray!50!white]

\tikzstyle{point}=[regular polygon,regular polygon sides=3,draw,scale=0.75,inner sep=-0.5pt,minimum width=9mm,fill=white,regular polygon rotate=180]
\tikzstyle{point nosep}=[regular polygon,regular polygon sides=3,draw,scale=0.75,inner sep=-2pt,minimum width=9mm,fill=white,regular polygon rotate=180]
\tikzstyle{copoint}=[regular polygon,regular polygon sides=3,draw,scale=0.75,inner sep=-0.5pt,minimum width=9mm,fill=white]
\tikzstyle{dpoint}=[point,doubled]
\tikzstyle{dcopoint}=[copoint,doubled]

\tikzstyle{pointgrow}=[shape=cornerpoint,kpoint common,scale=0.75,inner sep=3pt]
\tikzstyle{pointgrow dag}=[shape=cornercopoint,kpoint common,scale=0.75,inner sep=3pt]

\tikzstyle{wide copoint}=[fill=white,draw,shape=isosceles triangle,shape border rotate=90,isosceles triangle stretches=true,inner sep=0pt,minimum width=1.5cm,minimum height=6.12mm]
\tikzstyle{wide point}=[fill=white,draw,shape=isosceles triangle,shape border rotate=-90,isosceles triangle stretches=true,inner sep=0pt,minimum width=1.5cm,minimum height=6.12mm,yshift=-0.0mm]
\tikzstyle{wide point plus}=[fill=white,draw,shape=isosceles triangle,shape border rotate=-90,isosceles triangle stretches=true,inner sep=0pt,minimum width=1.74cm,minimum height=7mm,yshift=-0.0mm]

\tikzstyle{wide dpoint}=[fill=white,doubled,draw,shape=isosceles triangle,shape border rotate=-90,isosceles triangle stretches=true,inner sep=0pt,minimum width=1.5cm,minimum height=6.12mm,yshift=-0.0mm]

\tikzstyle{tinypoint}=[regular polygon,regular polygon sides=3,draw,scale=0.55,inner sep=-0.15pt,minimum width=6mm,fill=white,regular polygon rotate=180]

\tikzstyle{white point}=[point]
\tikzstyle{white dpoint}=[dpoint]
\tikzstyle{green point}=[white point] 
\tikzstyle{white copoint}=[copoint]
\tikzstyle{gray point}=[point,fill=gray!40!white]
\tikzstyle{gray dpoint}=[gray point,doubled]
\tikzstyle{red point}=[gray point] 
\tikzstyle{gray copoint}=[copoint,fill=gray!40!white]
\tikzstyle{gray dcopoint}=[gray copoint,doubled]

\tikzstyle{white point guide}=[regular polygon,regular polygon sides=3,font=\scriptsize,draw,scale=0.65,inner sep=-0.5pt,minimum width=9mm,fill=white,regular polygon rotate=180]

\tikzstyle{black point}=[point,fill=black,font=\color{white}]
\tikzstyle{black copoint}=[copoint,fill=black,font=\color{white}]

\tikzstyle{tiny gray point}=[tinypoint,fill=gray!40!white]

\tikzstyle{diredge}=[->]
\tikzstyle{ddiredge}=[<->]
\tikzstyle{rdiredge}=[<-]
\tikzstyle{thickdiredge}=[->, very thick]
\tikzstyle{pointer edge}=[->,very thick,gray]
\tikzstyle{pointer edge part}=[very thick,gray]
\tikzstyle{dashed edge}=[dashed]
\tikzstyle{thick dashed edge}=[very thick,dashed]
\tikzstyle{thick gray dashed edge}=[thick dashed edge,gray!40]
\tikzstyle{thick map edge}=[very thick,|->]

\makeatletter
\newcommand{\boxshape}[3]{%
\pgfdeclareshape{#1}{
\inheritsavedanchors[from=rectangle] 
\inheritanchorborder[from=rectangle]
\inheritanchor[from=rectangle]{center}
\inheritanchor[from=rectangle]{north}
\inheritanchor[from=rectangle]{south}
\inheritanchor[from=rectangle]{west}
\inheritanchor[from=rectangle]{east}
\backgroundpath{
\southwest \pgf@xa=\pgf@x \pgf@ya=\pgf@y
\northeast \pgf@xb=\pgf@x \pgf@yb=\pgf@y

\@tempdima=#2
\@tempdimb=#3

\pgfpathmoveto{\pgfpoint{\pgf@xa - 5pt + \@tempdima}{\pgf@ya}}
\pgfpathlineto{\pgfpoint{\pgf@xa - 5pt - \@tempdima}{\pgf@yb}}
\pgfpathlineto{\pgfpoint{\pgf@xb + 5pt + \@tempdimb}{\pgf@yb}}
\pgfpathlineto{\pgfpoint{\pgf@xb + 5pt - \@tempdimb}{\pgf@ya}}
\pgfpathlineto{\pgfpoint{\pgf@xa - 5pt + \@tempdima}{\pgf@ya}}
\pgfpathclose
}
}}

\boxshape{NEbox}{0pt}{3pt}
\boxshape{SEbox}{0pt}{-3pt}
\boxshape{NWbox}{5pt}{0pt}
\boxshape{SWbox}{-5pt}{0pt}
\boxshape{EBox}{-3pt}{3pt}
\boxshape{WBox}{3pt}{-3pt}
\makeatother

\tikzstyle{cloud}=[shape=cloud,draw,minimum width=1.5cm,minimum height=1.5cm]

\tikzstyle{map}=[draw,shape=NEbox,inner sep=1pt,minimum height=4mm,fill=white]
\tikzstyle{dashedmap}=[draw,dashed,shape=NEbox,inner sep=2pt,minimum height=6mm,fill=white]
\tikzstyle{mapdag}=[draw,shape=SEbox,inner sep=1pt,minimum height=4mm,fill=white]
\tikzstyle{mapadj}=[draw,shape=SEbox,inner sep=2pt,minimum height=6mm,fill=white]
\tikzstyle{maptrans}=[draw,shape=SWbox,inner sep=2pt,minimum height=6mm,fill=white]
\tikzstyle{mapconj}=[draw,shape=NWbox,inner sep=2pt,minimum height=6mm,fill=white]

\tikzstyle{medium map}=[draw,shape=NEbox,inner sep=2pt,minimum height=6mm,fill=white,minimum width=7mm]
\tikzstyle{medium map dag}=[draw,shape=SEbox,inner sep=2pt,minimum height=6mm,fill=white,minimum width=7mm]
\tikzstyle{medium map adj}=[draw,shape=SEbox,inner sep=2pt,minimum height=6mm,fill=white,minimum width=7mm]
\tikzstyle{medium map trans}=[draw,shape=SWbox,inner sep=2pt,minimum height=6mm,fill=white,minimum width=7mm]
\tikzstyle{medium map conj}=[draw,shape=NWbox,inner sep=2pt,minimum height=6mm,fill=white,minimum width=7mm]
\tikzstyle{semilarge map}=[draw,shape=NEbox,inner sep=2pt,minimum height=6mm,fill=white,minimum width=9.5mm]
\tikzstyle{semilarge map trans}=[draw,shape=SWbox,inner sep=2pt,minimum height=6mm,fill=white,minimum width=9.5mm]
\tikzstyle{semilarge map adj}=[draw,shape=SEbox,inner sep=2pt,minimum height=6mm,fill=white,minimum width=9.5mm]
\tikzstyle{semilarge map dag}=[draw,shape=SEbox,inner sep=2pt,minimum height=6mm,fill=white,minimum width=9.5mm]
\tikzstyle{semilarge map conj}=[draw,shape=NWbox,inner sep=2pt,minimum height=6mm,fill=white,minimum width=9.5mm]
\tikzstyle{large map}=[draw,shape=NEbox,inner sep=2pt,minimum height=6mm,fill=white,minimum width=12mm]
\tikzstyle{large map conj}=[draw,shape=NWbox,inner sep=2pt,minimum height=6mm,fill=white,minimum width=12mm]
\tikzstyle{very large map}=[draw,shape=NEbox,inner sep=2pt,minimum height=6mm,fill=white,minimum width=17mm]

\tikzstyle{medium dmap}=[draw,doubled,shape=NEbox,inner sep=2pt,minimum height=6mm,fill=white,minimum width=7mm]
\tikzstyle{medium dmap dag}=[draw,doubled,shape=SEbox,inner sep=2pt,minimum height=6mm,fill=white,minimum width=7mm]
\tikzstyle{medium dmap adj}=[draw,doubled,shape=SEbox,inner sep=2pt,minimum height=6mm,fill=white,minimum width=7mm]
\tikzstyle{medium dmap trans}=[draw,doubled,shape=SWbox,inner sep=2pt,minimum height=6mm,fill=white,minimum width=7mm]
\tikzstyle{medium dmap conj}=[draw,doubled,shape=NWbox,inner sep=2pt,minimum height=6mm,fill=white,minimum width=7mm]
\tikzstyle{semilarge dmap}=[draw,doubled,shape=NEbox,inner sep=2pt,minimum height=6mm,fill=white,minimum width=9.5mm]
\tikzstyle{semilarge dmap trans}=[draw,doubled,shape=SWbox,inner sep=2pt,minimum height=6mm,fill=white,minimum width=9.5mm]
\tikzstyle{semilarge dmap adj}=[draw,doubled,shape=SEbox,inner sep=2pt,minimum height=6mm,fill=white,minimum width=9.5mm]
\tikzstyle{semilarge dmap dag}=[draw,doubled,shape=SEbox,inner sep=2pt,minimum height=6mm,fill=white,minimum width=9.5mm]
\tikzstyle{semilarge dmap conj}=[draw,doubled,shape=NWbox,inner sep=2pt,minimum height=6mm,fill=white,minimum width=9.5mm]
\tikzstyle{large dmap}=[draw,doubled,shape=NEbox,inner sep=2pt,minimum height=6mm,fill=white,minimum width=12mm]
\tikzstyle{large dmap conj}=[draw,doubled,shape=NWbox,inner sep=2pt,minimum height=6mm,fill=white,minimum width=12mm]
\tikzstyle{large dmap trans}=[draw,doubled,shape=SWbox,inner sep=2pt,minimum height=6mm,fill=white,minimum width=12mm]
\tikzstyle{large dmap adj}=[draw,doubled,shape=SEbox,inner sep=2pt,minimum height=6mm,fill=white,minimum width=12mm]
\tikzstyle{large dmap dag}=[draw,doubled,shape=SEbox,inner sep=2pt,minimum height=6mm,fill=white,minimum width=12mm]
\tikzstyle{very large dmap}=[draw,doubled,shape=NEbox,inner sep=2pt,minimum height=6mm,fill=white,minimum width=19.5mm]

\tikzstyle{muxbox}=[draw,shape=rectangle,minimum height=3mm,minimum width=3mm,fill=white]
\tikzstyle{dmuxbox}=[muxbox,doubled]

\tikzstyle{box}=[draw,shape=rectangle,inner sep=2pt,minimum height=6mm,minimum width=6mm,fill=white]
\tikzstyle{dbox}=[draw,doubled,shape=rectangle,inner sep=2pt,minimum height=6mm,minimum width=6mm,fill=white]
\tikzstyle{dmap}=[draw,doubled,shape=NEbox,inner sep=2pt,minimum height=6mm,fill=white]
\tikzstyle{dmapdag}=[draw,doubled,shape=SEbox,inner sep=2pt,minimum height=6mm,fill=white]
\tikzstyle{dmapadj}=[draw,doubled,shape=SEbox,inner sep=2pt,minimum height=6mm,fill=white]
\tikzstyle{dmaptrans}=[draw,doubled,shape=SWbox,inner sep=2pt,minimum height=6mm,fill=white]
\tikzstyle{dmapconj}=[draw,doubled,shape=NWbox,inner sep=2pt,minimum height=6mm,fill=white]

\tikzstyle{ddmap}=[draw,doubled,dashed,shape=NEbox,inner sep=2pt,minimum height=6mm,fill=white]
\tikzstyle{ddmapdag}=[draw,doubled,dashed,shape=SEbox,inner sep=2pt,minimum height=6mm,fill=white]
\tikzstyle{ddmapadj}=[draw,doubled,dashed,shape=SEbox,inner sep=2pt,minimum height=6mm,fill=white]
\tikzstyle{ddmaptrans}=[draw,doubled,dashed,shape=SWbox,inner sep=2pt,minimum height=6mm,fill=white]
\tikzstyle{ddmapconj}=[draw,doubled,dashed,shape=NWbox,inner sep=2pt,minimum height=6mm,fill=white]

\boxshape{sNEbox}{0pt}{3pt}
\boxshape{sSEbox}{0pt}{-3pt}
\boxshape{sNWbox}{3pt}{0pt}
\boxshape{sSWbox}{-3pt}{0pt}
\tikzstyle{smap}=[draw,shape=sNEbox,fill=white]
\tikzstyle{smapdag}=[draw,shape=sSEbox,fill=white]
\tikzstyle{smapadj}=[draw,shape=sSEbox,fill=white]
\tikzstyle{smaptrans}=[draw,shape=sSWbox,fill=white]
\tikzstyle{smapconj}=[draw,shape=sNWbox,fill=white]

\tikzstyle{dsmap}=[draw,dashed,shape=sNEbox,fill=white]
\tikzstyle{dsmapdag}=[draw,dashed,shape=sSEbox,fill=white]
\tikzstyle{dsmaptrans}=[draw,dashed,shape=sSWbox,fill=white]
\tikzstyle{dsmapconj}=[draw,dashed,shape=sNWbox,fill=white]

\boxshape{mNEbox}{0pt}{10pt}
\boxshape{mSEbox}{0pt}{-10pt}
\boxshape{mNWbox}{10pt}{0pt}
\boxshape{mSWbox}{-10pt}{0pt}
\tikzstyle{mmap}=[draw,shape=mNEbox]
\tikzstyle{mmapdag}=[draw,shape=mSEbox]
\tikzstyle{mmaptrans}=[draw,shape=mSWbox]
\tikzstyle{mmapconj}=[draw,shape=mNWbox]

\tikzstyle{mmapgray}=[draw,fill=gray!40!white,shape=mNEbox]
\tikzstyle{smapgray}=[draw,fill=gray!40!white,shape=sNEbox]

\makeatletter

\pgfdeclareshape{cornerpoint}{
\inheritsavedanchors[from=rectangle] 
\inheritanchorborder[from=rectangle]
\inheritanchor[from=rectangle]{center}
\inheritanchor[from=rectangle]{north}
\inheritanchor[from=rectangle]{south}
\inheritanchor[from=rectangle]{west}
\inheritanchor[from=rectangle]{east}
\backgroundpath{
\southwest \pgf@xa=\pgf@x \pgf@ya=\pgf@y
\northeast \pgf@xb=\pgf@x \pgf@yb=\pgf@y

\pgfmathsetmacro{\pgf@shorten@left}{\pgfkeysvalueof{/tikz/shorten left}}
\pgfmathsetmacro{\pgf@shorten@right}{\pgfkeysvalueof{/tikz/shorten right}}

\pgfpathmoveto{\pgfpoint{0.5 * (\pgf@xa + \pgf@xb)}{\pgf@ya - 5pt}}
\pgfpathlineto{\pgfpoint{\pgf@xa - 8pt + \pgf@shorten@left}{\pgf@yb - 1.5 * \pgf@shorten@left}}
\pgfpathlineto{\pgfpoint{\pgf@xa - 8pt + \pgf@shorten@left}{\pgf@yb}}
\pgfpathlineto{\pgfpoint{\pgf@xb + 8pt - \pgf@shorten@right}{\pgf@yb}}
\pgfpathlineto{\pgfpoint{\pgf@xb + 8pt - \pgf@shorten@right}{\pgf@yb - 1.5 * \pgf@shorten@right}}
\pgfpathclose
}
}

\pgfdeclareshape{cornercopoint}{
\inheritsavedanchors[from=rectangle] 
\inheritanchorborder[from=rectangle]
\inheritanchor[from=rectangle]{center}
\inheritanchor[from=rectangle]{north}
\inheritanchor[from=rectangle]{south}
\inheritanchor[from=rectangle]{west}
\inheritanchor[from=rectangle]{east}
\backgroundpath{
\southwest \pgf@xa=\pgf@x \pgf@ya=\pgf@y
\northeast \pgf@xb=\pgf@x \pgf@yb=\pgf@y

\pgfmathsetmacro{\pgf@shorten@left}{\pgfkeysvalueof{/tikz/shorten left}}
\pgfmathsetmacro{\pgf@shorten@right}{\pgfkeysvalueof{/tikz/shorten right}}

\pgfpathmoveto{\pgfpoint{0.5 * (\pgf@xa + \pgf@xb)}{\pgf@yb + 5pt}}
\pgfpathlineto{\pgfpoint{\pgf@xa - 8pt + \pgf@shorten@left}{\pgf@ya + 1.5 * \pgf@shorten@left}}
\pgfpathlineto{\pgfpoint{\pgf@xa - 8pt + \pgf@shorten@left}{\pgf@ya}}
\pgfpathlineto{\pgfpoint{\pgf@xb + 8pt - \pgf@shorten@right}{\pgf@ya}}
\pgfpathlineto{\pgfpoint{\pgf@xb + 8pt - \pgf@shorten@right}{\pgf@ya + 1.5 * \pgf@shorten@right}}
\pgfpathclose
}
}

\makeatother

\pgfkeyssetvalue{/tikz/shorten left}{0pt}
\pgfkeyssetvalue{/tikz/shorten right}{0pt}

\tikzstyle{kpoint common}=[draw,fill=white,inner sep=1pt,minimum height=4mm]
\tikzstyle{kpoint sc}=[shape=cornerpoint,kpoint common]
\tikzstyle{kpoint adjoint sc}=[shape=cornercopoint,kpoint common]
\tikzstyle{kpoint}=[shape=cornerpoint,shorten left=5pt,kpoint common]
\tikzstyle{kpoint adjoint}=[shape=cornercopoint,shorten left=5pt,kpoint common]
\tikzstyle{kpoint conjugate}=[shape=cornerpoint,shorten right=5pt,kpoint common]
\tikzstyle{kpoint transpose}=[shape=cornercopoint,shorten right=5pt,kpoint common]
\tikzstyle{kpoint symm}=[shape=cornerpoint,shorten left=5pt,shorten right=5pt,kpoint common]

\tikzstyle{wide kpoint sc}=[shape=cornerpoint,kpoint common, minimum width=1 cm]
\tikzstyle{wide kpointdag sc}=[shape=cornercopoint,kpoint common, minimum width=1 cm]

\tikzstyle{black kpoint}=[shape=cornerpoint,shorten left=5pt,kpoint common,fill=black,font=\color{white}]

\tikzstyle{black kpoint sm}=[shape=cornerpoint,shorten left=5pt,kpoint common,fill=black,font=\color{white},scale=0.75]

\tikzstyle{black kpoint adjoint}=[shape=cornercopoint,shorten left=5pt,kpoint common,fill=black,font=\color{white}]
\tikzstyle{black kpointadj}=[shape=cornercopoint,shorten left=5pt,kpoint common,fill=black,font=\color{white}]

\tikzstyle{black kpointadj sm}=[shape=cornercopoint,shorten left=5pt,kpoint common,fill=black,font=\color{white},scale=0.75]

\tikzstyle{black dkpoint}=[shape=cornerpoint,shorten left=5pt,kpoint common,fill=black, doubled,font=\color{white}]
\tikzstyle{black dkpoint adjoint}=[shape=cornercopoint,shorten left=5pt,kpoint common,fill=black, doubled,font=\color{white}]
\tikzstyle{black dkpointadj}=[shape=cornercopoint,shorten left=5pt,kpoint common,fill=black, doubled,font=\color{white}]

\tikzstyle{black dkpoint sm}=[shape=cornerpoint,shorten left=5pt,kpoint common,fill=black, doubled,font=\color{white},scale=0.75]
\tikzstyle{black dkpointadj sm}=[shape=cornercopoint,shorten left=5pt,kpoint common,fill=black, doubled,font=\color{white},scale=0.75]

\tikzstyle{kpointdag}=[kpoint adjoint]
\tikzstyle{kpointadj}=[kpoint adjoint]
\tikzstyle{kpointconj}=[kpoint conjugate]
\tikzstyle{kpointtrans}=[kpoint transpose]

\tikzstyle{big kpoint}=[kpoint, minimum width=1.2 cm, minimum height=8mm, inner sep=4pt, text depth=3mm]

\tikzstyle{wide kpoint}=[kpoint, minimum width=1 cm, inner sep=2pt]
\tikzstyle{wide kpointdag}=[kpointdag, minimum width=1 cm, inner sep=2pt]
\tikzstyle{wide kpointconj}=[kpointconj, minimum width=1 cm, inner sep=2pt]
\tikzstyle{wide kpointtrans}=[kpointtrans, minimum width=1 cm, inner sep=2pt]

\tikzstyle{wider kpoint}=[kpoint, minimum width=1.25 cm, inner sep=2pt]
\tikzstyle{wider kpointdag}=[kpointdag, minimum width=1.25 cm, inner sep=2pt]
\tikzstyle{wider kpointconj}=[kpointconj, minimum width=1.25 cm, inner sep=2pt]
\tikzstyle{wider kpointtrans}=[kpointtrans, minimum width=1.25 cm, inner sep=2pt]

\tikzstyle{gray kpoint}=[kpoint,fill=gray!50!white]
\tikzstyle{gray kpointdag}=[kpointdag,fill=gray!50!white]
\tikzstyle{gray kpointadj}=[kpointadj,fill=gray!50!white]
\tikzstyle{gray kpointconj}=[kpointconj,fill=gray!50!white]
\tikzstyle{gray kpointtrans}=[kpointtrans,fill=gray!50!white]

\tikzstyle{gray dkpoint}=[kpoint,fill=gray!50!white,doubled]
\tikzstyle{gray dkpointdag}=[kpointdag,fill=gray!50!white,doubled]
\tikzstyle{gray dkpointadj}=[kpointadj,fill=gray!50!white,doubled]
\tikzstyle{gray dkpointconj}=[kpointconj,fill=gray!50!white,doubled]
\tikzstyle{gray dkpointtrans}=[kpointtrans,fill=gray!50!white,doubled]

\tikzstyle{white label}=[draw,fill=white,rectangle,inner sep=0.7 mm]
\tikzstyle{gray label}=[draw,fill=gray!50!white,rectangle,inner sep=0.7 mm]
\tikzstyle{black label}=[draw,fill=black,rectangle,inner sep=0.7 mm]

\tikzstyle{dkpoint}=[kpoint,doubled]
\tikzstyle{wide dkpoint}=[wide kpoint,doubled]
\tikzstyle{dkpointdag}=[kpoint adjoint,doubled]
\tikzstyle{wide dkpointdag}=[wide kpointdag,doubled]
\tikzstyle{dkcopoint}=[kpoint adjoint,doubled]
\tikzstyle{dkpointadj}=[kpoint adjoint,doubled]
\tikzstyle{dkpointconj}=[kpoint conjugate,doubled]
\tikzstyle{dkpointtrans}=[kpoint transpose,doubled]

\tikzstyle{kscalar}=[kpoint common, shape=EBox, inner xsep=-1pt, inner ysep=3pt,font=\small]
\tikzstyle{kscalarconj}=[kpoint common, shape=WBox, inner xsep=-1pt, inner ysep=3pt,font=\small]

\tikzstyle{spekpoint}=[kpoint sc,minimum height=5mm,inner sep=3pt]
\tikzstyle{spekcopoint}=[kpoint adjoint sc,minimum height=5mm,inner sep=3pt]

\tikzstyle{dspekpoint}=[spekpoint,doubled]
\tikzstyle{dspekcopoint}=[spekcopoint,doubled]

 \tikzstyle{upground}=[circuit ee IEC,thick,ground,rotate=90,scale=2.5]
 \tikzstyle{downground}=[circuit ee IEC,thick,ground,rotate=-90,scale=2.5]
 \tikzstyle{bigground}=[regular polygon,regular polygon sides=3,draw=gray,scale=0.50,inner sep=-0.5pt,minimum width=10mm,fill=gray]

\tikzstyle{arrs}=[-latex,font=\small,auto]
\tikzstyle{arrow plain}=[arrs]
\tikzstyle{arrow dashed}=[dashed,arrs]
\tikzstyle{arrow bold}=[very thick,arrs]
\tikzstyle{arrow hide}=[draw=white!0,-]
\tikzstyle{arrow reverse}=[latex-]
\tikzstyle{cdnode}=[]

\let\olddagger\dagger
\renewcommand{\dagger}{\ensuremath{\olddagger}\xspace}

\theoremstyle{definition}
\newtheorem{theorem}{Theorem}[section]

\newtheorem{proposition}[theorem]{Proposition}

\newtheorem{example*}[theorem]{Example*}
\newtheorem{examples*}[theorem]{Examples*}

\newtheorem{remark*}[theorem]{Remark*}

\theoremstyle{plain}

\usepackage{color}
\def\bR{\begin{color}{red}}
\def\bB{\begin{color}{blue}}
\def\bM{\begin{color}{magenta}}
\def\bC{\begin{color}{cyan}}
\def\bW{\begin{color}{white}}
\def\bBl{\begin{color}{black}}
\def\bG{\begin{color}{green}}
\def\bY{\begin{color}{yellow}}
\def\e{\end{color}\xspace}
\newcommand{\bit}{\begin{itemize}}
\newcommand{\eit}{\end{itemize}\par\noindent}
\newcommand{\ben}{\begin{enumerate}}
\newcommand{\een}{\end{enumerate}\par\noindent}
\newcommand{\beq}{\begin{equation}}
\newcommand{\eeq}{\end{equation}\par\noindent}
\newcommand{\beqa}{\begin{eqnarray*}}
\newcommand{\eeqa}{\end{eqnarray*}\par\noindent}
\newcommand{\beqn}{\begin{eqnarray}}
\newcommand{\eeqn}{\end{eqnarray}\par\noindent}

\def\jR{\begin{color}{DarkRed}}
\def\jB{\begin{color}{MidnightBlue}}
\def\jM{\begin{color}{magenta}}
\def\jC{\begin{color}{cyan}}
\def\jW{\begin{color}{white}}
\def\jBl{\begin{color}{black}}
\def\jG{\begin{color}{green}}
\def\jY{\begin{color}{yellow}}

\def\cR{\begin{color}{Crimson}}
\def\cB{\begin{color}{cyan}}
\def\cM{\begin{color}{magenta}}
\def\cC{\begin{color}{cyan}}
\def\cW{\begin{color}{white}}
\def\cBl{\begin{color}{black}}
\def\cG{\begin{color}{green}}
\def\cY{\begin{color}{yellow}}

\usepackage{amssymb}
\usepackage{epstopdf}
\usepackage{amsmath,amsthm}	
\usepackage{bm}
\usepackage{etex}
\usepackage{hyperref}
\newcommand{\Pe}{P_{\mathrm{Q}}}
\newcommand{\Ph}{P_{\mathrm{H}}}

\newcounter{procedurecount}
\newcommand{\protocoltitle}{}

\newenvironment{procedurelist}[1]{%
	\renewcommand{\protocoltitle}{#1}%
	\begin{center}%
		\setlength{\parskip}{0pt}%
		\hrule\vspace{2pt}\hrule\vspace{4pt}%
		\emph{\protocoltitle} \\[1ex] 
		\hrule
		\begin{list}{\arabic{procedurecount}.}{%
				\usecounter{procedurecount}%
				\setlength{\leftmargin}{1em}%
				\setlength{\itemindent}{0pt}%
				\setlength{\labelwidth}{0pt}%
				\setlength{\labelsep}{0.5em}%
				\setlength{\itemsep}{0pt}%
				\setlength{\parsep}{0pt}%
			}%
		}{%
		\end{list}%
		\vspace{4pt}\hrule\vspace{2pt}\hrule%
	\end{center}%
}
\begin{document}

        \title{Advantages of quantum communication revealed by the
reexamination of hyperbit theory limitations}
        
	\author{Giovanni Scala}\email{giovanni.scala@ug.edu.pl}
	\affiliation{International Centre for Theory of Quantum Technologies, Jana Ba\.zy\'nskiego 1A, University of Gdansk, 80-309 Gda\'nsk, Poland}
\affiliation{Dipartimento Interateneo di Fisica, Politecnico di Bari, 70126 Bari, Italy}
	
	\author{Seyed Arash Ghoreishi}
	\affiliation{Faculty of Informatics, Masaryk University, Botanická 68a, 602 00 Brno, Czech Republic} 
 \affiliation{Institute of Physics, Slovak Academy of Sciences, Dúbravská Cesta 9, 84511 Bratislava, Slovakia}

	\author{Marcin Paw\l owski}
	\affiliation{International Centre for Theory of Quantum Technologies, Jana Ba\.zy\'nskiego 1A, University of Gdansk, 80-309 Gda\'nsk, Poland}
	
\begin{abstract}

Paw\l owski and Winter's hyperbit theory, proposed in 2012, presented itself as an alternative to quantum theory, suggesting novel ways of redefining entanglement and classical communication paradigms. This research undertakes a meticulous reevaluation of hyperbit theory, uncovering significant operational constraints that question its equivalence with quantum mechanics. Crucially, the supposition that hyperbit theory and quantum theory are equivalent relies on the receiver having unattainable additional knowledge about the sender's laboratory, indicating that the work by Pawlowski and Winter is incorrect. This study accentuates the constraints of hyperbits in information processing and sheds light on the superiority of quantum communication, thereby advancing the investigation at the intersection of classical and quantum communication.

\end{abstract}

\maketitle
\section{Introduction}
Quantum resources can enhance information processing beyond classical limitations. In a communication scenario involving two distant parties, Alice and Bob, whose chosen settings of $\vec{x}$ and $\vec{y}$ produce classical output $s$ and $s^\prime$, then the joint probabilities $p(s,s^\prime|\vec{x},\vec{y})$ characterize the correlations between them. Two types of correlations can be considered: one where Alice and Bob share prior entanglement but do not communicate, and another where they do not share any prior entanglement but communicate directly (in its simplest case, one-way communication from Alice to Bob). The latter case corresponds to the prepare and measure scenario \cite{de2021general,tavakoli2021correlations,pauwels2022entanglement}. By combining these two cases, a more general scenario can be obtained, where Alice and Bob both share an entangled state and communicate with each other. This results in new types of quantum correlation, enabling new protocols like quantum teleportation \cite{bennett1993teleporting} and quantum dense coding \cite{bennett1992communication}. \\
From both a conceptual and applied standpoint it is crucial to examine whether entanglement-assisted classical communication \cite{EA} and another kind of transmission, are equivalent resources in communication tasks. Many quantum communication tasks can be completed using this equivalent and novel scenario \cite{pawlowski2010entanglement,hameedi2017complementarity}.

In 2012, Paw\l owski and Winter in Ref. \cite{Pawlowski2012} introduced a resource equivalent to entanglement and classical communication using the same communication protocols. This resource called \emph{hyperbit}, was introduced as an information quasi particle. Quasi particles are mathematical artifacts useful to describe the effective behavior of complex systems and make predictions by simplifying the calculations for the system as a whole. Examples are, phonons in the collective behavior of atoms, vacancies in solid-state physics, or as mediators, such as plasmons, magnons, polarons, and gluons just to mention a few of them \cite{LightMatter2020,Shen2022,Chandran2023,Woelfle2018,Rivera2020}. \\
The motivation behind the work of Paw\l owski and Winter comes from this background. Information theory, like all other physical theories, has its particles, bits, and qubits, and their quasi-particles are the hyperbits, so called because they are elements of a hyperball. In analogy with quantum theory, where states and measurements are vectors in some Euclidean ball \cite{rouhbakhsh2023geometric}, as it is easy to see for qubits in the Bloch sphere, there are difficulties in imagining how it works in higher dimensions. \\
For higher dimensions, generalized Bloch spheres have been defined \cite{zyczkowski2008quartic,uhlmann1996spheres,kurzynski2016three,sharma2021four,eltschka2021shape}. Within these frameworks, the formula for probabilities of experimental outcomes mirrors that for qubits. Here, the state is denoted by a vector $\vec{v}$, of length $|\vec{v}|$, in a $d$-dimensional Euclidean space, and the measurement outcomes $X=\pm1$ are depicted by the pair of unit vectors $\pm \vec{w}$. The expectation value is given by $ \langle X\rangle=\langle \vec{w},\vec{v}\rangle$. The primary distinction from the Bloch sphere lies in the dimension of the vectors signifying states and measurements. Therefore, a hyperbit corresponds to the vector representing a finite-dimensional system state.\\
On a broader note, such a hyperball embodies the set of preparations, while its dual signifies the measurement set. The choice of a rule for probability computation facilitates the construction of a probabilistic theory within a preparation-measurement scenario.\\

In this article, we discuss the shortcomings of the original theorem mentioned in Ref. \cite{Pawlowski2012}. A counterexample that disrupts the equivalence between hyperbit theory and quantum theory was previously identified in Ref. \cite{tavakoli2021correlations}. Here, we elucidate why that theorem is incorrect by highlighting physical constraints that were not initially considered. Our objective is to demonstrate that the operational limitations restrict the equivalence between hyperbit theory and quantum theory to merely sub-optimal communication protocols.

The paper is organized as follows. In Sec. \ref{sec:Hyperbit_theory} we revisit the aforementioned theorem within the theory of hyperbit (with its operational and diagrammatic procedures in the Appendix \ref{sec:GPTs}). In Sec. \ref{sec:PWprotocol} we show the unimposed physical restrictions of the original theorem. We clarify the limitation of the equivalence in Sec. \ref{sec:result}, subsequently highlighting the limits to use hyperbits in Sec. \ref{sec:proposals}. Finally, we offer a summation of our findings in the conclusions, paving the way for fresh interpretations of quantum superiority in communication tasks.

\section{Hyperbits for communications}\label{sec:Hyperbit_theory}
The main theorem of Ref. \cite{Pawlowski2012} establishes the equivalence of quantum theory and hyperbit theory\footnote{Here we recall hyperbit Theory for simplicity, specifically is only a fragment of a theory, since we describe only the set of preparations and effects, and the rule to compute probabilities without specifies other properties that fully define a theory which are the composition rule and the update state rule after the interaction with the effects.} in communication tasks:
\begin{center}
	\emph{For tasks where Bob gives binary answers,
	sending one hyperbit from Alice to Bob is equivalent to
	sharing any amount of entanglement and sending one classical bit.}
\end{center}
The theorem posits the existence of a theory, referred to as hyperbit theory $\mathcal{H}$, which has the capability to predict event probabilities analogous to Quantum Theory $\mathcal{Q}$, which employs the Born rule.

The hyperbit theory $\mathcal{H}$ finds its application particularly beneficial in scenarios where Alice and Bob, operating in separate laboratories, employ input $n$--bitstrings $\vec{a},\vec{b}\in\{0,1\}^{n}$. Alice utilizes these to choose one from her set of $2^n$ dichotomic measurements. Bob responds with $s\in\{-1,+1\}$ based on his communication with Alice and his input bitstring $\vec{b}$. Therefore, the theorem stipulates that the following expectation value can be achieved by both the theories $\mathcal{Q}$ and $\mathcal{H}$
\begin{equation}\label{eq:B=B}
	\langle B (\vec{a},\vec{b},r,\rho)\rangle_\mathcal{Q}=
	\langle B (\vec{a},\vec{b},r,\vec{x}_{\vec{a}})\rangle_\mathcal{H}
\end{equation}
where $\langle \cdot \rangle_\mathcal{Q}$ stands for the Born rule in $\mathcal{Q}$ and $\langle \cdot \rangle_\mathcal{H}$ is its analog rule in $\mathcal{H}$\footnote{In General Probabilistic Theories (GPTs), specifically within the framework of preparation-measurement scenarios, a fragment of a theory is characterized by its set of states, effects, and the rule employed to determine the probability of obtaining outcomes and expectation values (e.g. $\langle \cdot \rangle_\mathcal{H}$)  as a result of the effects exerted on the states \cite{Mazurek2021,Janotta2014}.}
. Equation \eqref{eq:B=B} yields
\begin{equation}\label{eq:sum=sum}
	\sum_{s\in\{-1,+1\}}s 	\Pe(s|\vec{a},\vec{b},r,\rho)=
	\sum_{s\in\{-1,+1\}}s \Ph(s|\vec{a},\vec{b},r\vec{x}_{\vec{a}})
\end{equation}
where $r$ in $\mathcal{H}$ is an unbiased dichotomic random variable shared by Alice and Bob, but in $\mathcal{Q}$ it corresponds to the classical bit of Alice's outcome that she sends to Bob
\footnote{Although $r$ serves different roles in the two theories---as Alice's output in quantum theory and as shared randomness in the hyperbit theory---this distinction will facilitate the introduction of a potential equivalence between the hyperbit and quantum theories. Then, the narrative in Eqs. \eqref{eq:diff} and \eqref{eq:Qresult} will be simplified.}
; $\vec{x}_{\vec{a}}$ is her prepared hyperbit which depends on the input bitstring $\vec{a}$ multiplied by $r$.\footnote{Notice that, in $\mathcal{H}$, $r$ is a free-cost random variable with vanishing expectation. Here, a copied random variable with vanishing expectation value does not carry on any information from Alice to Bob.} 

Defined by Eq. \eqref{eq:sum=sum}, we can say that $\mathcal{Q}$ is \emph{weak} or \emph{distributionally} equivalent to $\mathcal{H}$, i.e.,
\begin{equation}\label{PQ=PH}
	\Pe(s|\vec{a},\vec{b},r,\rho)\stackrel{w}{\simeq} \Ph(s|\vec{a},\vec{b},r\vec{x}_{\vec{a}}).
\end{equation}
The left-hand side (LHS) calculates the probability according to the Born rule. Meanwhile, the right-hand side (RHS) signifies the probability rule we intend to define for the hyperbit theory. Regardless of this rule's specifics, the theorem restated in Eqs. \eqref{eq:B=B}, \eqref{eq:sum=sum}, and \eqref{PQ=PH} can be expressed also via \emph{Penrose's graphical notation} \cite{coecke2018picturing,penrose1971applications},
	\begin{equation}\label{eq:equivalenceQH}
\begin{tikzpicture}
	\begin{pgfonlayer}{nodelayer}
		\node [style=none] (0) at (-0.7, 0) {};
		\node [style=none] (1) at (0.7, 0) {};
		\node [style=none] (2) at (0, -0.9) {};
		\node [style=none] (4) at (0, -0.35) {$\rho$};
		\node [style=none] (5) at (-0.5, 0) {};
		\node [style=none] (6) at (-0.5, 1) {};
		\node [style=none] (7) at (-2, 2) {};
		\node [style=none] (8) at (0, 2) {};
		\node [style=none] (9) at (-2, 1) {};
		\node [style=none] (10) at (0, 1) {};
		\node [style=none] (11) at (0.5, 2.75) {};
		\node [style=none] (12) at (0.5, 0) {};
		\node [style=none] (14) at (-1, 3.75) {};
		\node [style=none] (15) at (1.5, 3.75) {};
		\node [style=none] (16) at (-1, 2.75) {};
		\node [style=none] (17) at (1.5, 2.75) {};
		\node [style=none] (18) at (-0.5, 2) {};
		\node [style=none] (19) at (-0.5, 2.75) {};
		\node [style=none] (20) at (-0.25, 4.25) {};
		\node [style=none] (21) at (0.8, 4.25) {};
		\node [style=none] (22) at (0.25, 5) {};
		\node [style=none] (23) at (0.25, 4.25) {};
		\node [style=none] (24) at (0.25, 3.75) {};
		\node [style=none] (25) at (-1.025, 1.5) {\Alice};
		\node [style=none] (26) at (0.25, 3.25) {\Bob};
		\node [style=none] (27) at (0.25, 4.5) {$s$};
		\node [style=none] (28) at (0.75, 0) {};
		\node [style=none] (29) at (1.9, 0) {};
		\node [style=none] (30) at (1.325, -0.925) {};
		\node [style=none] (34) at (-2, 0) {};
		\node [style=none] (35) at (-0.775, 0) {};
		\node [style=none] (36) at (-1.35, -0.8) {};
		\node [style=none] (37) at (-1.3, 0) {};
		\node [style=none] (38) at (-1.3, 1) {};
		\node [style=none] (39) at (1.2, 0) {};
		\node [style=none] (40) at (1.2, 2.75) {};
		\node [style=none] (41) at (1.25, -0.25) {};
		\node [style=none] (42) at (1.35, -0.35) {$\vec{b}$};
		\node [style=none] (43) at (-1.5, -0.25) {};
		\node [style=none] (44) at (-1.35, -0.325) {$\vec{a}$};
		\node [style=none] (45) at (-0.85, 2.35) {$r$};
	\end{pgfonlayer}
	\begin{pgfonlayer}{edgelayer}
		\draw (0.center) to (1.center);
		\draw (0.center) to (2.center);
		\draw (1.center) to (2.center);
		\draw [style=qWire] (5.center) to (6.center);
		\draw (7.center) to (8.center);
		\draw (7.center) to (9.center);
		\draw (9.center) to (10.center);
		\draw (10.center) to (8.center);
		\draw [style=qWire] (11.center) to (12.center);
		\draw (14.center) to (15.center);
		\draw (14.center) to (16.center);
		\draw (16.center) to (17.center);
		\draw (17.center) to (15.center);
		\draw (19.center) to (18.center);
		\draw (20.center) to (21.center);
		\draw (20.center) to (22.center);
		\draw (21.center) to (22.center);
		\draw (23.center) to (24.center);
		\draw (28.center) to (29.center);
		\draw (28.center) to (30.center);
		\draw (29.center) to (30.center);
		\draw (34.center) to (35.center);
		\draw (34.center) to (36.center);
		\draw (35.center) to (36.center);
		\draw (37.center) to (38.center);
		\draw (40.center) to (39.center);
	\end{pgfonlayer}
\end{tikzpicture}}\quad\stackrel{w}{\simeq}\quad
\begin{tikzpicture}
	\begin{pgfonlayer}{nodelayer}
		\node [style=none] (0) at (-0.7, 0) {};
		\node [style=none] (1) at (0.7, 0) {};
		\node [style=none] (2) at (0, -0.9) {};
		\node [style=none] (4) at (0, -0.35) {$r$};
		\node [style=none] (5) at (-0.5, 0) {};
		\node [style=none] (6) at (-0.5, 1) {};
		\node [style=none] (7) at (-2, 2) {};
		\node [style=none] (8) at (0, 2) {};
		\node [style=none] (9) at (-2, 1) {};
		\node [style=none] (10) at (0, 1) {};
		\node [style=none] (11) at (0.5, 2.75) {};
		\node [style=none] (12) at (0.5, 0.5) {};
		\node [style=none] (14) at (-1, 3.75) {};
		\node [style=none] (15) at (1.5, 3.75) {};
		\node [style=none] (16) at (-1, 2.75) {};
		\node [style=none] (17) at (1.5, 2.75) {};
		\node [style=none] (18) at (-0.5, 2) {};
		\node [style=none] (19) at (-0.5, 2.75) {};
		\node [style=none] (20) at (-0.25, 4.25) {};
		\node [style=none] (21) at (0.8, 4.25) {};
		\node [style=none] (22) at (0.25, 5) {};
		\node [style=none] (23) at (0.25, 4.25) {};
		\node [style=none] (24) at (0.25, 3.75) {};
		\node [style=none] (25) at (-1.025, 1.5) {\Alice};
		\node [style=none] (26) at (0.25, 3.25) {\Bob};
		\node [style=none] (27) at (0.25, 4.5) {$s$};
		\node [style=none] (28) at (0.75, 0) {};
		\node [style=none] (29) at (1.9, 0) {};
		\node [style=none] (30) at (1.325, -0.925) {};
		\node [style=none] (34) at (-2, 0) {};
		\node [style=none] (35) at (-0.775, 0) {};
		\node [style=none] (36) at (-1.35, -0.8) {};
		\node [style=none] (37) at (-1.3, 0) {};
		\node [style=none] (38) at (-1.3, 1) {};
		\node [style=none] (39) at (1.2, 0) {};
		\node [style=none] (40) at (1.2, 2.75) {};
		\node [style=none] (41) at (1.25, -0.25) {};
		\node [style=none] (42) at (1.35, -0.35) {$\vec{b}$};
		\node [style=none] (43) at (-1.5, -0.25) {};
		\node [style=none] (44) at (-1.35, -0.325) {$\vec{a}$};
		\node [dot] (45) at (-0.5, 0.5) {};
	\end{pgfonlayer}
	\begin{pgfonlayer}{edgelayer}
		\draw (0.center) to (1.center);
		\draw (0.center) to (2.center);
		\draw (1.center) to (2.center);
		\draw [style=none] (5.center) to (6.center);
		\draw (7.center) to (8.center);
		\draw (7.center) to (9.center);
		\draw (9.center) to (10.center);
		\draw (10.center) to (8.center);
		\draw [style=none] (11.center) to (12.center);
		\draw (14.center) to (15.center);
		\draw (14.center) to (16.center);
		\draw (16.center) to (17.center);
		\draw (17.center) to (15.center);
		\draw [style=hyperbit] (19.center) to (18.center);
		\draw (20.center) to (21.center);
		\draw (20.center) to (22.center);
		\draw (21.center) to (22.center);
		\draw (23.center) to (24.center);
		\draw (28.center) to (29.center);
		\draw (28.center) to (30.center);
		\draw (29.center) to (30.center);
		\draw (34.center) to (35.center);
		\draw (34.center) to (36.center);
		\draw (35.center) to (36.center);
		\draw (37.center) to (38.center);
		\draw (40.center) to (39.center);
		\draw (45) to (12.center);
	\end{pgfonlayer}
\end{tikzpicture}}.
	\end{equation}
Here, the boxes labeled by $\mathcal{A}$ and $\mathcal{B}$ encapsulate the internal operational procedures that Alice and Bob undertake in their spatially separated laboratories (refer to the Appendix \ref{sec:GPTs} for further details).

The LHS represents the probability of Bob obtaining outcome $s$ when measuring the observable $B_{\vec{b},r}$ with PVM elements $\Pi_{\vec{b},r}^{\pm}$, chosen based on the input bitstring $\vec{b}$ and Alice's outcome $r$. Alice measures $A_{\vec{a}}$ with PVM elements $P_{\vec{a}}^{\pm}$,
\begin{equation}
\Pe(s|\vec{a},\vec{b},r,\rho)=\mathrm{Tr}\left(P_{\vec{a}}^A\otimes \Pi_{\vec{b},r}^s\rho\right).
\end{equation}
The double wire in the LHS denotes that Alice and Bob share the quantum state $\rho$ and Alice sends the classical bit $r$ to Bob.

In contrast, on the RHS, Alice and Bob share only a classical copy $r$ and the thicker wire represents the hyperbit sent from Alice to Bob. We proceed now to define the analog of Born rule to compute $\Ph$ for the hyperbit theory.

\subsection{Tsirelson isomorphism}

In communication tasks where Bob's answer $s\in\{-1,+1\}$ is a dichotomic outcome, its expectation value $E[s]$ fully characterizes the probability distribution 
$\Ph(s=\pm1|\vec{a},\vec{b},r\vec{x}_{\vec{a}})\equiv\pi_{\vec{b},r}^\pm$,
\begin{equation}\label{eq:ppm}
    \pi_{\vec{b},r}^\pm=\frac{1\pm E[s]}{2}.
\end{equation}
Notice that,  in $\mathcal{Q}$, $\pi_{\vec{b},r}^\pm=\mathrm{Tr}\rho\Pi_{\vec{b},r}^\pm$, but in $\mathcal{H}$ we have not defined yet the analogous of the Born rule. In order to do that, we introduce
 Tsirelson's isomorphism $\mathcal{J}_\rho$
\cite{Tsirelson1987} which, in essence, maps observables into vectors:
\begin{align}
	\mathcal{J}_\rho(A_k,B_m)&=(\vec{x}_k,\vec{y}_m)\\
	(A_k,B_m)=&\mathcal{J}^{-1}_\rho(\vec{x}_k,\vec{y}_m)
\end{align}

where $A_k, B_m$ are Hermitian operators satisfying $-\bm{1}\le A_k,B_m\le \bm{1}$, $[A_k,B_m]=0$, and the vectors satisfy $|\vec{x}_k|,|\vec{y}_m|\le1$. This relationship is explicitly characterized by
\begin{equation}
\langle \vec{x}_k|\vec{y}_m\rangle=\mathrm{Tr}\left( A_k\otimes B_m \rho\right).
\end{equation}
In this context the aforementioned equivalence is formulated: Alice's \emph{steering measurement} and a single classical bit of communication are equivalent to Bob receiving the hyperbit $\vec{x}_k$.
The isomorphism then defines the rule for calculating probabilities and expectation values in $\mathcal{H}$, akin to the Born rule in $\mathcal{Q}$. Hence the expectation value of the probability distribution in Eq. \eqref{eq:ppm} is opportunely calculated as $E[s]=\langle \vec{x}_k|\vec{y}_m\rangle$ represents the expectation value of measurement $\vec{y}_m$ given $\vec{x}_k$.

In other words, for $\rho=T^\dagger T$, the isomorphism can be visualized as
\begin{equation}
\begin{tikzpicture}
	\begin{pgfonlayer}{nodelayer}
		\node [style=none] (0) at (-1, 0) {};
		\node [style=none] (1) at (1, 0) {};
		\node [style=none] (2) at (0, -1) {};
		\node [style=none] (3) at (0, -1) {};
		\node [style=none] (16) at (0, 1.5) {};
		\node [style=none] (17) at (1, 1.5) {};
		\node [style=none] (18) at (0, 0.5) {};
		\node [style=none] (19) at (1, 0.5) {};
		\node [style=none] (20) at (-1.25, 1.5) {};
		\node [style=none] (21) at (-0.25, 1.5) {};
		\node [style=none] (22) at (-1.25, 0.5) {};
		\node [style=none] (23) at (-0.25, 0.5) {};
		\node [style=none] (24) at (0.5, 1.5) {};
		\node [style=none] (25) at (1, 0.5) {};
		\node [style=none] (26) at (2, 1.5) {};
		\node [style=none] (29) at (0.5, -1.05) {};
		\node [style=none] (30) at (2, -1.05) {};
		\node [style=none] (33) at (-0.475, -1.1) {};
		\node [style=none] (34) at (1.5, -1.05) {};
		\node [style=none] (35) at (-0.5, 0) {};
		\node [style=none] (36) at (-0.5, 1.5) {};
		\node [style=none] (37) at (1.5, 1.5) {};
		\node [style=none] (42) at (-0.5, -0.5) {};
		\node [style=none] (43) at (0.5, -0.5) {};
		\node [style=none] (45) at (0, -0.5) {$\rho$};
		\node [style=none] (47) at (0.5, 0) {};
		\node [style=none] (48) at (0.5, 0.5) {};
		\node [style=none] (49) at (-0.75, 1) {$A_k$};
		\node [style=none] (50) at (0.5, 1) {$B_m$};
		\node [style=none] (51) at (-0.75, 0) {};
		\node [style=none] (52) at (-0.75, 0.5) {};
	\end{pgfonlayer}
	\begin{pgfonlayer}{edgelayer}
		\draw (0.center) to (1.center);
		\draw (1.center) to (3.center);
		\draw (3.center) to (2.center);
		\draw (2.center) to (0.center);
		\draw (16.center) to (17.center);
		\draw (17.center) to (19.center);
		\draw (19.center) to (18.center);
		\draw (18.center) to (16.center);
		\draw (20.center) to (21.center);
		\draw (21.center) to (23.center);
		\draw (23.center) to (22.center);
		\draw (22.center) to (20.center);
		\draw [bend left=270, looseness=1.75] (26.center) to (24.center);
		\draw [bend left=90, looseness=1.75] (30.center) to (29.center);
		\draw (26.center) to (30.center);
		\draw [bend left=90, looseness=1.25] (34.center) to (33.center);
		\draw (37.center) to (34.center);
		\draw [bend left=90, looseness=1.25] (36.center) to (37.center);
		\draw (48.center) to (47.center);
		\draw (52.center) to (51.center);
		\draw (33.center) to (42.center);
		\draw (29.center) to (43.center);
	\end{pgfonlayer}
\end{tikzpicture}}=%
\begin{tikzpicture}
	\begin{pgfonlayer}{nodelayer}
		\node [style=none] (0) at (-1.25, 1.5) {};
		\node [style=none] (1) at (0.75, 1.5) {};
		\node [style=none] (2) at (-1.25, 0.5) {};
		\node [style=none] (3) at (0.75, 0.5) {};
		\node [style=none] (12) at (-1.25, -0.5) {};
		\node [style=none] (13) at (0.75, -0.5) {};
		\node [style=none] (14) at (-1.25, -1.5) {};
		\node [style=none] (15) at (0.75, -1.5) {};
		\node [style=none] (16) at (-0.25, 3) {};
		\node [style=none] (17) at (0.75, 3) {};
		\node [style=none] (18) at (-0.25, 2) {};
		\node [style=none] (19) at (0.75, 2) {};
		\node [style=none] (20) at (-1.25, -2) {};
		\node [style=none] (21) at (-0.25, -2) {};
		\node [style=none] (22) at (-1.25, -3) {};
		\node [style=none] (23) at (-0.25, -3) {};
		\node [style=none] (24) at (0.25, 3) {};
		\node [style=none] (25) at (0.75, 2) {};
		\node [style=none] (26) at (1.75, 3) {};
		\node [style=none] (27) at (0.25, -1.5) {};
		\node [style=none] (29) at (0.25, -3.05) {};
		\node [style=none] (30) at (1.75, -3.05) {};
		\node [style=none] (33) at (-0.725, -3.1) {};
		\node [style=none] (34) at (1.25, -3.05) {};
		\node [style=none] (35) at (-0.75, 1.5) {};
		\node [style=none] (36) at (-0.75, 3) {};
		\node [style=none] (37) at (1.25, 3) {};
		\node [style=none] (38) at (-0.75, -2) {};
		\node [style=none] (39) at (-0.75, -1.5) {};
		\node [style=none] (40) at (-0.75, -0.5) {};
		\node [style=none] (41) at (0.25, -0.5) {};
		\node [style=none] (42) at (-0.75, 0.5) {};
		\node [style=none] (43) at (0.25, 0.5) {};
		\node [style=none] (44) at (-0.25, -1) {};
		\node [style=none] (45) at (-0.25, 1) {$T^\dagger$};
		\node [style=none] (46) at (-0.25, -1) {$T$};
		\node [style=none] (47) at (0.25, 1.5) {};
		\node [style=none] (48) at (0.25, 2) {};
		\node [style=none] (49) at (-0.75, -2.5) {$A_k$};
		\node [style=none] (50) at (0.25, 2.5) {$B_m$};
		\node [style=none] (51) at (-1.75, 0) {};
		\node [style=none] (52) at (2.25, 0) {};
	\end{pgfonlayer}
	\begin{pgfonlayer}{edgelayer}
		\draw (0.center) to (1.center);
		\draw (1.center) to (3.center);
		\draw (3.center) to (2.center);
		\draw (2.center) to (0.center);
		\draw (12.center) to (13.center);
		\draw (13.center) to (15.center);
		\draw (15.center) to (14.center);
		\draw (14.center) to (12.center);
		\draw (16.center) to (17.center);
		\draw (17.center) to (19.center);
		\draw (19.center) to (18.center);
		\draw (18.center) to (16.center);
		\draw (20.center) to (21.center);
		\draw (21.center) to (23.center);
		\draw (23.center) to (22.center);
		\draw (22.center) to (20.center);
		\draw [bend left=270, looseness=1.75] (26.center) to (24.center);
		\draw (27.center) to (29.center);
		\draw [bend left=90, looseness=1.75] (30.center) to (29.center);
		\draw (26.center) to (30.center);
		\draw [bend left=90, looseness=1.25] (34.center) to (33.center);
		\draw (36.center) to (35.center);
		\draw (37.center) to (34.center);
		\draw [bend left=90, looseness=1.25] (36.center) to (37.center);
		\draw (39.center) to (38.center);
		\draw (43.center) to (41.center);
		\draw (42.center) to (40.center);
		\draw (48.center) to (47.center);
		\draw [dashed, in=360, out=180] (52.center) to (51.center);
	\end{pgfonlayer}
\end{tikzpicture}}=%
\begin{tikzpicture}
	\begin{pgfonlayer}{nodelayer}
		\node [style=none] (53) at (-1, -0.375) {};
		\node [style=none] (54) at (1, -0.375) {};
		\node [style=none] (55) at (0, -1.375) {};
		\node [style=none] (56) at (-1, 0.375) {};
		\node [style=none] (57) at (1, 0.375) {};
		\node [style=none] (58) at (0, 1.375) {};
		\node [style=none] (59) at (0, -0.775) {$\vec{x}_k$};
		\node [style=none] (60) at (0.025, 0.725) {$\vec{y}_m$};
		\node [style=none] (61) at (0, 0.375) {};
		\node [style=none] (62) at (0, -0.375) {};
	\end{pgfonlayer}
	\begin{pgfonlayer}{edgelayer}
		\draw (53.center) to (54.center);
		\draw (54.center) to (55.center);
		\draw (53.center) to (55.center);
		\draw (56.center) to (57.center);
		\draw (57.center) to (58.center);
		\draw (56.center) to (58.center);
		\draw (61.center) to (62.center);
	\end{pgfonlayer}
\end{tikzpicture}}.
\end{equation}
In this representation, the cyclic wire signifies the contraction of indices, i.e., the trace operation. The lower triangle depicts the input state, while the upper triangle represents the result of the effect. The dotted line helps to distinguish the preparation (lower side) from the measurement (upper side).

\subsubsection{Properties of the Tsirelson isomorphism}
The homomorphism (perfect correlated input choices) maps the identity to the identity,
\begin{align}\label{eq:x1=y1}
	1=&\mathrm{Tr}\left(\bm{1}\otimes\bm{1}\rho
	\right)=\langle \vec{x}_{\bm{1}}|\vec{y}_{\bm{1}}\rangle
	=
	|\vec{x}_{\bm{1}}||\vec{y}_{\bm{1}}|\cos\theta\nonumber\\
	&\Longrightarrow \vec{x}_{\bm{1}}=\vec{y}_{\bm{1}},\quad|\vec{x}_{\bm{1}}|=1.
\end{align}
For dichotomic measurements with equal probability on the outcomes related to $\{P_{\vec{a}}^{+1},P_{\vec{a}}^{-1}\}$ we have

\begin{equation}\label{eq:AliceUniform}
    \Biggl\{
    \begin{aligned}
        A_{\vec{a}} &= P_{\vec{a}}^{+1} - P_{\vec{a}}^{-1}, \\
        \boldsymbol{1} &= P_{\vec{a}}^{+1} + P_{\vec{a}}^{-1},
    \end{aligned}
    \xrightarrow{\mathcal{J}_\rho(A_{\vec{a}}, \cdot)}
    \Biggl\{
    \begin{aligned}
        \vec{x}_{\vec{a}} &= \vec{x}_{\vec{a},+1} - \vec{x}_{\vec{a},-1}, \\
        \vec{x}_{\boldsymbol{1}} &= \vec{x}_{\vec{a},+1} + \vec{x}_{\vec{a},-1}.
    \end{aligned}
\end{equation}

\begin{figure}
	\centering
	\includegraphics[width=0.8\linewidth]{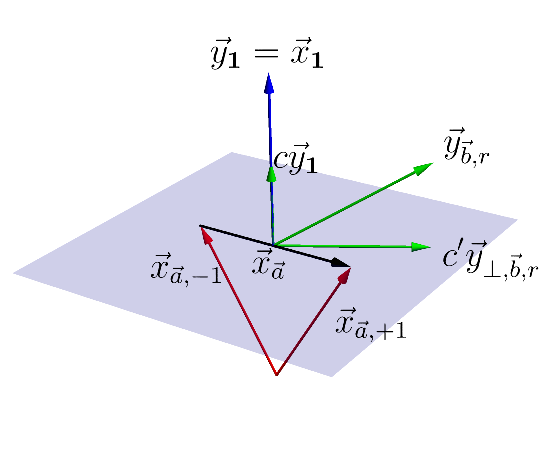}
	\caption{Geometrical representation in the affine space for the Tsirelson isomorphism for Alice's choice of $A_{\vec{a}}$ related to $\{\vec{x}_{\vec{a},+1},\vec{x}_{\vec{a},-1}\}$ with norm of the sum $|\vec{x}_{\vec{a}}|\le 1$ and Bob's choice $\vec{y}_{\vec{b},r}$ with $|\vec{y}_{\vec{b},r}|\le1$ decomposed along $\vec{y}_{\bm{1}}$, and $\vec{y}_{\perp,\vec{b},r}$ orthogonal to $\vec{y}_{\bm{1}}$. }
	\label{fig:vectors}
\end{figure}

Without any loss of generality, suppose that Alice has unbiased expectation value $\langle A_{\Vec{a}}\rangle=0$, then, going through the isomorphism, the sum of two vectors, e.g. $\vec{x}_{\vec{a},\pm 1}$,  with the same module is always orthogonal to the difference, as in Fig. \ref{fig:vectors}, and with Eq. \eqref{eq:x1=y1} it yields
\begin{align}\label{eq:PAequal}
	\left\langle \vec{x}_{\vec{a}}|\vec{y}_{\boldsymbol{1}}\right\rangle =0\Longrightarrow\left\langle \vec{x}_{\vec{a},+1}|\vec{y}_{\boldsymbol{1}}\right\rangle &=\left\langle \vec{x}_{\vec{a},-1}|\vec{y}_{\boldsymbol{1}}\right\rangle\nonumber\\
	|\vec{x}_{\vec{a},+1}|\cos\theta_+&=|\vec{x}_{\vec{a},-1}|\cos\theta_- .
\end{align}
Therefore $\theta_-=\theta_+$. Then

\begin{align}
\mathrm{Tr}\left(P_{\vec{a}^r}\otimes\boldsymbol{1}\rho\right)
	=&\mathrm{Tr}\left(\frac{\boldsymbol{1}+rA_{\vec{a}}}{2}\otimes\boldsymbol{1}\rho\right)\nonumber\\
	=&\frac{1}{2}+r\left\langle \vec{x}_{\vec{a}}|\vec{y}_{\boldsymbol{1}}\right\rangle =\frac{1}{2},\label{eq:diff}
\end{align}

and combined with Eq. \eqref{eq:PAequal} gives
\begin{align}
	\left\langle \vec{x}_{\vec{a},1}|\vec{y}_{\boldsymbol{1}}\right\rangle +\left\langle \vec{x}_{\vec{a},-1}|\vec{y}_{\boldsymbol{1}}\right\rangle
	=&2\left\langle \vec{x}_{\vec{a},\pm1}|\vec{y}_{\boldsymbol{1}}\right\rangle =1\nonumber\\
	\Longrightarrow& \left|\vec{x}_{\vec{a},\pm1}\right|\cos\theta=\frac{1}{2}.
\end{align}

In the case of Bob's choices the probability distribution of the outcomes is not uniform as in Eq. \eqref{eq:AliceUniform}; therefore,
\begin{equation}\label{eq:BobNotUniform}
    \Biggl\{
    \begin{aligned}
        y_{\vec{b},r}  &=\alpha y_{\vec{b},r,+1}-\beta y_{\vec{b},r,-1}\\
		\vec{y}_{\boldsymbol{1}} &=y_{\vec{b},r,+1}+y_{\vec{b},r,-1}
        \end{aligned}
    \longrightarrow
    \langle y_{\vec{b},r}|\vec{y}_{\boldsymbol{1}}\rangle =\alpha-\beta.
\end{equation}
In Fig. \ref{fig:vectors}, $\vec{y}_{\vec{b},r}$ is decomposed in $c\,\vec{y}_{\bm{1}}$ along the axes parallel to $\vec{y}_{\bm{1}}$ with $0\le c\le 1$ and $c^\prime N \vec{y}_{\perp,\vec{b},r}$ orthogonal to $\vec{y}_{\bm{1}}$.
The crucial assumption for our communication tasks is that all possible choices of Alice $A_{\vec{a}}$ must be with equal probability outcomes\footnote{It should be noted that this condition comes from the assumption of the expectation value of the Alice outcome $r\in\{-1,+1\}$ being zero. This condition does not reduce the generality of the protocol since it is obtainable by some postprocessing (in Alice’s case)
	or preprocessing (in Bob’s) and can be incorporated into the
	measurement operators.} as commented along Eq. \eqref{eq:AliceUniform}, and the distribution of Bob's probability outcomes depends on the values of $\alpha$ and $\beta$ in Eq. \eqref{eq:BobNotUniform} going from $-1$ to $1$. This assumption allows copying and sharing, as a free resource, a random bit $r$, equal to Alice's output with vanishing expectation value in the Hyperbit Theory as represented by a circle on the RHS of Eq. \eqref{eq:equivalenceQH}.
\subsection{Expectation value in Quantum Theory}
Now we are ready to compute the LHS of Eq. \eqref{eq:B=B}, the expectation value of Bob's answer in quantum theory $\mathcal{Q}$, as the expectation value of the observable $B_{\vec{b},r}$ on the steering state $\rho_{\vec{a},r}$ via $A_{\vec{a}}=P_{\vec{a}}^{+1}-P_{\vec{a}}^{-1}$. With Eq. \eqref{eq:diff} and tracing the Alice's Hilbert space we found
\begin{equation}
	\rho_{\vec{a},r}=\frac{\mathrm{Tr_A}\left(P_{\vec{a}}^r\otimes\bm{1}\rho\right)}{\mathrm{Tr}\left(P_{\vec{a}}^r\otimes\bm{1}\rho\right)}=2\, \mathrm{Tr_A}\left(P_{\vec{a}}^r\otimes\bm{1}\rho\right)
\end{equation}
 such that, summing on Eq. \eqref{eq:AliceUniform} and then using  Tsirelson isomorphism,
\begin{align}
	\langle B(\vec{a},\vec{b},r,\rho)&\rangle _{\mathcal{Q}}= \mathrm{Tr}_{\mathrm{B}}B_{\vec{b},r}\rho_{\vec{a},r}=2\mathrm{Tr}P_{\vec{a}}^{r}\otimes B_{\vec{b},r}\rho\nonumber \\
	= & \mathrm{Tr}\left[\left(\boldsymbol{1}+r\hat{A}_{\vec{a}}\right)\otimes B_{\vec{b},r}\rho\right]\nonumber \\
	= & \left\langle x_{\boldsymbol{1}}+r\vec{x}_{\vec{a}}|\vec{y}_{\vec{b},r}\right\rangle \nonumber\\
	= & \langle x_{\boldsymbol{1}}+r\vec{x}_{\vec{a}}|c(\vec{b},r)y_{\bm{1}}+c^{\prime}(\vec{b},r)\vec{y}_{\perp\vec{b},r}\rangle \nonumber\\
	= & c(\vec{b},r)+c^{\prime}(\vec{b},r)N(\vec{b},r)\left\langle r\vec{x}_{\vec{a}}|\vec{\tilde{y}}_{\perp\vec{b},r}\right\rangle .\label{eq:Qresult}
\end{align}
It is important to note that the vector \(x_{\bm{1}} + A \vec{x}_{\vec{a}}\)  possesses a \textit{length} exceeding 1. Consequently, it does not qualify as a hyperbit, as its length precludes the possibility of measurement with normalized outcome probabilities. This necessitates the application of postprocessing procedures to ensure that hyperbits attain a normalized length suitable for measurement and necessary for the equivalence $\mathcal{H}\equiv\mathcal{Q}$. 
Moreover, it is particularly evident from the second to the third line of Eq. \eqref{eq:Qresult} the application of the Tsirelson isomorphism that represents the passage from $\mathcal{Q}$ to $\mathcal{H}$, hence the change of the meaning of the variable $r$, as Alice's output to shared randomness [see Eq. \eqref{eq:diff}], despite the equivalence at the operational level. This approaches a potential equivalence for these theories, despite their different interpretations.

Now, according to Fig. \ref{fig:vectors}, we defined  $c(\vec{b},r)=\langle y_{\boldsymbol{1}}|\vec{y}_{\vec{b},r}\rangle $
and $c^{\prime}(\vec{b},r)N(\vec{b},r)=\langle \vec{\tilde{y}}_{\perp\vec{b},r}|\vec{y}_{\vec{b},r}\rangle $
with $\vec{\tilde{y}}_{\perp\vec{b},r}$ the normalized $\vec{y}_{\perp\vec{b},r}$, which is the projection of $\vec{y}_{\vec{b},r}$ on
the subspace orthogonal to $\vec{y}_{\boldsymbol{1}}$ and $N(\vec{b},r)$ the normalization factor. We expressed $c$, $c^\prime$, and $N$ with the dependence on $\vec{b}$ and the random bit $r$ to stress the dependence on Bob's resources. In addition to that, the constraint from the Tsirelson isomorphism gives
\begin{equation}
	|\vec{y}_{\vec{b},r}|\le 1 \Longrightarrow
	[c^{\prime}(\vec{b},r)N(\vec{b},r)]^2+c^2(\vec{b},r)\le 1.
\end{equation}
For a lighter notation we denoted:
\begin{align}\label{eq:xyz}
	x=c^{\prime}(\vec{b},r)N(\vec{b},r),\quad
	y=c(\vec{b},r),\quad
	z=\langle r\vec{x}_{\vec{a}}|\vec{\tilde{y}}_{\perp\vec{b},r}\rangle.
\end{align}
Equation \eqref{eq:Qresult} reads as a three-variable function $t$:
\begin{equation}\label{eq:fQ}
	\langle B(\vec{a},\vec{b},r,\rho)\rangle _{\mathcal{Q}}=t(x,y,z)=y+xz
\end{equation}
in the domain
\begin{equation}\label{eq:cylinder}
	\mathcal{C}=\left\{(x,y,z)\in \mathbb{R}^3\big\vert\ x^2+y^2\le 1, |t(x,y,z)|\le 1,|z|\le 1\right\}.
 \end{equation}
  Notice that, Bob's answer is $-1$ or $+1$; therefore, this induces the constraint $|t(x,y,z)|\le 1$.
 In conclusion, the expectation value of Bob's answer in $\mathcal{Q}$ is represented by the function $t$ in its domain $\mathcal{C}$. 
\subsection{Expectation value in Hyperbit theory}
To show the equivalence between the quantum theory $\mathcal{Q}$ and the hyperbit theory $\mathcal{H}$, each value of the function $t(x,y,z)$ for any selection of $(x,y,z) \in \mathcal{C}$, derived from the Born rule in $\mathcal{Q}$, must be also obtained within the hyperbit theory through meaningful operational procedures. First, the Tsirelson isomorphism defines the operational procedure to obtain the expectation value of Bob's answer as $E[s]=\langle r\vec{x}_{\vec{a}}|\vec{\tilde{y}}_{\perp\vec{b},r}\rangle$, which must correspond to the expectation value in Eq. \eqref{eq:ppm} of the output $s$ distributed in $\{-1,+1\}$ by $\pi_{\vec{b},r}^{\pm1}$ as

\begin{equation}
E[s]=\pi_{\vec{b},r}^{+1}-\pi_{\vec{b},r}^{-1},\quad \text{ with }\sum_{s=\pm1}\pi_{\vec{b},r}^{s}=1.\label{eq:z}
\end{equation}

The expectation value of Bob's answer $s$ is obtained, similarly to the Born rule in $\mathcal{Q}$, by measuring with the effect $\vec{\tilde{y}}_{\perp\vec{b},A}$ on the received hyperbit $r\vec{x}_{\vec{a}}$, as shown in the lighter notation in Eq. \eqref{eq:xyz}; thus $z=E[s]$. 
Now, we are ready to reformulate the equivalence.
\begin{center}
    \emph{$\mathcal{H}$ is equivalent to $\mathcal{Q}$ if and only if the most general postprocessing operational procedures on Bob's random answer $s$ admit a function $g=t$ for all $(x,y,z) \in \mathcal{C}$.}    
\end{center}
Notice that, being $s$ random, there is no \textit{a priori} information about its expectation value $z$. To characterize the most general postprocessing procedure we observe that the function $g$ can be represented as a point of a tetrahedron obtainable as a convex combination of all the possible deterministic functions which are
\begin{align}\label{eq:vertex_4hedron}
f_1\left(s\right)= +1,\quad
f_2\left(s\right)= -1,\quad
f_3\left(s\right)= s,\quad
f_4\left(s\right)= -s
\end{align}
where $f_1$ and $f_2$ are the deterministic discarding operations that replace the outcome into $s\to +1$ and $s\to -1$ respectively; $f_3$ is \textit{do nothing} operation and $f_4$ flips Bob's outcome. Therefore, the most general operational procedure that constructs $g=\langle B (\vec{a},\vec{b},r,\vec{x}_{\vec{a}})\rangle_\mathcal{H}$ is
\begin{equation}\label{eq:g}
g=\sum_{s=\pm 1}\sum_{i=1}^4 k_i f_i(s) \pi_{\vec{b},r}^s,\quad \sum_{i=1}^4 k_i=1,\, k_i\ge0
\end{equation}
which yields,  from Eqs. \eqref{eq:z} and \eqref{eq:vertex_4hedron}
\begin{equation}\label{eq:g2}
g(\vec{k},z)=k_1-k_2+(k_3-k_4)z.
\end{equation}
In conclusion, Eq. \eqref{eq:g2} represents the expectation value of Bob's answer in $\mathcal{H}$ as a function $g$ of the weights $\vec{z}=(k_1,\dots,k_4)$ and the expectation value $z$.

Before showing the limitations concerning the equivalence of Eq. \eqref{eq:B=B}, let us revisit the protocol in Ref.\cite{Pawlowski2012}.

\section{Paw\l owski--Winter protocol}\label{sec:PWprotocol}
Paw\l owski and Winter in Ref. \cite{Pawlowski2012} formulate the following operational procedure in hyperbit theory $\mathcal{H}$ intending to demonstrate its equivalence to the shared entanglement scenario in the quantum theory $\mathcal{Q}$.
\begin{procedurelist}{PW Protocol}
	\item  Alice and Bob copy and share a random bit $r$ with vanishing expectation value.
	\item  Alice receives the input string $\vec{a}$ and selects the measurement $x_{\vec{a}}$ and sends to Bob the hyperbit multiplied by $r$, that is $\vec{x}_{\vec{a},r}=r\vec{x}_{\vec{a}}$.
	\item  Bob receives the input bitstring $\vec{b}$ and then measures and computes the expectation value on the hyperbit with $\vec{\tilde{y}}_{\perp,\vec{b},r}$.
	\item  The expectation value $\langle B(\vec{a},\vec{b},r,\vec{x}_{\vec{a}})\rangle_\mathcal{H}$ can be obtained by the following postprocessing:
	\begin{itemize}
		\item [4.1] With probability $|c(\vec{b},r)|=|y|$, as in Eq. \eqref{eq:xyz}, Bob discards his outcome and outputs $\mathrm{sign}(c(\vec{b},r))$.
		\item [4.2] With probability $1-|c(\Vec{b},r)|$ he tosses a coin that has probability $q$ for the outcome ``head" and probability $1-q$ for the outcome ``tail."
		\item [4.3] If the outcome of the coin is ``head", then Bob flips his outcome; otherwise, with ``tail", his outcome remains invariant.
	\end{itemize}
\end{procedurelist}

\begin{figure}
	\centering
 \includegraphics[width=0.7\linewidth]{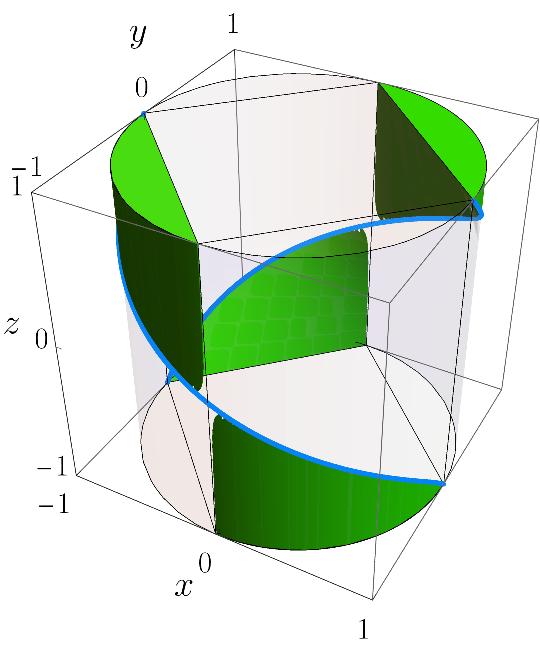}
 \includegraphics[width=0.7\linewidth]{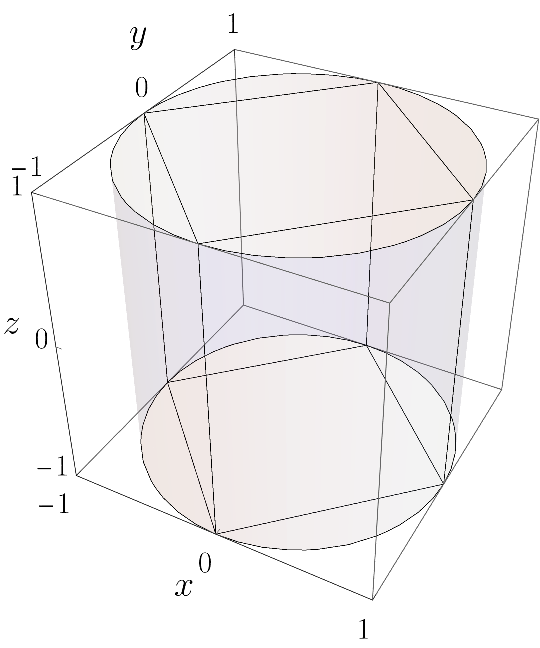}
	\caption{On the lower panel: a white parallelepiped representing the region $\mathcal{D}$ and a white cylinder. 
    On the upper panel: by removing the four green regions, characterized by $1\le |t(x,y,z)|\le \sqrt{2}$, from the white cylinder the region $\mathcal{C}$ explicitly written in Eq. \eqref{eq:cylinder} is obtained. It contains $\mathcal{D}$.  The blue helices on the surface of the cylinder and parametric equation $\{\cos \tau,\sin(\pm\tau),\pm \frac{1-\sin \tau}{\cos \tau}\}$ for $\tau\in[0,\pi]$ 
     correspond to the points for $t=1$. These helices illustrate that, for each point $z$, it is possible to identify a violation of the equivalence as the PW protocol, as well as, its widest generalization being valid only in the region $\mathcal{D}$.}
	\label{fig:marcinregion}
\end{figure}

Now the protocol that realizes $g=\langle B (\vec{a},\vec{b},A,\vec{x}_{\vec{a}})\rangle_{\mathcal{H}}$ mathematically reads
\begin{align}
	\langle B (\vec{a},\vec{b},r,\vec{x}_{\vec{a}})&\rangle_{\mathcal{H}}=\nonumber\\
 \sum_{s=\pm 1}\pi_{\vec{b},r}^s\biggl[
 |y|&\left(\frac{1+y/|y|}{2}f_1(s)+\left(1-\frac{1+y/|y|}{2}\right)f_2(s)\right)\nonumber\\
 +(1-&|y|)\left(qf_4(s)+(1-q)f_3(s)\right)\biggr]
\end{align}
or equivalently, introducing the \textit{probabilistic discarding operation} $\mathbb{D}_{c(r,N)}$ with probability $c(A,N)=|y|$ (in lighter notation) that replaces $s$ with $\mathrm{sign}(y)$ and with probability $1-|y|$, $s$ is not discarded; then the \textit{probabilistic flipping operation}  $\mathbb{F}_q$ happens with probability $q$,
\begin{align}
	\langle B (\vec{a},\vec{b},r,\vec{x}_{\vec{a}})\rangle_{\mathcal{H}}&=(\mathbb{F}_q\circ\mathbb{D}_{|y|}\circ E)(s)\nonumber\\
	 =&\mathbb{F}_q[|y|\mathrm{sign}(y)+(1-|y|)z]\nonumber\\
	 =&\mathbb{F}_q[y+(1-|y|)z]\nonumber\\
	=&y+(1-|y|)z(1-2q).	
\end{align}
Explicitly, first it acts as $E(s)=\pi_{\vec{b},r}^+-\pi_{\vec{b},r}^-=z$. Then we apply the probabilistic discarding operation, such that with probability $|y|$ gives $\mathrm{sgn(y)}$ and with probability $1-|y|$ we toss a coin where we apply the probabilistic flip operation; therefore, we obtain
\begin{equation*}
    |y|\mathrm{sgn}(y)+(1-|y|)\mathbb{F}_q(z)
\end{equation*}
Now, with probability $q$, we have $z\mapsto -z$ and with probability $1-q$, $z\mapsto z$. Therefore, $\mathbb{F}_q(z)=q(-z)+(1-q)z=(1-2q)z$. Finally,  $y=|y|\mathrm{sgn}(y)$.
Now, by imposing Eq. \eqref{eq:B=B}, and with  Eq. \eqref{eq:xyz}-\eqref{eq:z}, one has
\begin{equation}\label{eq:qPW}
	q=\frac{1}{2}\left(1 - \frac{x}{1-|y|}\right).
\end{equation}
We observe that the PW protocol makes sense if and only if the flipping probability lies $0\le q\le 1$, i.e.,
\begin{equation}
	|c|\pm Nc^\prime\le 1, \qquad \langle r\vec{x}_{\vec{a}},\vec{\tilde{y}}_{\vec{b},r}\rangle\in [-1,1]
\end{equation} 
that with lighter notation is
\begin{equation}\label{eq:PWconstraints}
	\mathcal{D}=\{(x,y,z)\in \mathbb{R}^3|\,-1\le z \le 1, \,\, |y|\pm x\le 1\}.
\end{equation} 
In its current formulation, the hyperbit theory is equivalent to the quantum theory only within the subregion $\mathcal{D}\subset \mathcal{C}$ as depicted in Fig. \ref{fig:marcinregion}. 
For points in $\mathcal{C}\setminus\mathcal{D}$ the equivalence of Eq. \eqref{eq:B=B}, i.e. $g=t$ iff $q\not\in[0,1]$ which is impossible, as noticed in Ref. \cite{tavakoli2021correlations}. 
Therefore, to establish an equivalence between quantum theory and hyperbit theory for communication tasks, other operational procedures $g=\langle B (\vec{a},\vec{b},r,\vec{x}_{\vec{a}})\rangle_\mathcal{H}$ must be considered. These new procedures from Eq. \eqref{eq:g} must satisfy \eqref{eq:B=B}; hence $g(\{k_i,q_i,y_i\},z)=t(x,y,z)$, since $t(x,y,z)=\langle B (\vec{a},\vec{b},A,\rho)\rangle_\mathcal{Q}$ also for all $(x,y,z)\in \mathcal{C}\setminus\mathcal{D}$.

The purpose of this paper is to show that even by considering all possible operational procedures in Eq. \eqref{eq:g} obtained by fixing $k_i$'s the equivalence $\mathcal{Q}\equiv \mathcal{H}$ is unattainable under the assumption that Bob has no access to Alice's laboratory.

\section{Main Result}\label{sec:result}
We show the lack of equivalence between $\mathcal{Q}$ and $\mathcal{H}$ and the operational limitation of using hyperbits starting from the following proposition that corrects the main theorem stated in Sec. \ref{sec:Hyperbit_theory} of Ref \cite{Pawlowski2012}.

\begin{proposition}\label{prop1}
     For tasks where Bob gives binary answers, sending one hyperbit from Alice to Bob is equivalent to sharing any amount of entanglement and sending one classical bit if and only if Bob can adjust his strategy by knowing \textit{a priori} the expectation value on the hyperbit.
\end{proposition}
    
\begin{proof}
Let us consider the reformulation of the equivalence $\mathcal{H}\simeq \mathcal{Q}$ as in the paragraph above Eq. \eqref{eq:vertex_4hedron}. In other words, within the framework of $\mathcal{H}$ the most general Bob's postprocessing procedure admits a function $g=t$ for all $(x,y,z)\in \mathcal{C}$, where $g$, $t$, and $\mathcal{C}$ are respectively defined in Eqs. \eqref{eq:g2}, \eqref{eq:fQ}, and \eqref{eq:cylinder}. Therefore, $g=t$, from Eqs. \eqref{eq:g2}, and \eqref{eq:fQ}, reads as  
    \begin{equation}\label{eq:proof_g_t_equal}
        k_1-k_2+(k_3-k_4)z=y+xz,\quad\forall (x,y,z)\in \mathcal{C}.
    \end{equation} 
    Now if $\mathcal{H}\simeq \mathcal{Q}$ is trivial to show that Bob knows \textit{a priori} the expectation value of the hyperbit $z$, indeed from Eq. \eqref{eq:proof_g_t_equal} immediately he can determine 
    \begin{equation}
        k_1 - k_2 - y + (k_3 - k_4 - x)z = 0\Longrightarrow
        z=\frac{y-k_1 + k_2 }{k_3 - k_4 - x}.
    \end{equation}
 The other direction is more interesting. To prove it, we simply observe from Eq. \eqref{eq:xyz} that $x$ and $y$ are functions of  Bob's input $\vec{b}$ and $r$; hence if Bob also knows the value of $z$ he can freely select $k_i$'s values (with \(\sum_i k_i = 1\) and \(k_i \ge 0\))  as part of his strategy such that Eq. \eqref{eq:proof_g_t_equal} holds; thus $\mathcal{H}\simeq \mathcal{Q}$. For $(x,y,z)\in \mathcal{D}$ using the PW protocol once obtain $g=t$. For $(x,y,z)\in \mathcal{C}/ \mathcal{D}$ a numerical evidence is in Fig. \ref{fig:numerical}.
 
\begin{figure}
    \centering
    \includegraphics[width=0.7\linewidth]{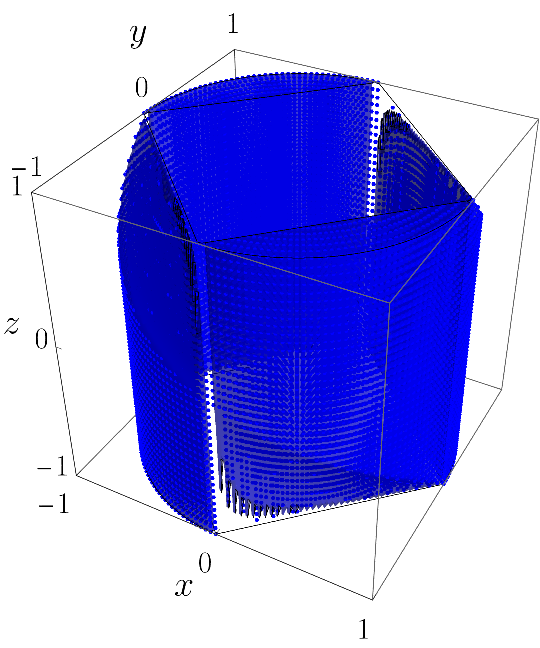}\caption{Bob's \textit{a priori} knowledge of $z$ allows to Bob to adjust this strategy such that $g=t$ for all $(x,y,z)\in \mathcal{C}/\mathcal{D}$.}
    \label{fig:numerical}
\end{figure} 

\end{proof}

At this point, we are confronted with the question of whether the equivalence \(\mathcal{Q}\simeq\mathcal{H}\) established in proposition \ref{prop1} is operationally meaningful or is required to drop the hypothesis of Bob's \textit{a priori} knowledge of the expectation value of the hyperbit. The answer hinges on understanding which resources are operationally accessible to Bob. From Eqs.~\eqref{eq:xyz}, it is evident that the variables \(x\) and \(y\) depend solely on \(\vec{b}\) and the random variable \(r\), both of which are readily accessible resources in Bob's laboratory. However, for Bob to know the value of \(z\), he requires \textit{a priori} knowledge of the outcome from the effect \(\vec{\tilde{y}}_{\perp,\vec{b},r}\) on the preparation \(r\vec{x}_{\vec{a}}\). 
 This knowledge enables him to adjust his strategy by selecting appropriate \(k_i\) values to satisfy Eq.~\eqref{eq:B=B}. 
 
 Clearly, \(\vec{\tilde{y}}_{\perp,\vec{b},r}\) is known to Bob, as it is for \(x\) and \(y\), but he needs to fix the postprocessing procedure (depicted in the box \(\mathcal{B}\)) even before receiving the hyperbit \(r\vec{x}_{\vec{a}}\). However, once the strategy is fixed, it must satisfy Eq. \eqref{eq:proof_g_t_equal} for all the values of $z$. In other words, the equivalence between quantum and hyperbit theory must always hold regardless of the specific value of \(z\) (the expectation value of the hyperbit Bob receives). This creates an obstacle to establishing equivalence between hyperbit Theory and quantum theory without the \textit{a priori} knowledge of $z$. It means that Eq.~\eqref{eq:proof_g_t_equal} must be satisfied as follows (subject to the postprocessing probability constraints):
\begin{equation}\label{eq:fixks}
    k_1 - k_2 = y, \quad k_3 - k_4 = x, \quad \sum_i k_i = 1, \quad k_i \geq 0.
\end{equation}
Then, we observe that the equivalence $\mathcal{Q}\simeq\mathcal{H}$ is only possible within the \(\mathcal{D}\) region.\footnote{Remarkably, the parallelepiped \(\mathcal{D}\) is the largest region in \(\mathcal{C}\) shape invariant in the $xy$ plane along the \(z\) axis and PW protocol is obtained from Eq.~\eqref{eq:g} with 
\begin{equation}
    q = \frac{1}{2} \left(1 - \frac{k_3 - k_4}{1 - |k_1 - k_2|}\right).
\end{equation}}
The operational limitation necessitates reexamining and adjusting a kind of negation of Proposition \ref{prop1}. Specifically, if Bob does not know the value of \(z\) (meaning the \(k_i\) values are chosen in advance to fix his strategy), then there will be instances where \(t(x,y,z) \neq g(k_i,z)\), as it happens within the region \(\mathcal{C}/\mathcal{D}\) shown in Fig.~\ref{fig:marcinregion}.

The subsequent proposition is the kind of negation we are looking for. It formally states this discussion highlighting the importance of distinguishing between the resources that are operationally accessible \textit{a priori} to Bob.
\begin{proposition}
    Suppose that the operationally accessible resources \textit{a priori} are those below the dashed line, i.e., those available before the communication starts. Then the following equivalence is impossible:
    \begin{equation}\label{eq:equivalenceQH_2}
\begin{tikzpicture}
	\begin{pgfonlayer}{nodelayer}
		\node [style=none] (0) at (-0.7, -0.85) {};
		\node [style=none] (1) at (0.7, -0.85) {};
		\node [style=none] (2) at (0, -1.75) {};
		\node [style=none] (4) at (0, -1.2) {$\rho$};
		\node [style=none] (5) at (-0.5, -0.85) {};
		\node [style=none] (6) at (-0.5, 0.15) {};
		\node [style=none] (7) at (-2, 1.15) {};
		\node [style=none] (8) at (0, 1.15) {};
		\node [style=none] (9) at (-2, 0.15) {};
		\node [style=none] (10) at (0, 0.15) {};
		\node [style=none] (11) at (0.5, 1.9) {};
		\node [style=none] (12) at (0.5, -0.85) {};
		\node [style=none] (14) at (-1, 2.9) {};
		\node [style=none] (15) at (1.5, 2.9) {};
		\node [style=none] (16) at (-1, 1.9) {};
		\node [style=none] (17) at (1.5, 1.9) {};
		\node [style=none] (18) at (-0.5, 1.15) {};
		\node [style=none] (19) at (-0.5, 1.9) {};
		\node [style=none] (20) at (-0.25, 3.4) {};
		\node [style=none] (21) at (0.8, 3.4) {};
		\node [style=none] (22) at (0.25, 4.15) {};
		\node [style=none] (23) at (0.25, 3.4) {};
		\node [style=none] (24) at (0.25, 2.9) {};
		\node [style=none] (25) at (-1.025, 0.65) {\Alice};
		\node [style=none] (26) at (0.25, 2.4) {\Bob};
		\node [style=none] (27) at (0.25, 3.65) {$s$};
		\node [style=none] (28) at (0.75, -0.85) {};
		\node [style=none] (29) at (1.9, -0.85) {};
		\node [style=none] (30) at (1.325, -1.775) {};
		\node [style=none] (34) at (-2, -0.85) {};
		\node [style=none] (35) at (-0.775, -0.85) {};
		\node [style=none] (36) at (-1.35, -1.65) {};
		\node [style=none] (37) at (-1.3, -0.85) {};
		\node [style=none] (38) at (-1.3, 0.15) {};
		\node [style=none] (39) at (1.2, -0.85) {};
		\node [style=none] (40) at (1.2, 1.9) {};
		\node [style=none] (41) at (1.25, -1.1) {};
		\node [style=none] (42) at (1.35, -1.2) {$\vec{b}$};
		\node [style=none] (43) at (-1.5, -1.1) {};
		\node [style=none] (44) at (-1.35, -1.175) {$\vec{a}$};
		\node [style=none] (45) at (-0.85, 1.5) {$A$};
		\node [style=none] (46) at (-4, 0) {};
		\node [style=none] (48) at (2.5, 0) {};
		\node [style=none] (49) at (-3.25, -0.425) {\textit{a priori}};
	\end{pgfonlayer}
	\begin{pgfonlayer}{edgelayer}
		\draw (0.center) to (1.center);
		\draw (0.center) to (2.center);
		\draw (1.center) to (2.center);
		\draw [style=qWire] (5.center) to (6.center);
		\draw (7.center) to (8.center);
		\draw (7.center) to (9.center);
		\draw (9.center) to (10.center);
		\draw (10.center) to (8.center);
		\draw [style=qWire] (11.center) to (12.center);
		\draw (14.center) to (15.center);
		\draw (14.center) to (16.center);
		\draw (16.center) to (17.center);
		\draw (17.center) to (15.center);
		\draw (19.center) to (18.center);
		\draw (20.center) to (21.center);
		\draw (20.center) to (22.center);
		\draw (21.center) to (22.center);
		\draw (23.center) to (24.center);
		\draw (28.center) to (29.center);
		\draw (28.center) to (30.center);
		\draw (29.center) to (30.center);
		\draw (34.center) to (35.center);
		\draw (34.center) to (36.center);
		\draw (35.center) to (36.center);
		\draw (37.center) to (38.center);
		\draw (40.center) to (39.center);
		\draw [dashed] (46.center) to (48.center);
	\end{pgfonlayer}
\end{tikzpicture}}\quad\stackrel{w}{\simeq}\quad
\begin{tikzpicture}
	\begin{pgfonlayer}{nodelayer}
		\node [style=none] (0) at (-0.7, -0.875) {};
		\node [style=none] (1) at (0.7, -0.875) {};
		\node [style=none] (2) at (0, -1.775) {};
		\node [style=none] (4) at (0, -1.225) {$A$};
		\node [style=none] (5) at (-0.5, -0.875) {};
		\node [style=none] (6) at (-0.5, 0.125) {};
		\node [style=none] (7) at (-2, 1.125) {};
		\node [style=none] (8) at (0, 1.125) {};
		\node [style=none] (9) at (-2, 0.125) {};
		\node [style=none] (10) at (0, 0.125) {};
		\node [style=none] (11) at (0.5, 1.875) {};
		\node [style=none] (12) at (0.5, -0.375) {};
		\node [style=none] (14) at (-1, 2.875) {};
		\node [style=none] (15) at (1.5, 2.875) {};
		\node [style=none] (16) at (-1, 1.875) {};
		\node [style=none] (17) at (1.5, 1.875) {};
		\node [style=none] (18) at (-0.5, 1.125) {};
		\node [style=none] (19) at (-0.5, 1.875) {};
		\node [style=none] (20) at (-0.25, 3.375) {};
		\node [style=none] (21) at (0.8, 3.375) {};
		\node [style=none] (22) at (0.25, 4.125) {};
		\node [style=none] (23) at (0.25, 3.375) {};
		\node [style=none] (24) at (0.25, 2.875) {};
		\node [style=none] (25) at (-1.025, 0.625) {\Alice};
		\node [style=none] (26) at (0.25, 2.375) {\Bob};
		\node [style=none] (27) at (0.25, 3.625) {$s$};
		\node [style=none] (28) at (0.75, -0.875) {};
		\node [style=none] (29) at (1.9, -0.875) {};
		\node [style=none] (30) at (1.325, -1.8) {};
		\node [style=none] (34) at (-2, -0.875) {};
		\node [style=none] (35) at (-0.775, -0.875) {};
		\node [style=none] (36) at (-1.35, -1.675) {};
		\node [style=none] (37) at (-1.3, -0.875) {};
		\node [style=none] (38) at (-1.3, 0.125) {};
		\node [style=none] (39) at (1.2, -0.875) {};
		\node [style=none] (40) at (1.2, 1.875) {};
		\node [style=none] (41) at (1.25, -1.125) {};
		\node [style=none] (42) at (1.35, -1.225) {$\vec{b}$};
		\node [style=none] (43) at (-1.5, -1.125) {};
		\node [style=none] (44) at (-1.35, -1.2) {$\vec{a}$};
		\node [dot] (45) at (-0.5, -0.375) {};
		\node [style=none] (46) at (-2.4, 0) {};
		\node [style=none] (47) at (2.5, 0) {};
	\end{pgfonlayer}
	\begin{pgfonlayer}{edgelayer}
		\draw (0.center) to (1.center);
		\draw (0.center) to (2.center);
		\draw (1.center) to (2.center);
		\draw [style=none] (5.center) to (6.center);
		\draw (7.center) to (8.center);
		\draw (7.center) to (9.center);
		\draw (9.center) to (10.center);
		\draw (10.center) to (8.center);
		\draw [style=none] (11.center) to (12.center);
		\draw (14.center) to (15.center);
		\draw (14.center) to (16.center);
		\draw (16.center) to (17.center);
		\draw (17.center) to (15.center);
		\draw [style=hyperbit] (19.center) to (18.center);
		\draw (20.center) to (21.center);
		\draw (20.center) to (22.center);
		\draw (21.center) to (22.center);
		\draw (23.center) to (24.center);
		\draw (28.center) to (29.center);
		\draw (28.center) to (30.center);
		\draw (29.center) to (30.center);
		\draw (34.center) to (35.center);
		\draw (34.center) to (36.center);
		\draw (35.center) to (36.center);
		\draw (37.center) to (38.center);
		\draw (40.center) to (39.center);
		\draw (45) to (12.center);
		\draw [dashed] (46.center) to (47.center);
	\end{pgfonlayer}
\end{tikzpicture}}.
	\end{equation}
\end{proposition}

\begin{proof}
    It is enough to observe that \eqref{eq:equivalenceQH_2} implies that Bob's strategy (the values of $k_i$'s) is fixed for all values of $z$; therefore the only way to satisfy the equivalence $\mathcal{H}\simeq \mathcal{Q}$ formally written in Eq. \eqref{eq:proof_g_t_equal} is via Eq. \eqref{eq:fixks}. But this is true only for $(x,y,z)\in \mathcal{D}$. Since there are still points $(x,y,z)\in\mathcal{C}/\mathcal{D}$, this proves that, without \textit{a priori} Bob's knowledge of $z$, in \textit{reductio ad absurdum} $\mathcal{H}\not\simeq \mathcal{Q}$.
\end{proof}

We thus infer that, in the absence of \textit{a priori} knowledge of \(z\), there is no \(z\)-invariant postprocessing strategy for Bob that yields an output \(\langle B (\vec{a},\vec{b},r,\vec{x}_{\vec{a}})\rangle_\mathcal{H}\) capable of reproducing all possible Bob responses \(\langle B (\vec{a},\vec{b},A,\rho)\rangle_\mathcal{Q}\).

In the context of quantum theory, a physical interpretation suggests that, if Bob had access to the information of \(z\), that is, if he knew Alice's preparation \textit{a priori}, the communication tasks would be rendered trivial as Alice's laboratory would essentially become accessible to Bob. Observe that, in such a case, Bob would have three \textit{a priori} inputs:
\begin{equation}\label{eq:equivalenceQH_3}
\begin{tikzpicture}
	\begin{pgfonlayer}{nodelayer}
		\node [style=none] (0) at (-0.7, -0.85) {};
		\node [style=none] (1) at (0.7, -0.85) {};
		\node [style=none] (2) at (0, -1.75) {};
		\node [style=none] (4) at (0, -1.2) {$\rho$};
		\node [style=none] (5) at (-0.5, -0.85) {};
		\node [style=none] (6) at (-0.5, 0.15) {};
		\node [style=none] (7) at (-2, 1.15) {};
		\node [style=none] (8) at (0, 1.15) {};
		\node [style=none] (9) at (-2, 0.15) {};
		\node [style=none] (10) at (0, 0.15) {};
		\node [style=none] (11) at (0.5, 1.9) {};
		\node [style=none] (12) at (0.5, -0.85) {};
		\node [style=none] (14) at (-1, 2.9) {};
		\node [style=none] (15) at (1.5, 2.9) {};
		\node [style=none] (16) at (-1, 1.9) {};
		\node [style=none] (17) at (1.5, 1.9) {};
		\node [style=none] (18) at (-0.5, 1.15) {};
		\node [style=none] (19) at (-0.5, 1.9) {};
		\node [style=none] (20) at (-0.25, 3.4) {};
		\node [style=none] (21) at (0.8, 3.4) {};
		\node [style=none] (22) at (0.25, 4.15) {};
		\node [style=none] (23) at (0.25, 3.4) {};
		\node [style=none] (24) at (0.25, 2.9) {};
		\node [style=none] (25) at (-1.025, 0.65) {\Alice};
		\node [style=none] (26) at (0.25, 2.4) {\Bob};
		\node [style=none] (27) at (0.25, 3.65) {$s$};
		\node [style=none] (28) at (0.75, -0.85) {};
		\node [style=none] (29) at (1.9, -0.85) {};
		\node [style=none] (30) at (1.325, -1.775) {};
		\node [style=none] (34) at (-2, -0.85) {};
		\node [style=none] (35) at (-0.775, -0.85) {};
		\node [style=none] (36) at (-1.35, -1.65) {};
		\node [style=none] (37) at (-1.3, -0.85) {};
		\node [style=none] (38) at (-1.3, 0.15) {};
		\node [style=none] (39) at (1.2, -0.85) {};
		\node [style=none] (40) at (1.2, 1.9) {};
		\node [style=none] (41) at (1.25, -1.1) {};
		\node [style=none] (42) at (1.35, -1.2) {$\vec{b}$};
		\node [style=none] (43) at (-1.5, -1.1) {};
		\node [style=none] (44) at (-1.35, -1.175) {$\vec{a}$};
		\node [style=none] (45) at (-0.85, 1.5) {$A$};
		\node [style=none] (46) at (-4, 0) {};
		\node [style=none] (48) at (2.5, 0) {};
		\node [style=none] (49) at (-3.25, -0.425) {\textit{a priori}};
	\end{pgfonlayer}
	\begin{pgfonlayer}{edgelayer}
		\draw (0.center) to (1.center);
		\draw (0.center) to (2.center);
		\draw (1.center) to (2.center);
		\draw [style=qWire] (5.center) to (6.center);
		\draw (7.center) to (8.center);
		\draw (7.center) to (9.center);
		\draw (9.center) to (10.center);
		\draw (10.center) to (8.center);
		\draw [style=qWire] (11.center) to (12.center);
		\draw (14.center) to (15.center);
		\draw (14.center) to (16.center);
		\draw (16.center) to (17.center);
		\draw (17.center) to (15.center);
		\draw (19.center) to (18.center);
		\draw (20.center) to (21.center);
		\draw (20.center) to (22.center);
		\draw (21.center) to (22.center);
		\draw (23.center) to (24.center);
		\draw (28.center) to (29.center);
		\draw (28.center) to (30.center);
		\draw (29.center) to (30.center);
		\draw (34.center) to (35.center);
		\draw (34.center) to (36.center);
		\draw (35.center) to (36.center);
		\draw (37.center) to (38.center);
		\draw (40.center) to (39.center);
		\draw [dashed] (46.center) to (48.center);
	\end{pgfonlayer}
\end{tikzpicture}}\quad\stackrel{w}{\simeq}\quad
\begin{tikzpicture}
	\begin{pgfonlayer}{nodelayer}
		\node [style=none] (0) at (-0.7, -2.65) {};
		\node [style=none] (1) at (0.7, -2.65) {};
		\node [style=none] (2) at (0, -3.55) {};
		\node [style=none] (4) at (0, -3) {$A$};
		\node [style=none] (5) at (-0.5, -2.65) {};
		\node [style=none] (6) at (-0.5, -1.65) {};
		\node [style=none] (7) at (-2, -0.65) {};
		\node [style=none] (8) at (0, -0.65) {};
		\node [style=none] (9) at (-2, -1.65) {};
		\node [style=none] (10) at (0, -1.65) {};
		\node [style=none] (11) at (0.5, 0.1) {};
		\node [style=none] (12) at (0.5, -2.15) {};
		\node [style=none] (14) at (-1, 1.1) {};
		\node [style=none] (15) at (1.5, 1.1) {};
		\node [style=none] (16) at (-1, 0.1) {};
		\node [style=none] (17) at (1.5, 0.1) {};
		\node [style=none] (18) at (-0.5, -0.65) {};
		\node [style=none] (19) at (-0.5, 0.1) {};
		\node [style=none] (20) at (-0.25, 1.6) {};
		\node [style=none] (21) at (0.8, 1.6) {};
		\node [style=none] (22) at (0.25, 2.35) {};
		\node [style=none] (23) at (0.25, 1.6) {};
		\node [style=none] (24) at (0.25, 1.1) {};
		\node [style=none] (25) at (-1.025, -1.15) {\Alice};
		\node [style=none] (26) at (0.25, 0.6) {\Bob};
		\node [style=none] (27) at (0.25, 1.85) {$s$};
		\node [style=none] (28) at (0.75, -2.65) {};
		\node [style=none] (29) at (1.9, -2.65) {};
		\node [style=none] (30) at (1.325, -3.575) {};
		\node [style=none] (34) at (-2, -2.65) {};
		\node [style=none] (35) at (-0.775, -2.65) {};
		\node [style=none] (36) at (-1.35, -3.45) {};
		\node [style=none] (37) at (-1.3, -2.65) {};
		\node [style=none] (38) at (-1.3, -1.65) {};
		\node [style=none] (39) at (1.2, -2.65) {};
		\node [style=none] (40) at (1.2, 0.1) {};
		\node [style=none] (41) at (1.25, -2.9) {};
		\node [style=none] (42) at (1.35, -3) {$\vec{b}$};
		\node [style=none] (43) at (-1.5, -2.9) {};
		\node [style=none] (44) at (-1.35, -2.975) {$\vec{a}$};
		\node [dot] (45) at (-0.5, -2.15) {};
		\node [style=none] (46) at (-2.4, 0) {};
		\node [style=none] (47) at (2.5, 0) {};
	\end{pgfonlayer}
	\begin{pgfonlayer}{edgelayer}
		\draw (0.center) to (1.center);
		\draw (0.center) to (2.center);
		\draw (1.center) to (2.center);
		\draw [style=none] (5.center) to (6.center);
		\draw (7.center) to (8.center);
		\draw (7.center) to (9.center);
		\draw (9.center) to (10.center);
		\draw (10.center) to (8.center);
		\draw [style=none] (11.center) to (12.center);
		\draw (14.center) to (15.center);
		\draw (14.center) to (16.center);
		\draw (16.center) to (17.center);
		\draw (17.center) to (15.center);
		\draw [style=hyperbit] (19.center) to (18.center);
		\draw (20.center) to (21.center);
		\draw (20.center) to (22.center);
		\draw (21.center) to (22.center);
		\draw (23.center) to (24.center);
		\draw (28.center) to (29.center);
		\draw (28.center) to (30.center);
		\draw (29.center) to (30.center);
		\draw (34.center) to (35.center);
		\draw (34.center) to (36.center);
		\draw (35.center) to (36.center);
		\draw (37.center) to (38.center);
		\draw (40.center) to (39.center);
		\draw (45) to (12.center);
		\draw [dashed] (46.center) to (47.center);
	\end{pgfonlayer}
\end{tikzpicture}}
\end{equation}
which contrasts the scenario of Eq.~\ref{eq:equivalenceQH_2}, where Bob has only two \textit{a priori} inputs. In this situation, the hyperbit conveys the information of \(\vec{x}_{\vec{a}}\), enabling Bob to adapt his strategy according to his \(z\)-dependent software program. Consequently, we have \(\langle B (\vec{a},\vec{b},A,\rho)\rangle_\mathcal{Q} = \langle B (\vec{a},\vec{b},r,\vec{x}_{\vec{a}})\rangle_\mathcal{H}\). Note that Eq. \eqref{eq:equivalenceQH_3} diagrammatically reformulates Proposition \ref{prop1}.

The resources available to Bob within quantum theory and hyperbit theory enhance our comprehension of quantum theory.  hyperbit theory, originally developed to simplify calculations, still remains categorized as a foil theory \cite{chiribella2016quantum}. Foil theories are purposefully designed to highlight the distinctive features within quantum mechanics. This categorization helps in delineating and deepening our appreciation of the specific and unique aspects of quantum theory.

\subsection{Related papers}\label{sec:proposals}

In Ref.~\cite{tavakoli2021correlations}, a finite gap was identified that exposed a specific type of inequivalence between the entanglement-assisted (EA) bit value and the hyperbit value, primarily due to the constraints of the PW protocol. This means that only a counterexample was found that disproves the equivalence $\mathcal{Q}\simeq\mathcal{H}$ without any explanation at the level of the interpretation. Our results, however, suggest that even when the most general strategy of Eq.~\eqref{eq:g} is employed, under the assumption that Bob cannot access Alice's laboratory (i.e., Bob does not know the expectation value \(z\)), the equivalence is limited to the \(\mathcal{D}\) region, as observed in the PW protocol (see Fig.~\ref{fig:marcinregion}). If the focus is solely on the violation of an inequality based on the upper bound of a linear functional, such as the witness used in \cite{tavakoli2021correlations}, one can only observe the discrepancy between quantum theory and hyperbit theory without understanding why the equivalence cannot be achieved or the underlying assumption that leads to such a \textit{no go} result. In our analysis, we pinpoint the no go result in diagram~\eqref{eq:equivalenceQH_2} and the equivalence of diagram~\eqref{eq:equivalenceQH_3} in cases where the communication is trivial. Both these diagrams are more informative than diagram \eqref{eq:equivalenceQH}.

\section{Conclusion}\label{sec:conclusions}
In this paper, we explored the Tsirelson isomorphism with the potential to simplify computations and offer a physical interpretation via the hyperbit concept. Yet, given the hypotheses that ground this isomorphism, such an interpretation could be perceived as a general probabilistic theory. Consequently, we demonstrated that, for communication tasks covered by the theorem in Ref. \cite{Pawlowski2012}, postprocessing operations on hyperbits and their associated dual effects exhibit equivalence in a confined region given an input-independent protocol. This finding served to clarify the previously discussed equivalence \(\mathcal{H}\simeq \mathcal{Q}\)~\cite{Pawlowski2012,tavakoli2021correlations}.
Further, we examined the conditions leading to disrupting this equivalence based on certain characteristics of post-processing and accessibility within the hyperbit theory.

Upon the hypotheses of Tsirelson's theorem, such as the commutativity among settings, further research might be directed toward formulating a CHSH-like correlator in hyperbit terms as it implies a form of ``quantumness" certification that may only involve a single correlation term similar to those defining the CHSH functional \cite{CHSH,Gigena2022}.
In addition to these insights, our geometrical construction lends support to the idea that quantum mechanics may carry more information about Alice's laboratory, even when Bob has no direct access, as in the case of hyperbits. Our proof shows that if Bob already knows the value of \(z\), a protocol satisfying \(g=t\) in Eq. \eqref{eq:proof_g_t_equal} always exists. When identifying \((x,y)\) as expectation values of some operational procedure, the linear trade-off of Eq. \eqref{eq:PWconstraints} is known to be noncontextual~\cite{Catani2022a,Catani2022b}, and the quadratic one of Eq. \eqref{eq:cylinder} is quantum. Such quantum superiority might be linked with contextuality akin to a random access code~\cite{horodecki2010contextuality}, opening new avenues for understanding and exploring the complex world of quantum communication.

\begin{acknowledgements}
    We acknowledge C. Raj whi contributed at the early stages of the project, and P. J. Cavalcanti for the discussion on the GPTs. G.S. is supported by QuantERA/2/2020, an ERA-Net co-fund in Quantum Technologies, under the eDICT project. A.G. is supported by funding from QuantERA, an ERA-Net cofund in Quantum Technologies, under the project eDICT. AG further benefited from support by project DESCOM (VEGA 2/0183/21). M.P. acknowledges support by NCN SHENG Grant NO.  UMO-2018/30/Q/ST2/00625.
\end{acknowledgements}

\bibliography{hyperbib}

\begin{thebibliography}{31}%
\makeatletter
\providecommand \@ifxundefined [1]{%
 \@ifx{#1\undefined}
}%
\providecommand \@ifnum [1]{%
 \ifnum #1\expandafter \@firstoftwo
 \else \expandafter \@secondoftwo
 \fi
}%
\providecommand \@ifx [1]{%
 \ifx #1\expandafter \@firstoftwo
 \else \expandafter \@secondoftwo
 \fi
}%
\providecommand \natexlab [1]{#1}%
\providecommand \enquote  [1]{``#1''}%
\providecommand \bibnamefont  [1]{#1}%
\providecommand \bibfnamefont [1]{#1}%
\providecommand \citenamefont [1]{#1}%
\providecommand \href@noop [0]{\@secondoftwo}%
\providecommand \href [0]{\begingroup \@sanitize@url \@href}%
\providecommand \@href[1]{\@@startlink{#1}\@@href}%
\providecommand \@@href[1]{\endgroup#1\@@endlink}%
\providecommand \@sanitize@url [0]{\catcode `\\12\catcode `\$12\catcode
  `\&12\catcode `\#12\catcode `\^12\catcode `\_12\catcode `\%12\relax}%
\providecommand \@@startlink[1]{}%
\providecommand \@@endlink[0]{}%
\providecommand \url  [0]{\begingroup\@sanitize@url \@url }%
\providecommand \@url [1]{\endgroup\@href {#1}{\urlprefix }}%
\providecommand \urlprefix  [0]{URL }%
\providecommand \Eprint [0]{\href }%
\providecommand \doibase [0]{https://doi.org/}%
\providecommand \selectlanguage [0]{\@gobble}%
\providecommand \bibinfo  [0]{\@secondoftwo}%
\providecommand \bibfield  [0]{\@secondoftwo}%
\providecommand \translation [1]{[#1]}%
\providecommand \BibitemOpen [0]{}%
\providecommand \bibitemStop [0]{}%
\providecommand \bibitemNoStop [0]{.\EOS\space}%
\providecommand \EOS [0]{\spacefactor3000\relax}%
\providecommand \BibitemShut  [1]{\csname bibitem#1\endcsname}%
\let\auto@bib@innerbib\@empty
\bibitem [{\citenamefont {de~Gois}\ \emph {et~al.}(2021)\citenamefont
  {de~Gois}, \citenamefont {Moreno}, \citenamefont {Nery}, \citenamefont
  {Brito}, \citenamefont {Chaves},\ and\ \citenamefont
  {Rabelo}}]{de2021general}%
  \BibitemOpen
  \bibfield  {author} {\bibinfo {author} {\bibfnamefont {C.}~\bibnamefont
  {de~Gois}}, \bibinfo {author} {\bibfnamefont {G.}~\bibnamefont {Moreno}},
  \bibinfo {author} {\bibfnamefont {R.}~\bibnamefont {Nery}}, \bibinfo {author}
  {\bibfnamefont {S.}~\bibnamefont {Brito}}, \bibinfo {author} {\bibfnamefont
  {R.}~\bibnamefont {Chaves}},\ and\ \bibinfo {author} {\bibfnamefont
  {R.}~\bibnamefont {Rabelo}},\ }\bibfield  {title} {\bibinfo {title} {General
  method for classicality certification in the prepare and measure scenario},\
  }\href {https://doi.org/https://doi.org/10.1103/PRXQuantum.2.030311}
  {\bibfield  {journal} {\bibinfo  {journal} {PRX Quantum}\ }\textbf {\bibinfo
  {volume} {2}},\ \bibinfo {pages} {030311} (\bibinfo {year}
  {2021})}\BibitemShut {NoStop}%
\bibitem [{\citenamefont {Tavakoli}\ \emph {et~al.}(2021)\citenamefont
  {Tavakoli}, \citenamefont {Pauwels}, \citenamefont {Woodhead},\ and\
  \citenamefont {Pironio}}]{tavakoli2021correlations}%
  \BibitemOpen
  \bibfield  {author} {\bibinfo {author} {\bibfnamefont {A.}~\bibnamefont
  {Tavakoli}}, \bibinfo {author} {\bibfnamefont {J.}~\bibnamefont {Pauwels}},
  \bibinfo {author} {\bibfnamefont {E.}~\bibnamefont {Woodhead}},\ and\
  \bibinfo {author} {\bibfnamefont {S.}~\bibnamefont {Pironio}},\ }\bibfield
  {title} {\bibinfo {title} {Correlations in entanglement-assisted
  prepare-and-measure scenarios},\ }\href
  {https://doi.org/https://doi.org/10.1103/PRXQuantum.2.040357} {\bibfield
  {journal} {\bibinfo  {journal} {PRX Quantum}\ }\textbf {\bibinfo {volume}
  {2}},\ \bibinfo {pages} {040357} (\bibinfo {year} {2021})}\BibitemShut
  {NoStop}%
\bibitem [{\citenamefont {Pauwels}\ \emph {et~al.}(2022)\citenamefont
  {Pauwels}, \citenamefont {Tavakoli}, \citenamefont {Woodhead},\ and\
  \citenamefont {Pironio}}]{pauwels2022entanglement}%
  \BibitemOpen
  \bibfield  {author} {\bibinfo {author} {\bibfnamefont {J.}~\bibnamefont
  {Pauwels}}, \bibinfo {author} {\bibfnamefont {A.}~\bibnamefont {Tavakoli}},
  \bibinfo {author} {\bibfnamefont {E.}~\bibnamefont {Woodhead}},\ and\
  \bibinfo {author} {\bibfnamefont {S.}~\bibnamefont {Pironio}},\ }\bibfield
  {title} {\bibinfo {title} {Entanglement in prepare-and-measure scenarios:
  many questions, a few answers},\ }\href
  {https://doi.org/https://doi.org/10.1088/1367-2630/ac724a} {\bibfield
  {journal} {\bibinfo  {journal} {New Journal of Physics}\ }\textbf {\bibinfo
  {volume} {24}},\ \bibinfo {pages} {063015} (\bibinfo {year}
  {2022})}\BibitemShut {NoStop}%
\bibitem [{\citenamefont {Bennett}\ \emph {et~al.}(1993)\citenamefont
  {Bennett}, \citenamefont {Brassard}, \citenamefont {Cr{\'e}peau},
  \citenamefont {Jozsa}, \citenamefont {Peres},\ and\ \citenamefont
  {Wootters}}]{bennett1993teleporting}%
  \BibitemOpen
  \bibfield  {author} {\bibinfo {author} {\bibfnamefont {C.~H.}\ \bibnamefont
  {Bennett}}, \bibinfo {author} {\bibfnamefont {G.}~\bibnamefont {Brassard}},
  \bibinfo {author} {\bibfnamefont {C.}~\bibnamefont {Cr{\'e}peau}}, \bibinfo
  {author} {\bibfnamefont {R.}~\bibnamefont {Jozsa}}, \bibinfo {author}
  {\bibfnamefont {A.}~\bibnamefont {Peres}},\ and\ \bibinfo {author}
  {\bibfnamefont {W.~K.}\ \bibnamefont {Wootters}},\ }\bibfield  {title}
  {\bibinfo {title} {Teleporting an unknown quantum state via dual classical
  and einstein-podolsky-rosen channels},\ }\href
  {https://doi.org/10.1103/PhysRevLett.70.1895} {\bibfield  {journal} {\bibinfo
   {journal} {Physical review letters}\ }\textbf {\bibinfo {volume} {70}},\
  \bibinfo {pages} {1895} (\bibinfo {year} {1993})}\BibitemShut {NoStop}%
\bibitem [{\citenamefont {Bennett}\ and\ \citenamefont
  {Wiesner}(1992)}]{bennett1992communication}%
  \BibitemOpen
  \bibfield  {author} {\bibinfo {author} {\bibfnamefont {C.~H.}\ \bibnamefont
  {Bennett}}\ and\ \bibinfo {author} {\bibfnamefont {S.~J.}\ \bibnamefont
  {Wiesner}},\ }\bibfield  {title} {\bibinfo {title} {Communication via one-and
  two-particle operators on einstein-podolsky-rosen states},\ }\href
  {https://doi.org/https://doi.org/10.1103/PhysRevLett.69.2881} {\bibfield
  {journal} {\bibinfo  {journal} {Physical review letters}\ }\textbf {\bibinfo
  {volume} {69}},\ \bibinfo {pages} {2881} (\bibinfo {year}
  {1992})}\BibitemShut {NoStop}%
\bibitem [{\citenamefont {Chaturvedi}\ \emph {et~al.}(2017)\citenamefont
  {Chaturvedi}, \citenamefont {Pawlowski},\ and\ \citenamefont
  {Horodecki}}]{EA}%
  \BibitemOpen
  \bibfield  {author} {\bibinfo {author} {\bibfnamefont {A.}~\bibnamefont
  {Chaturvedi}}, \bibinfo {author} {\bibfnamefont {M.}~\bibnamefont
  {Pawlowski}},\ and\ \bibinfo {author} {\bibfnamefont {K.}~\bibnamefont
  {Horodecki}},\ }\bibfield  {title} {\bibinfo {title} {Random access codes and
  nonlocal resources},\ }\href {https://doi.org/10.1103/PhysRevA.96.022125}
  {\bibfield  {journal} {\bibinfo  {journal} {Phys. Rev. A}\ }\textbf {\bibinfo
  {volume} {96}},\ \bibinfo {pages} {022125} (\bibinfo {year}
  {2017})}\BibitemShut {NoStop}%
\bibitem [{\citenamefont {Paw{\l}owski}\ and\ \citenamefont
  {{\.Z}ukowski}(2010)}]{pawlowski2010entanglement}%
  \BibitemOpen
  \bibfield  {author} {\bibinfo {author} {\bibfnamefont {M.}~\bibnamefont
  {Paw{\l}owski}}\ and\ \bibinfo {author} {\bibfnamefont {M.}~\bibnamefont
  {{\.Z}ukowski}},\ }\bibfield  {title} {\bibinfo {title}
  {Entanglement-assisted random access codes},\ }\href
  {https://doi.org/https://doi.org/10.1103/PhysRevA.81.042326} {\bibfield
  {journal} {\bibinfo  {journal} {Physical Review A}\ }\textbf {\bibinfo
  {volume} {81}},\ \bibinfo {pages} {042326} (\bibinfo {year}
  {2010})}\BibitemShut {NoStop}%
\bibitem [{\citenamefont {Hameedi}\ \emph {et~al.}(2017)\citenamefont
  {Hameedi}, \citenamefont {Saha}, \citenamefont {Mironowicz}, \citenamefont
  {Paw{\l}owski},\ and\ \citenamefont
  {Bourennane}}]{hameedi2017complementarity}%
  \BibitemOpen
  \bibfield  {author} {\bibinfo {author} {\bibfnamefont {A.}~\bibnamefont
  {Hameedi}}, \bibinfo {author} {\bibfnamefont {D.}~\bibnamefont {Saha}},
  \bibinfo {author} {\bibfnamefont {P.}~\bibnamefont {Mironowicz}}, \bibinfo
  {author} {\bibfnamefont {M.}~\bibnamefont {Paw{\l}owski}},\ and\ \bibinfo
  {author} {\bibfnamefont {M.}~\bibnamefont {Bourennane}},\ }\bibfield  {title}
  {\bibinfo {title} {Complementarity between entanglement-assisted and quantum
  distributed random access code},\ }\href
  {https://doi.org/https://doi.org/10.1103/PhysRevA.95.052345} {\bibfield
  {journal} {\bibinfo  {journal} {Physical Review A}\ }\textbf {\bibinfo
  {volume} {95}},\ \bibinfo {pages} {052345} (\bibinfo {year}
  {2017})}\BibitemShut {NoStop}%
\bibitem [{\citenamefont {Paw{\l}owski}\ and\ \citenamefont
  {Winter}(2012)}]{Pawlowski2012}%
  \BibitemOpen
  \bibfield  {author} {\bibinfo {author} {\bibfnamefont {M.}~\bibnamefont
  {Paw{\l}owski}}\ and\ \bibinfo {author} {\bibfnamefont {A.}~\bibnamefont
  {Winter}},\ }\bibfield  {title} {\bibinfo {title} {Hyperbits: The information
  quasiparticles},\ }\href {https://doi.org/10.1103/physreva.85.022331}
  {\bibfield  {journal} {\bibinfo  {journal} {Physical Review A}\ }\textbf
  {\bibinfo {volume} {85}},\ \bibinfo {pages} {022331} (\bibinfo {year}
  {2012})}\BibitemShut {NoStop}%
\bibitem [{\citenamefont {Scala}\ \emph {et~al.}(2020)\citenamefont {Scala},
  \citenamefont {Pepe}, \citenamefont {Facchi}, \citenamefont {Pascazio},\ and\
  \citenamefont {S{\l}owik}}]{LightMatter2020}%
  \BibitemOpen
  \bibfield  {author} {\bibinfo {author} {\bibfnamefont {G.}~\bibnamefont
  {Scala}}, \bibinfo {author} {\bibfnamefont {F.~V.}\ \bibnamefont {Pepe}},
  \bibinfo {author} {\bibfnamefont {P.}~\bibnamefont {Facchi}}, \bibinfo
  {author} {\bibfnamefont {S.}~\bibnamefont {Pascazio}},\ and\ \bibinfo
  {author} {\bibfnamefont {K.}~\bibnamefont {S{\l}owik}},\ }\bibfield  {title}
  {\bibinfo {title} {Light interaction with extended quantum systems in
  dispersive media},\ }\href
  {https://doi.org/https://doi.org/10.1088/1367-2630/abd204} {\bibfield
  {journal} {\bibinfo  {journal} {New Journal of Physics}\ }\textbf {\bibinfo
  {volume} {22}},\ \bibinfo {pages} {123047} (\bibinfo {year}
  {2020})}\BibitemShut {NoStop}%
\bibitem [{\citenamefont {Shen}\ \emph {et~al.}(2022)\citenamefont {Shen},
  \citenamefont {Zhang}, \citenamefont {Shi}, \citenamefont {Du}, \citenamefont
  {Zayats},\ and\ \citenamefont {Yuan}}]{Shen2022}%
  \BibitemOpen
  \bibfield  {author} {\bibinfo {author} {\bibfnamefont {Y.}~\bibnamefont
  {Shen}}, \bibinfo {author} {\bibfnamefont {Q.}~\bibnamefont {Zhang}},
  \bibinfo {author} {\bibfnamefont {P.}~\bibnamefont {Shi}}, \bibinfo {author}
  {\bibfnamefont {L.}~\bibnamefont {Du}}, \bibinfo {author} {\bibfnamefont
  {A.~V.}\ \bibnamefont {Zayats}},\ and\ \bibinfo {author} {\bibfnamefont
  {X.}~\bibnamefont {Yuan}},\ }\bibfield  {title} {\bibinfo {title}
  {Topological quasiparticles of light: Optical skyrmions and beyond},\ }\href
  {https://doi.org/10.48550/arXiv.2205.10329} {\bibfield  {journal} {\bibinfo
  {journal} {arXiv:2205.10329}\ } (\bibinfo {year} {2022})}\BibitemShut
  {NoStop}%
\bibitem [{\citenamefont {Chandran}\ \emph {et~al.}(2023)\citenamefont
  {Chandran}, \citenamefont {Iadecola}, \citenamefont {Khemani},\ and\
  \citenamefont {Moessner}}]{Chandran2023}%
  \BibitemOpen
  \bibfield  {author} {\bibinfo {author} {\bibfnamefont {A.}~\bibnamefont
  {Chandran}}, \bibinfo {author} {\bibfnamefont {T.}~\bibnamefont {Iadecola}},
  \bibinfo {author} {\bibfnamefont {V.}~\bibnamefont {Khemani}},\ and\ \bibinfo
  {author} {\bibfnamefont {R.}~\bibnamefont {Moessner}},\ }\bibfield  {title}
  {\bibinfo {title} {Quantum many-body scars: A quasiparticle perspective},\
  }\href {https://doi.org/10.1146/annurev-conmatphys-031620-101617} {\bibfield
  {journal} {\bibinfo  {journal} {Annual Review of Condensed Matter Physics}\
  }\textbf {\bibinfo {volume} {14}},\ \bibinfo {pages} {443} (\bibinfo {year}
  {2023})}\BibitemShut {NoStop}%
\bibitem [{\citenamefont {Wölfle}(2018)}]{Woelfle2018}%
  \BibitemOpen
  \bibfield  {author} {\bibinfo {author} {\bibfnamefont {P.}~\bibnamefont
  {Wölfle}},\ }\bibfield  {title} {\bibinfo {title} {Quasiparticles in
  condensed matter systems},\ }\href
  {https://doi.org/https://doi.org/10.1088/1361-6633/aa9bc4} {\bibfield
  {journal} {\bibinfo  {journal} {Reports on Progress in Physics}\ }\textbf
  {\bibinfo {volume} {81}},\ \bibinfo {pages} {032501} (\bibinfo {year}
  {2018})}\BibitemShut {NoStop}%
\bibitem [{\citenamefont {Rivera}\ and\ \citenamefont
  {Kaminer}(2020)}]{Rivera2020}%
  \BibitemOpen
  \bibfield  {author} {\bibinfo {author} {\bibfnamefont {N.}~\bibnamefont
  {Rivera}}\ and\ \bibinfo {author} {\bibfnamefont {I.}~\bibnamefont
  {Kaminer}},\ }\bibfield  {title} {\bibinfo {title} {Light{\textendash}matter
  interactions with photonic quasiparticles},\ }\href
  {https://doi.org/10.1038/s42254-020-0224-2} {\bibfield  {journal} {\bibinfo
  {journal} {Nature Reviews Physics}\ }\textbf {\bibinfo {volume} {2}},\
  \bibinfo {pages} {538} (\bibinfo {year} {2020})}\BibitemShut {NoStop}%
\bibitem [{\citenamefont {Rouhbakhsh~N}\ and\ \citenamefont
  {Ghoreishi}(2023)}]{rouhbakhsh2023geometric}%
  \BibitemOpen
  \bibfield  {author} {\bibinfo {author} {\bibfnamefont {M.}~\bibnamefont
  {Rouhbakhsh~N}}\ and\ \bibinfo {author} {\bibfnamefont {S.~A.}\ \bibnamefont
  {Ghoreishi}},\ }\bibfield  {title} {\bibinfo {title} {Geometric bloch vector
  solution to minimum-error discriminations of mixed qubit states},\ }\href
  {https://doi.org/10.1007/s11128-023-04080-4} {\bibfield  {journal} {\bibinfo
  {journal} {Quantum Information Processing}\ }\textbf {\bibinfo {volume}
  {22}},\ \bibinfo {pages} {323} (\bibinfo {year} {2023})}\BibitemShut
  {NoStop}%
\bibitem [{\citenamefont {{\.Z}yczkowski}(2008)}]{zyczkowski2008quartic}%
  \BibitemOpen
  \bibfield  {author} {\bibinfo {author} {\bibfnamefont {K.}~\bibnamefont
  {{\.Z}yczkowski}},\ }\bibfield  {title} {\bibinfo {title} {Quartic quantum
  theory: an extension of the standard quantum mechanics},\ }\href
  {https://doi.org/https://doi.org/10.1088/1751-8113/41/35/355302} {\bibfield
  {journal} {\bibinfo  {journal} {Journal of Physics A: Mathematical and
  Theoretical}\ }\textbf {\bibinfo {volume} {41}},\ \bibinfo {pages} {355302}
  (\bibinfo {year} {2008})}\BibitemShut {NoStop}%
\bibitem [{\citenamefont {Uhlmann}(1996)}]{uhlmann1996spheres}%
  \BibitemOpen
  \bibfield  {author} {\bibinfo {author} {\bibfnamefont {A.}~\bibnamefont
  {Uhlmann}},\ }\bibfield  {title} {\bibinfo {title} {Spheres and hemispheres
  as quantum state spaces},\ }\href
  {https://doi.org/https://doi.org/10.1016/0393-0440(95)00004-6} {\bibfield
  {journal} {\bibinfo  {journal} {Journal of Geometry and Physics}\ }\textbf
  {\bibinfo {volume} {18}},\ \bibinfo {pages} {76} (\bibinfo {year}
  {1996})}\BibitemShut {NoStop}%
\bibitem [{\citenamefont {Kurzy{\'n}ski}\ \emph {et~al.}(2016)\citenamefont
  {Kurzy{\'n}ski}, \citenamefont {Ko{\l}odziejski}, \citenamefont {Laskowski},\
  and\ \citenamefont {Markiewicz}}]{kurzynski2016three}%
  \BibitemOpen
  \bibfield  {author} {\bibinfo {author} {\bibfnamefont {P.}~\bibnamefont
  {Kurzy{\'n}ski}}, \bibinfo {author} {\bibfnamefont {A.}~\bibnamefont
  {Ko{\l}odziejski}}, \bibinfo {author} {\bibfnamefont {W.}~\bibnamefont
  {Laskowski}},\ and\ \bibinfo {author} {\bibfnamefont {M.}~\bibnamefont
  {Markiewicz}},\ }\bibfield  {title} {\bibinfo {title} {Three-dimensional
  visualization of a qutrit},\ }\href
  {https://doi.org/https://doi.org/10.1103/PhysRevA.93.062126} {\bibfield
  {journal} {\bibinfo  {journal} {Physical Review A}\ }\textbf {\bibinfo
  {volume} {93}},\ \bibinfo {pages} {062126} (\bibinfo {year}
  {2016})}\BibitemShut {NoStop}%
\bibitem [{\citenamefont {Sharma}\ and\ \citenamefont
  {Ghosh}(2021)}]{sharma2021four}%
  \BibitemOpen
  \bibfield  {author} {\bibinfo {author} {\bibfnamefont {G.}~\bibnamefont
  {Sharma}}\ and\ \bibinfo {author} {\bibfnamefont {S.}~\bibnamefont {Ghosh}},\
  }\bibfield  {title} {\bibinfo {title} {Four-dimensional bloch sphere
  representation of qutrits using heisenberg-weyl operators},\ }\href
  {https://doi.org/10.48550/arXiv.2101.06408} {\bibfield  {journal} {\bibinfo
  {journal} {arXiv:2101.06408}\ } (\bibinfo {year} {2021})}\BibitemShut
  {NoStop}%
\bibitem [{\citenamefont {Eltschka}\ \emph {et~al.}(2021)\citenamefont
  {Eltschka}, \citenamefont {Huber}, \citenamefont {Morelli},\ and\
  \citenamefont {Siewert}}]{eltschka2021shape}%
  \BibitemOpen
  \bibfield  {author} {\bibinfo {author} {\bibfnamefont {C.}~\bibnamefont
  {Eltschka}}, \bibinfo {author} {\bibfnamefont {M.}~\bibnamefont {Huber}},
  \bibinfo {author} {\bibfnamefont {S.}~\bibnamefont {Morelli}},\ and\ \bibinfo
  {author} {\bibfnamefont {J.}~\bibnamefont {Siewert}},\ }\bibfield  {title}
  {\bibinfo {title} {The shape of higher-dimensional state space: Bloch-ball
  analog for a qutrit},\ }\href
  {https://doi.org/https://doi.org/10.22331/q-2021-06-29-485} {\bibfield
  {journal} {\bibinfo  {journal} {Quantum}\ }\textbf {\bibinfo {volume} {5}},\
  \bibinfo {pages} {485} (\bibinfo {year} {2021})}\BibitemShut {NoStop}%
\bibitem [{\citenamefont {Mazurek}\ \emph {et~al.}(2021)\citenamefont
  {Mazurek}, \citenamefont {Pusey}, \citenamefont {Resch},\ and\ \citenamefont
  {Spekkens}}]{Mazurek2021}%
  \BibitemOpen
  \bibfield  {author} {\bibinfo {author} {\bibfnamefont {M.~D.}\ \bibnamefont
  {Mazurek}}, \bibinfo {author} {\bibfnamefont {M.~F.}\ \bibnamefont {Pusey}},
  \bibinfo {author} {\bibfnamefont {K.~J.}\ \bibnamefont {Resch}},\ and\
  \bibinfo {author} {\bibfnamefont {R.~W.}\ \bibnamefont {Spekkens}},\
  }\bibfield  {title} {\bibinfo {title} {Experimentally bounding deviations
  from quantum theory in the landscape of generalized probabilistic theories},\
  }\href {https://doi.org/https://doi.org/10.1103/prxquantum.2.020302}
  {\bibfield  {journal} {\bibinfo  {journal} {{PRX} Quantum}\ }\textbf
  {\bibinfo {volume} {2}},\ \bibinfo {pages} {020302} (\bibinfo {year}
  {2021})}\BibitemShut {NoStop}%
\bibitem [{\citenamefont {Janotta}\ and\ \citenamefont
  {Hinrichsen}(2014)}]{Janotta2014}%
  \BibitemOpen
  \bibfield  {author} {\bibinfo {author} {\bibfnamefont {P.}~\bibnamefont
  {Janotta}}\ and\ \bibinfo {author} {\bibfnamefont {H.}~\bibnamefont
  {Hinrichsen}},\ }\bibfield  {title} {\bibinfo {title} {Generalized
  probability theories: what determines the structure of quantum theory?},\
  }\href {https://doi.org/10.1088/1751-8113/47/32/323001} {\bibfield  {journal}
  {\bibinfo  {journal} {Journal of Physics A: Mathematical and Theoretical}\
  }\textbf {\bibinfo {volume} {47}},\ \bibinfo {pages} {323001} (\bibinfo
  {year} {2014})}\BibitemShut {NoStop}%
\bibitem [{\citenamefont {Coecke}\ and\ \citenamefont
  {Kissinger}(2018)}]{coecke2018picturing}%
  \BibitemOpen
  \bibfield  {author} {\bibinfo {author} {\bibfnamefont {B.}~\bibnamefont
  {Coecke}}\ and\ \bibinfo {author} {\bibfnamefont {A.}~\bibnamefont
  {Kissinger}},\ }\bibfield  {title} {\bibinfo {title} {Picturing quantum
  processes: A first course on quantum theory and diagrammatic reasoning},\
  }in\ \href {https://doi.org/https://doi.org/10.1007/978-3-319-91376-6_6}
  {\emph {\bibinfo {booktitle} {Diagrammatic Representation and Inference: 10th
  International Conference, Diagrams 2018, Edinburgh, UK, June 18-22, 2018,
  Proceedings 10}}}\ (\bibinfo {organization} {Springer},\ \bibinfo {year}
  {2018})\ pp.\ \bibinfo {pages} {28--31}\BibitemShut {NoStop}%
\bibitem [{\citenamefont {Penrose}(1971)}]{penrose1971applications}%
  \BibitemOpen
  \bibfield  {author} {\bibinfo {author} {\bibfnamefont {R.}~\bibnamefont
  {Penrose}},\ }\bibfield  {title} {\bibinfo {title} {Applications of negative
  dimensional tensors},\ }\href@noop {} {\bibfield  {journal} {\bibinfo
  {journal} {Combinatorial mathematics and its applications}\ }\textbf
  {\bibinfo {volume} {1}},\ \bibinfo {pages} {221} (\bibinfo {year}
  {1971})}\BibitemShut {NoStop}%
\bibitem [{\citenamefont {Tsirel{\textquotesingle}son}(1987)}]{Tsirelson1987}%
  \BibitemOpen
  \bibfield  {author} {\bibinfo {author} {\bibfnamefont {B.~S.}\ \bibnamefont
  {Tsirel{\textquotesingle}son}},\ }\bibfield  {title} {\bibinfo {title}
  {Quantum analogues of the bell inequalities. the case of two spatially
  separated domains},\ }\href {https://doi.org/10.1007/bf01663472} {\bibfield
  {journal} {\bibinfo  {journal} {Journal of Soviet Mathematics}\ }\textbf
  {\bibinfo {volume} {36}},\ \bibinfo {pages} {557} (\bibinfo {year}
  {1987})}\BibitemShut {NoStop}%
\bibitem [{\citenamefont {Chiribella}\ and\ \citenamefont
  {Spekkens}(2016)}]{chiribella2016quantum}%
  \BibitemOpen
  \bibfield  {author} {\bibinfo {author} {\bibfnamefont {G.}~\bibnamefont
  {Chiribella}}\ and\ \bibinfo {author} {\bibfnamefont {R.~W.}\ \bibnamefont
  {Spekkens}},\ }\href
  {https://doi.org/https://doi.org/10.1007/978-94-017-7303-4} {\emph {\bibinfo
  {title} {Quantum theory: informational foundations and foils}}}\ (\bibinfo
  {publisher} {Springer},\ \bibinfo {year} {2016})\BibitemShut {NoStop}%
\bibitem [{\citenamefont {Clauser}\ \emph {et~al.}(1969)\citenamefont
  {Clauser}, \citenamefont {Horne}, \citenamefont {Shimony},\ and\
  \citenamefont {Holt}}]{CHSH}%
  \BibitemOpen
  \bibfield  {author} {\bibinfo {author} {\bibfnamefont {J.~F.}\ \bibnamefont
  {Clauser}}, \bibinfo {author} {\bibfnamefont {M.~A.}\ \bibnamefont {Horne}},
  \bibinfo {author} {\bibfnamefont {A.}~\bibnamefont {Shimony}},\ and\ \bibinfo
  {author} {\bibfnamefont {R.~A.}\ \bibnamefont {Holt}},\ }\bibfield  {title}
  {\bibinfo {title} {Proposed experiment to test local hidden-variable
  theories},\ }\href {https://doi.org/10.1103/PhysRevLett.23.880} {\bibfield
  {journal} {\bibinfo  {journal} {Phys. Rev. Lett.}\ }\textbf {\bibinfo
  {volume} {23}},\ \bibinfo {pages} {880} (\bibinfo {year} {1969})}\BibitemShut
  {NoStop}%
\bibitem [{\citenamefont {Gigena}\ \emph {et~al.}(2023)\citenamefont {Gigena},
  \citenamefont {Scala},\ and\ \citenamefont {Mandarino}}]{Gigena2022}%
  \BibitemOpen
  \bibfield  {author} {\bibinfo {author} {\bibfnamefont {N.}~\bibnamefont
  {Gigena}}, \bibinfo {author} {\bibfnamefont {G.}~\bibnamefont {Scala}},\ and\
  \bibinfo {author} {\bibfnamefont {A.}~\bibnamefont {Mandarino}},\ }\bibfield
  {title} {\bibinfo {title} {Revisited aspects of the local set in {CHSH} bell
  scenario},\ }\bibfield  {journal} {\bibinfo  {journal} {International Journal
  of Quantum Information}\ }\textbf {\bibinfo {volume} {21}},\ \href
  {https://doi.org/10.1142/s0219749923400051} {10.1142/s0219749923400051}
  (\bibinfo {year} {2023})\BibitemShut {NoStop}%
\bibitem [{\citenamefont {Catani}\ \emph {et~al.}(2023)\citenamefont {Catani},
  \citenamefont {Leifer}, \citenamefont {Scala}, \citenamefont {Schmid},\ and\
  \citenamefont {Spekkens}}]{Catani2022a}%
  \BibitemOpen
  \bibfield  {author} {\bibinfo {author} {\bibfnamefont {L.}~\bibnamefont
  {Catani}}, \bibinfo {author} {\bibfnamefont {M.}~\bibnamefont {Leifer}},
  \bibinfo {author} {\bibfnamefont {G.}~\bibnamefont {Scala}}, \bibinfo
  {author} {\bibfnamefont {D.}~\bibnamefont {Schmid}},\ and\ \bibinfo {author}
  {\bibfnamefont {R.~W.}\ \bibnamefont {Spekkens}},\ }\bibfield  {title}
  {\bibinfo {title} {Aspects of the phenomenology of interference that are
  genuinely nonclassical},\ }\href
  {https://doi.org/10.1103/PhysRevA.108.022207} {\bibfield  {journal} {\bibinfo
   {journal} {Phys. Rev. A}\ }\textbf {\bibinfo {volume} {108}},\ \bibinfo
  {pages} {022207} (\bibinfo {year} {2023})}\BibitemShut {NoStop}%
\bibitem [{\citenamefont {Catani}\ \emph {et~al.}(2022)\citenamefont {Catani},
  \citenamefont {Leifer}, \citenamefont {Scala}, \citenamefont {Schmid},\ and\
  \citenamefont {Spekkens}}]{Catani2022b}%
  \BibitemOpen
  \bibfield  {author} {\bibinfo {author} {\bibfnamefont {L.}~\bibnamefont
  {Catani}}, \bibinfo {author} {\bibfnamefont {M.}~\bibnamefont {Leifer}},
  \bibinfo {author} {\bibfnamefont {G.}~\bibnamefont {Scala}}, \bibinfo
  {author} {\bibfnamefont {D.}~\bibnamefont {Schmid}},\ and\ \bibinfo {author}
  {\bibfnamefont {R.~W.}\ \bibnamefont {Spekkens}},\ }\bibfield  {title}
  {\bibinfo {title} {What is nonclassical about uncertainty relations?},\
  }\href {https://doi.org/10.1103/PhysRevLett.129.240401} {\bibfield  {journal}
  {\bibinfo  {journal} {Phys. Rev. Lett.}\ }\textbf {\bibinfo {volume} {129}},\
  \bibinfo {pages} {240401} (\bibinfo {year} {2022})}\BibitemShut {NoStop}%
\bibitem [{\citenamefont {Horodecki}\ \emph {et~al.}(2010)\citenamefont
  {Horodecki}, \citenamefont {Horodecki}, \citenamefont {Horodecki},
  \citenamefont {Horodecki}, \citenamefont {Pawlowski},\ and\ \citenamefont
  {Bourennane}}]{horodecki2010contextuality}%
  \BibitemOpen
  \bibfield  {author} {\bibinfo {author} {\bibfnamefont {K.}~\bibnamefont
  {Horodecki}}, \bibinfo {author} {\bibfnamefont {M.}~\bibnamefont
  {Horodecki}}, \bibinfo {author} {\bibfnamefont {P.}~\bibnamefont
  {Horodecki}}, \bibinfo {author} {\bibfnamefont {R.}~\bibnamefont
  {Horodecki}}, \bibinfo {author} {\bibfnamefont {M.}~\bibnamefont
  {Pawlowski}},\ and\ \bibinfo {author} {\bibfnamefont {M.}~\bibnamefont
  {Bourennane}},\ }\href
  {https://doi.org/https://doi.org/10.48550/arXiv.1006.0468} {\bibinfo {title}
  {Contextuality offers device-independent security}} (\bibinfo {year}
  {2010}),\ \Eprint {https://arxiv.org/abs/1006.0468} {arXiv:1006.0468
  [quant-ph]} \BibitemShut {NoStop}%
\end{thebibliography}%

\appendix
\section{Operational features}\label{sec:GPTs}
In this appendix, we explain the details of Eq. \ref{eq:equivalenceQH} that summarize the whole paper.
Let us start from the quantum protocol to compute the probability that Bob's outcome is $s$:
\begin{equation}\label{eq:equivalenceQH2}
\begin{tikzpicture}
	\begin{pgfonlayer}{nodelayer}
		\node [style=none] (0) at (-0.7, 0) {};
		\node [style=none] (1) at (0.7, 0) {};
		\node [style=none] (2) at (0, -0.9) {};
		\node [style=none] (4) at (0, -0.35) {$\rho$};
		\node [style=none] (5) at (-0.5, 0) {};
		\node [style=none] (6) at (-0.5, 1) {};
		\node [style=none] (7) at (-2, 2) {};
		\node [style=none] (8) at (0, 2) {};
		\node [style=none] (9) at (-2, 1) {};
		\node [style=none] (10) at (0, 1) {};
		\node [style=none] (11) at (0.5, 2.75) {};
		\node [style=none] (12) at (0.5, 0) {};
		\node [style=none] (14) at (-1, 3.75) {};
		\node [style=none] (15) at (1.5, 3.75) {};
		\node [style=none] (16) at (-1, 2.75) {};
		\node [style=none] (17) at (1.5, 2.75) {};
		\node [style=none] (18) at (-0.5, 2) {};
		\node [style=none] (19) at (-0.5, 2.75) {};
		\node [style=none] (20) at (-0.25, 4.25) {};
		\node [style=none] (21) at (0.8, 4.25) {};
		\node [style=none] (22) at (0.25, 5) {};
		\node [style=none] (23) at (0.25, 4.25) {};
		\node [style=none] (24) at (0.25, 3.75) {};
		\node [style=none] (25) at (-1.025, 1.5) {\Alice};
		\node [style=none] (26) at (0.25, 3.25) {\Bob};
		\node [style=none] (27) at (0.25, 4.5) {$s$};
		\node [style=none] (28) at (0.75, 0) {};
		\node [style=none] (29) at (1.9, 0) {};
		\node [style=none] (30) at (1.325, -0.925) {};
		\node [style=none] (34) at (-2, 0) {};
		\node [style=none] (35) at (-0.775, 0) {};
		\node [style=none] (36) at (-1.35, -0.8) {};
		\node [style=none] (37) at (-1.3, 0) {};
		\node [style=none] (38) at (-1.3, 1) {};
		\node [style=none] (39) at (1.2, 0) {};
		\node [style=none] (40) at (1.2, 2.75) {};
		\node [style=none] (41) at (1.25, -0.25) {};
		\node [style=none] (42) at (1.35, -0.35) {$\vec{b}$};
		\node [style=none] (43) at (-1.5, -0.25) {};
		\node [style=none] (44) at (-1.35, -0.325) {$\vec{a}$};
		\node [style=none] (45) at (-0.85, 2.35) {$r$};
	\end{pgfonlayer}
	\begin{pgfonlayer}{edgelayer}
		\draw (0.center) to (1.center);
		\draw (0.center) to (2.center);
		\draw (1.center) to (2.center);
		\draw [style=qWire] (5.center) to (6.center);
		\draw (7.center) to (8.center);
		\draw (7.center) to (9.center);
		\draw (9.center) to (10.center);
		\draw (10.center) to (8.center);
		\draw [style=qWire] (11.center) to (12.center);
		\draw (14.center) to (15.center);
		\draw (14.center) to (16.center);
		\draw (16.center) to (17.center);
		\draw (17.center) to (15.center);
		\draw (19.center) to (18.center);
		\draw (20.center) to (21.center);
		\draw (20.center) to (22.center);
		\draw (21.center) to (22.center);
		\draw (23.center) to (24.center);
		\draw (28.center) to (29.center);
		\draw (28.center) to (30.center);
		\draw (29.center) to (30.center);
		\draw (34.center) to (35.center);
		\draw (34.center) to (36.center);
		\draw (35.center) to (36.center);
		\draw (37.center) to (38.center);
		\draw (40.center) to (39.center);
	\end{pgfonlayer}
\end{tikzpicture}}.
\end{equation}
Alice's lab is labeled as $\mathcal{A}$ and depending on the input bitstring $\vec{a}$ she selects one of the possible dicothomic measurements $A_{\vec{a}^\prime}$,
\begin{equation}
\begin{tikzpicture}
	\begin{pgfonlayer}{nodelayer}
		\node [style=none] (0) at (-1, 0.5) {};
		\node [style=none] (1) at (1, 0.5) {};
		\node [style=none] (2) at (-1, -0.5) {};
		\node [style=none] (3) at (1, -0.5) {};
		\node [style=none] (4) at (-0.5, -1.5) {};
		\node [style=none] (5) at (0.25, -1.5) {};
		\node [style=none] (6) at (-0.5, -0.5) {};
		\node [style=none] (7) at (0.25, -0.5) {};
		\node [style=none] (8) at (0, 1.5) {};
		\node [style=none] (9) at (0, 0.5) {};
		\node [style=none] (10) at (0, 0) {\Alice};
	\end{pgfonlayer}
	\begin{pgfonlayer}{edgelayer}
		\draw (0.center) to (2.center);
		\draw (0.center) to (1.center);
		\draw (1.center) to (3.center);
		\draw (3.center) to (2.center);
		\draw (6.center) to (4.center);
		\draw [style=qWire] (7.center) to (5.center);
		\draw (9.center) to (8.center);
	\end{pgfonlayer}
\end{tikzpicture}}=\sum_{\vec{a}^\prime}
\begin{tikzpicture}
	\begin{pgfonlayer}{nodelayer}
		\node [style=none] (61) at (-0.75, 0.8) {};
		\node [style=none] (62) at (-1.5, -0.45) {};
		\node [style=none] (63) at (0, -0.45) {};
		\node [style=none] (64) at (-0.75, -1.45) {};
		\node [style=none] (65) at (-0.75, -0.45) {};
		\node [style=none] (66) at (-0.75, 0.05) {};
		\node [style=none] (67) at (-0.75, 0.1) {$\vec{a}^\prime$};
		\node [style=none] (68) at (0.25, 0.55) {};
		\node [style=none] (69) at (0.25, -0.45) {};
		\node [style=none] (70) at (1.75, -0.45) {};
		\node [style=none] (71) at (1.75, 0.55) {};
		\node [style=none] (72) at (1, 1.55) {};
		\node [style=none] (73) at (1, -1.45) {};
		\node [style=none] (75) at (1, 0.55) {};
		\node [style=none] (76) at (0.75, -0.45) {};
		\node [style=none] (77) at (0.75, -0.45) {};
		\node [style=none] (78) at (0.75, -0.45) {};
		\node [style=none] (79) at (1, -0.45) {};
		\node [style=none] (81) at (1, 0.05) {$A_{\vec{a}^\prime}$};
	\end{pgfonlayer}
	\begin{pgfonlayer}{edgelayer}
		\draw (61.center) to (62.center);
		\draw (61.center) to (63.center);
		\draw (63.center) to (62.center);
		\draw (65.center) to (64.center);
		\draw (68.center) to (69.center);
		\draw (68.center) to (71.center);
		\draw (71.center) to (70.center);
		\draw (70.center) to (69.center);
		\draw (72.center) to (75.center);
		\draw [style=qWire] (79.center) to (73.center);
	\end{pgfonlayer}
\end{tikzpicture}}.
\end{equation}
In the sum only one term is connected to the input so that the cardinality of the input and output \emph{legs} on the LHS is the same as the one on the RHS by the action of the selector $\vec{a}^\prime$, indeed
\begin{equation}
	\delta_{\vec{a}^\prime \vec{a}}=%
\begin{tikzpicture}
	\begin{pgfonlayer}{nodelayer}
		\node [style=none] (61) at (0, 1.7) {};
		\node [style=none] (62) at (-0.75, 0.45) {};
		\node [style=none] (63) at (0.75, 0.45) {};
		\node [style=none] (64) at (0, -0.55) {};
		\node [style=none] (65) at (0, 0.45) {};
		\node [style=none] (66) at (0, 0.95) {};
		\node [style=none] (67) at (0, 0.975) {$\vec{a}^\prime$};
		\node [style=none] (68) at (0, -1.825) {};
		\node [style=none] (69) at (-0.75, -0.575) {};
		\node [style=none] (70) at (0.75, -0.575) {};
		\node [style=none] (72) at (0, -1.1) {$\vec{a}$};
	\end{pgfonlayer}
	\begin{pgfonlayer}{edgelayer}
		\draw (61.center) to (62.center);
		\draw (61.center) to (63.center);
		\draw (63.center) to (62.center);
		\draw (65.center) to (64.center);
		\draw (68.center) to (69.center);
		\draw (68.center) to (70.center);
		\draw (70.center) to (69.center);
	\end{pgfonlayer}
\end{tikzpicture}}.
\end{equation}
The spectral representation of Alice's dicothomic measurement $A_{\vec{a}^\prime}$ with PVM elements $\{P_{\vec{a}^\prime}^+,P_{\vec{a}^\prime}^-\}$ is
\begin{equation}	%
\begin{tikzpicture}
	\begin{pgfonlayer}{nodelayer}
		\node [style=none] (0) at (-0.75, 0.55) {};
		\node [style=none] (1) at (-0.75, -0.45) {};
		\node [style=none] (2) at (0.75, -0.45) {};
		\node [style=none] (3) at (0.75, 0.55) {};
		\node [style=none] (4) at (0, 1.55) {};
		\node [style=none] (5) at (0, -1.45) {};
		\node [style=none] (6) at (0, 0.55) {};
		\node [style=none] (7) at (-0.25, -0.45) {};
		\node [style=none] (8) at (-0.25, -0.45) {};
		\node [style=none] (9) at (-0.25, -0.45) {};
		\node [style=none] (10) at (0, -0.45) {};
		\node [style=none] (11) at (0, 0.05) {$A_{\vec{a}^\prime}$};
	\end{pgfonlayer}
	\begin{pgfonlayer}{edgelayer}
		\draw (0.center) to (1.center);
		\draw (0.center) to (3.center);
		\draw (3.center) to (2.center);
		\draw (2.center) to (1.center);
		\draw (4.center) to (6.center);
		\draw [style=qWire] (10.center) to (5.center);
	\end{pgfonlayer}
\end{tikzpicture}}=%
\begin{tikzpicture}
	\begin{pgfonlayer}{nodelayer}
		\node [style=none] (16) at (-2, 0.3) {};
		\node [style=none] (17) at (-2.75, 1.075) {};
		\node [style=none] (18) at (-1.25, 1.075) {};
		\node [style=none] (19) at (-3, -1.275) {};
		\node [style=none] (20) at (-1, -1.275) {};
		\node [style=none] (21) at (-2, 0.775) {$+$};
		\node [style=none] (22) at (-2, -0.725) {$P_{\vec{a'}}^{+}$};
		\node [style=none] (23) at (1.75, 0.3) {};
		\node [style=none] (24) at (1, 1.075) {};
		\node [style=none] (25) at (2.5, 1.075) {};
		\node [style=none] (26) at (0.75, -1.275) {};
		\node [style=none] (27) at (2.75, -1.275) {};
		\node [style=none] (28) at (1.75, 0.775) {$-$};
		\node [style=none] (29) at (1.75, -0.725) {$P_{\vec{a'}}^{-}$};
		\node [style=none] (30) at (0, 0) {$+$};
		\node [style=none] (35) at (-2, -1.75) {};
		\node [style=none] (36) at (-2, -1.275) {};
		\node [style=none] (37) at (-2, 1.075) {};
		\node [style=none] (38) at (-2, 1.55) {};
		\node [style=none] (75) at (1.75, -1.75) {};
		\node [style=none] (76) at (1.75, -1.275) {};
		\node [style=none] (77) at (1.75, 1.075) {};
		\node [style=none] (78) at (1.75, 1.5) {};
	\end{pgfonlayer}
	\begin{pgfonlayer}{edgelayer}
		\draw (17.center) to (18.center);
		\draw (18.center) to (16.center);
		\draw (17.center) to (16.center);
		\draw (16.center) to (19.center);
		\draw (19.center) to (20.center);
		\draw (16.center) to (20.center);
		\draw (24.center) to (25.center);
		\draw (25.center) to (23.center);
		\draw (24.center) to (23.center);
		\draw (23.center) to (26.center);
		\draw (26.center) to (27.center);
		\draw (23.center) to (27.center);
		\draw [style=qWire] (36.center) to (35.center);
		\draw (37.center) to (38.center);
		\draw [style=qWire] (76.center) to (75.center);
		\draw (78.center) to (77.center);
	\end{pgfonlayer}
\end{tikzpicture}},
\end{equation}
where only one channel is activated so that the number of legs on both sides remains the same.
After the measurement, Alice outputs the outcome $A\in\{-1,+1\}$ and sends it to Bob's lab labeled as $\mathcal{B}$. Here, Bob will receive two \textit{a priori} inputs, i.e., before the communication start: the bitstring $\vec{b}$ and the shared entangled state $\rho$ represented with a double wire. The third Bob's input is received during the communication, namely, Alice's outcome that selects one of his dichotomic observables $B_{\vec{A^\prime,\vec{b}^\prime}}$. Specifically,
\begin{equation}
\begin{tikzpicture}
	\begin{pgfonlayer}{nodelayer}
		\node [style=none] (0) at (-1, 0.5) {};
		\node [style=none] (1) at (1, 0.5) {};
		\node [style=none] (2) at (-1, -0.5) {};
		\node [style=none] (3) at (1, -0.5) {};
		\node [style=none] (4) at (-0.75, -1.5) {};
		\node [style=none] (5) at (0, -1.5) {};
		\node [style=none] (6) at (-0.75, -0.5) {};
		\node [style=none] (7) at (0, -0.5) {};
		\node [style=none] (8) at (0, 1.5) {};
		\node [style=none] (9) at (0, 0.5) {};
		\node [style=none] (10) at (0, 0) {$\mathcal{B}$};
		\node [style=none] (11) at (0.75, -1.5) {};
		\node [style=none] (12) at (0.75, -0.5) {};
	\end{pgfonlayer}
	\begin{pgfonlayer}{edgelayer}
		\draw (0.center) to (2.center);
		\draw (0.center) to (1.center);
		\draw (1.center) to (3.center);
		\draw (3.center) to (2.center);
		\draw (6.center) to (4.center);
		\draw [style=qWire] (7.center) to (5.center);
		\draw (9.center) to (8.center);
		\draw (12.center) to (11.center);
	\end{pgfonlayer}
\end{tikzpicture}}=\sum_{\vec{b}^\prime}\sum_{A^\prime=\pm 1}%
\begin{tikzpicture}
	\begin{pgfonlayer}{nodelayer}
		\node [style=none] (15) at (-2, 0.75) {};
		\node [style=none] (16) at (-2.75, -0.5) {};
		\node [style=none] (17) at (-1.25, -0.5) {};
		\node [style=none] (18) at (-2, -1.5) {};
		\node [style=none] (19) at (-2, -0.5) {};
		\node [style=none] (20) at (-2, 0) {};
		\node [style=none] (21) at (-2, -0.125) {$A'$};
		\node [style=none] (22) at (-0.75, 0.75) {};
		\node [style=none] (23) at (-0.75, -0.5) {};
		\node [style=none] (24) at (1, -0.5) {};
		\node [style=none] (25) at (1, 0.75) {};
		\node [style=none] (26) at (0, 1.5) {};
		\node [style=none] (27) at (0, -1.5) {};
		\node [style=none] (29) at (0, 0.75) {};
		\node [style=none] (30) at (-0.25, -0.5) {};
		\node [style=none] (31) at (-0.25, -0.5) {};
		\node [style=none] (32) at (-0.25, -0.5) {};
		\node [style=none] (33) at (0, -0.5) {};
		\node [style=none] (35) at (0.1, 0) {$B_{A^\prime,\vec{b}^\prime}$};
		\node [style=none] (38) at (2.25, 0.75) {};
		\node [style=none] (39) at (1.5, -0.5) {};
		\node [style=none] (40) at (3, -0.5) {};
		\node [style=none] (41) at (2.25, -1.25) {};
		\node [style=none] (42) at (2.25, -0.5) {};
		\node [style=none] (44) at (2.25, -0.05) {$\vec{b'}$};
	\end{pgfonlayer}
	\begin{pgfonlayer}{edgelayer}
		\draw (15.center) to (16.center);
		\draw (15.center) to (17.center);
		\draw (17.center) to (16.center);
		\draw (19.center) to (18.center);
		\draw (22.center) to (23.center);
		\draw (22.center) to (25.center);
		\draw (25.center) to (24.center);
		\draw (24.center) to (23.center);
		\draw (26.center) to (29.center);
		\draw [style=qWire] (33.center) to (27.center);
		\draw (38.center) to (39.center);
		\draw (38.center) to (40.center);
		\draw (40.center) to (39.center);
		\draw (42.center) to (41.center);
	\end{pgfonlayer}
\end{tikzpicture}}
\end{equation}
where the possible dichotomic measurements output $s\in\{-1,+1\}$, 
\begin{equation}
\begin{tikzpicture}
	\begin{pgfonlayer}{nodelayer}
		\node [style=none] (0) at (-1, 1) {};
		\node [style=none] (1) at (-1, -1) {};
		\node [style=none] (2) at (1, -1) {};
		\node [style=none] (3) at (1, 1) {};
		\node [style=none] (4) at (0, 1.75) {};
		\node [style=none] (5) at (0, -1.825) {};
		\node [style=none] (6) at (0, 1) {};
		\node [style=none] (10) at (0, -1) {};
		\node [style=none] (11) at (0, -0.05) {$B_{\vec{b}^\prime,A^\prime}$};
	\end{pgfonlayer}
	\begin{pgfonlayer}{edgelayer}
		\draw (0.center) to (1.center);
		\draw (0.center) to (3.center);
		\draw (3.center) to (2.center);
		\draw (2.center) to (1.center);
		\draw (4.center) to (6.center);
		\draw [style=qWire] (10.center) to (5.center);
	\end{pgfonlayer}
\end{tikzpicture}}=%
\begin{tikzpicture}
	\begin{pgfonlayer}{nodelayer}
		\node [style=none] (16) at (-2, 0.4) {};
		\node [style=none] (17) at (-2.75, 1.3) {};
		\node [style=none] (18) at (-1.25, 1.3) {};
		\node [style=none] (19) at (-3.5, -1.275) {};
		\node [style=none] (20) at (-0.5, -1.275) {};
		\node [style=none] (21) at (-2, 0.875) {$+$};
		\node [style=none] (22) at (-1.875, -0.725) {$P_{A^\prime,\vec{b}^\prime}^{+}$};
		\node [style=none] (23) at (1.75, 0.375) {};
		\node [style=none] (24) at (1, 1.3) {};
		\node [style=none] (25) at (2.5, 1.3) {};
		\node [style=none] (26) at (0.25, -1.275) {};
		\node [style=none] (27) at (3.25, -1.275) {};
		\node [style=none] (28) at (1.75, 0.875) {$-$};
		\node [style=none] (29) at (1.875, -0.725) {$P_{A^\prime,\vec{b}^\prime}^{-}$};
		\node [style=none] (30) at (0, -0.025) {$+$};
		\node [style=none] (35) at (-2, -1.825) {};
		\node [style=none] (36) at (-2, -1.275) {};
		\node [style=none] (37) at (-2, 1.3) {};
		\node [style=none] (38) at (-2, 1.8) {};
		\node [style=none] (75) at (1.75, -1.825) {};
		\node [style=none] (76) at (1.75, -1.275) {};
		\node [style=none] (77) at (1.75, 1.3) {};
		\node [style=none] (78) at (1.75, 1.75) {};
	\end{pgfonlayer}
	\begin{pgfonlayer}{edgelayer}
		\draw (17.center) to (18.center);
		\draw (18.center) to (16.center);
		\draw (17.center) to (16.center);
		\draw (16.center) to (19.center);
		\draw (19.center) to (20.center);
		\draw (16.center) to (20.center);
		\draw (24.center) to (25.center);
		\draw (25.center) to (23.center);
		\draw (24.center) to (23.center);
		\draw (23.center) to (26.center);
		\draw (26.center) to (27.center);
		\draw (23.center) to (27.center);
		\draw [style=qWire] (36.center) to (35.center);
		\draw (37.center) to (38.center);
		\draw [style=qWire] (76.center) to (75.center);
		\draw (78.center) to (77.center);
	\end{pgfonlayer}
\end{tikzpicture}}.
\end{equation}
Notice that the spectral decomposition of the measurement $B_{\vec{b}^\prime,A^\prime}$ leads to a different decomposition of the vector $\vec{y}_{\vec{b}^\prime,A^\prime}$ than the decomposition in Fig. \ref{fig:vectors}, which is more similar to the one of the vector $\vec{x}_{\vec{a}}$.

\begin{figure}
    \centering
    \includegraphics[width=0.7\linewidth]{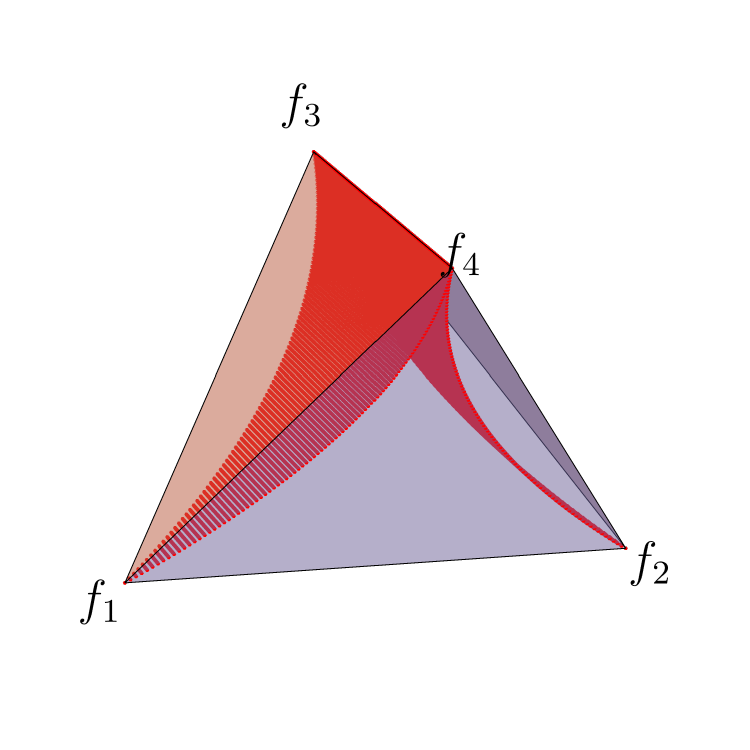}
    \caption{Each point in the tetrahedron identifies a PW protocol at given $y\in[-1,1]$ and $q\in[0,1]$ that respectively characterize the probabilistic discarding and the probabilistic flip operations obtained as convex combinations of the function $f_i$ of Eq.~\eqref{eq:vertex_4hedron}.}
    \label{fig:PWprotocols}
\end{figure}

As regards the RHS of Eq.~\eqref{eq:equivalenceQH2} about the fragment of hyperbit theory, we have
\begin{equation}
\begin{tikzpicture}
	\begin{pgfonlayer}{nodelayer}
		\node [style=none] (0) at (-0.7, 0) {};
		\node [style=none] (1) at (0.7, 0) {};
		\node [style=none] (2) at (0, -0.9) {};
		\node [style=none] (4) at (0, -0.35) {$r$};
		\node [style=none] (5) at (-0.5, 0) {};
		\node [style=none] (6) at (-0.5, 1) {};
		\node [style=none] (7) at (-2, 2) {};
		\node [style=none] (8) at (0, 2) {};
		\node [style=none] (9) at (-2, 1) {};
		\node [style=none] (10) at (0, 1) {};
		\node [style=none] (11) at (0.5, 2.75) {};
		\node [style=none] (12) at (0.5, 0.5) {};
		\node [style=none] (14) at (-1, 3.75) {};
		\node [style=none] (15) at (1.5, 3.75) {};
		\node [style=none] (16) at (-1, 2.75) {};
		\node [style=none] (17) at (1.5, 2.75) {};
		\node [style=none] (18) at (-0.5, 2) {};
		\node [style=none] (19) at (-0.5, 2.75) {};
		\node [style=none] (20) at (-0.25, 4.25) {};
		\node [style=none] (21) at (0.8, 4.25) {};
		\node [style=none] (22) at (0.25, 5) {};
		\node [style=none] (23) at (0.25, 4.25) {};
		\node [style=none] (24) at (0.25, 3.75) {};
		\node [style=none] (25) at (-1.025, 1.5) {\Alice};
		\node [style=none] (26) at (0.25, 3.25) {\Bob};
		\node [style=none] (27) at (0.25, 4.5) {$s$};
		\node [style=none] (28) at (0.75, 0) {};
		\node [style=none] (29) at (1.9, 0) {};
		\node [style=none] (30) at (1.325, -0.925) {};
		\node [style=none] (34) at (-2, 0) {};
		\node [style=none] (35) at (-0.775, 0) {};
		\node [style=none] (36) at (-1.35, -0.8) {};
		\node [style=none] (37) at (-1.3, 0) {};
		\node [style=none] (38) at (-1.3, 1) {};
		\node [style=none] (39) at (1.2, 0) {};
		\node [style=none] (40) at (1.2, 2.75) {};
		\node [style=none] (41) at (1.25, -0.25) {};
		\node [style=none] (42) at (1.35, -0.35) {$\vec{b}$};
		\node [style=none] (43) at (-1.5, -0.25) {};
		\node [style=none] (44) at (-1.35, -0.325) {$\vec{a}$};
		\node [dot] (45) at (-0.5, 0.5) {};
	\end{pgfonlayer}
	\begin{pgfonlayer}{edgelayer}
		\draw (0.center) to (1.center);
		\draw (0.center) to (2.center);
		\draw (1.center) to (2.center);
		\draw [style=none] (5.center) to (6.center);
		\draw (7.center) to (8.center);
		\draw (7.center) to (9.center);
		\draw (9.center) to (10.center);
		\draw (10.center) to (8.center);
		\draw [style=none] (11.center) to (12.center);
		\draw (14.center) to (15.center);
		\draw (14.center) to (16.center);
		\draw (16.center) to (17.center);
		\draw (17.center) to (15.center);
		\draw [style=hyperbit] (19.center) to (18.center);
		\draw (20.center) to (21.center);
		\draw (20.center) to (22.center);
		\draw (21.center) to (22.center);
		\draw (23.center) to (24.center);
		\draw (28.center) to (29.center);
		\draw (28.center) to (30.center);
		\draw (29.center) to (30.center);
		\draw (34.center) to (35.center);
		\draw (34.center) to (36.center);
		\draw (35.center) to (36.center);
		\draw (37.center) to (38.center);
		\draw (40.center) to (39.center);
		\draw (45) to (12.center);
	\end{pgfonlayer}
\end{tikzpicture}}.
\end{equation}
It is clear from this scheme that Alice and Bob shared only a classical random variable $r$ with a vanishing expectation value and the communication is powered by the hyperbit that Alice sends to Bob represented by a thicker output line. With a slight abuse of notation, we refer to the Alice operational procedure again with the label $\mathcal{A}$, where now she has two classical inputs and the hyperbit output:
\begin{equation}
\begin{tikzpicture}
	\begin{pgfonlayer}{nodelayer}
		\node [style=none] (7) at (-1, 0.5) {};
		\node [style=none] (8) at (1, 0.5) {};
		\node [style=none] (9) at (-1, -0.5) {};
		\node [style=none] (10) at (1, -0.5) {};
		\node [style=none] (11) at (0, -1.5) {};
		\node [style=none] (12) at (0, -0.5) {};
		\node [style=none] (18) at (0, 0.5) {};
		\node [style=none] (19) at (0, 1.25) {};
		\node [style=none] (25) at (-0.025, 0) {\Alice};
	\end{pgfonlayer}
	\begin{pgfonlayer}{edgelayer}
		\draw (7.center) to (8.center);
		\draw (7.center) to (9.center);
		\draw (9.center) to (10.center);
		\draw (10.center) to (8.center);
		\draw [style=none] (11.center) to (12.center);
		\draw [style=hyperbit] (19.center) to (18.center);
	\end{pgfonlayer}
\end{tikzpicture}}=\sum_{\vec{a}^\prime}\sum_{r^\prime=\pm 1}%
\begin{tikzpicture}
	\begin{pgfonlayer}{nodelayer}
		\node [style=none] (61) at (2.5, 0.75) {};
		\node [style=none] (62) at (1.75, -0.5) {};
		\node [style=none] (63) at (3.25, -0.5) {};
		\node [style=none] (64) at (2.5, -1.5) {};
		\node [style=none] (65) at (2.5, -0.5) {};
		\node [style=none] (66) at (2.5, 0) {};
		\node [style=none] (67) at (2.5, 0) {$\vec{a}^\prime$};
		\node [style=none] (245) at (5, 0.75) {};
		\node [style=none] (246) at (4.25, -0.5) {};
		\node [style=none] (247) at (5.75, -0.5) {};
		\node [style=none] (248) at (5, -1.5) {};
		\node [style=none] (249) at (5, -0.5) {};
		\node [style=none] (250) at (5, 0) {};
		\node [style=none] (251) at (5, 0) {$r^\prime$};
		\node [style=none] (252) at (2.5, 1.75) {};
		\node [style=none] (253) at (4.9, 1.75) {};
		\node [style=none] (254) at (3.7, -0.15) {};
		\node [style=none] (255) at (3.7, 1.325) {$r^\prime\vec{x}_{\vec{a}}$};
		\node [style=none] (256) at (3.7, 1.75) {};
		\node [style=none] (257) at (3.7, 2.5) {};
	\end{pgfonlayer}
	\begin{pgfonlayer}{edgelayer}
		\draw (61.center) to (62.center);
		\draw (61.center) to (63.center);
		\draw (63.center) to (62.center);
		\draw (65.center) to (64.center);
		\draw (245.center) to (246.center);
		\draw (245.center) to (247.center);
		\draw (247.center) to (246.center);
		\draw (249.center) to (248.center);
		\draw (252.center) to (253.center);
		\draw (252.center) to (254.center);
		\draw (253.center) to (254.center);
		\draw [style=hyperbit] (257.center) to (256.center);
	\end{pgfonlayer}
\end{tikzpicture}}
\end{equation}
where we select as a shared copy the unbiased Alice's output such that $r^\prime =A$. Once Bob receives the hyperbit multiplied by the random copied variable $A\vec{x}_{\vec{a}}$ his strategy is fixed: the expectation value associated with the effect $\vec{y}_{\perp,\vec{b},A}$ inputs into a preestablished postprocess. All the protocols can be visualized as a point in the tetrahedron with vertices $f_i$ of Eq.~\eqref{eq:vertex_4hedron} as in Fig~\ref{fig:PWprotocols}.
. 
As an example, we specialize in the PW protocol. Then Bob's strategy is the one presented in the PW protocol of Sec. \ref{sec:PWprotocol},
\begin{equation}\label{eq:Bob-box}
\begin{tikzpicture}
	\begin{pgfonlayer}{nodelayer}
		\node [style=none] (11) at (-0.025, -0.5) {};
		\node [style=none] (12) at (-0.025, -1.25) {};
		\node [style=none] (14) at (-1.25, 0.5) {};
		\node [style=none] (15) at (1.25, 0.5) {};
		\node [style=none] (16) at (-1.25, -0.5) {};
		\node [style=none] (17) at (1.25, -0.5) {};
		\node [style=none] (18) at (-0.75, -1.25) {};
		\node [style=none] (19) at (-0.75, -0.5) {};
		\node [style=none] (23) at (0, 1) {};
		\node [style=none] (24) at (0, 0.5) {};
		\node [style=none] (26) at (0, 0) {\Bob};
		\node [style=none] (39) at (0.95, -1.25) {};
		\node [style=none] (40) at (0.95, -0.5) {};
	\end{pgfonlayer}
	\begin{pgfonlayer}{edgelayer}
		\draw [style=none] (11.center) to (12.center);
		\draw (14.center) to (15.center);
		\draw (14.center) to (16.center);
		\draw (16.center) to (17.center);
		\draw (17.center) to (15.center);
		\draw [style=hyperbit] (19.center) to (18.center);
		\draw (23.center) to (24.center);
		\draw (40.center) to (39.center);
	\end{pgfonlayer}
\end{tikzpicture}}=%
\begin{tikzpicture}
	\begin{pgfonlayer}{nodelayer}
		\node [style=none] (0) at (1.25, -1.25) {};
		\node [style=none] (1) at (1.25, -0.5) {};
		\node [style=none] (2) at (0.5, -0.5) {};
		\node [style=none] (3) at (2, -0.5) {};
		\node [style=none] (4) at (0.75, -0.25) {};
		\node [style=none] (5) at (1.75, -0.25) {};
		\node [style=none] (6) at (1, 0) {};
		\node [style=none] (7) at (1.5, 0) {};
		\node [style=none] (8) at (1.25, 0.25) {};
		\node [style=none] (9) at (0.5, 1.25) {};
		\node [style=none] (10) at (2, 1.25) {};
		\node [style=none] (11) at (1.25, 0.95) {$c$};
		\node [style=none] (12) at (1.25, 2.5) {};
		\node [style=none] (13) at (1.25, 1.25) {};
	\end{pgfonlayer}
	\begin{pgfonlayer}{edgelayer}
		\draw (0.center) to (1.center);
		\draw (2.center) to (3.center);
		\draw (4.center) to (5.center);
		\draw (6.center) to (7.center);
		\draw (8.center) to (10.center);
		\draw (8.center) to (9.center);
		\draw (9.center) to (10.center);
		\draw (12.center) to (13.center);
	\end{pgfonlayer}
\end{tikzpicture}}+(1-|c|)\sum_{A^\prime,\vec{b}^\prime}	%
\begin{tikzpicture}
	\begin{pgfonlayer}{nodelayer}
		\node [style=none] (34) at (-5.25, -2.25) {};
		\node [style=none] (35) at (-3.25, 0) {};
		\node [style=none] (36) at (-1.5, -2.25) {};
		\node [style=none] (38) at (-3.25, -1.75) {$\tilde{y}_{{\perp}{\vec{b}^\prime,A^\prime}}$};
		\node [style=none] (40) at (-3.25, -3) {};
		\node [style=none] (42) at (-3.25, -2.25) {};
		\node [style=none] (43) at (-1, -2.25) {};
		\node [style=none] (44) at (0, -1) {};
		\node [style=none] (45) at (1, -2.25) {};
		\node [style=none] (46) at (0, -1.75) {$A'$};
		\node [style=none] (47) at (0, -3) {};
		\node [style=none] (48) at (0, -2.25) {};
		\node [style=none] (49) at (1.25, -2.25) {};
		\node [style=none] (50) at (2.25, -1) {};
		\node [style=none] (51) at (3.25, -2.25) {};
		\node [style=none] (52) at (2.25, -1.75) {$\vec{b'}$};
		\node [style=none] (53) at (2.25, -3) {};
		\node [style=none] (54) at (2.25, -2.25) {};
		\node [style=none] (55) at (-1.25, -0.5) {};
		\node [style=none] (56) at (-1.25, -0.5) {$y_{\perp,\vec{b',A'}}$};
		\node [style=none] (57) at (-1.25, -1.75) {};
		\node [style=none] (58) at (0.75, 0.5) {};
		\node [style=none] (59) at (-3.25, 0.5) {};
		\node [style=none] (60) at (-1.25, 0.5) {};
		\node [style=none] (61) at (-1.75, 2.5) {};
		\node [style=none] (62) at (-0.75, 2.5) {};
		\node [style=none] (63) at (-0.75, 1.5) {};
		\node [style=none] (64) at (-1.75, 1.5) {};
		\node [style=none] (65) at (-1.25, 3.25) {};
		\node [style=none] (66) at (-1.25, 1.5) {};
		\node [style=none] (67) at (-1.25, 2.5) {};
		\node [style=none] (68) at (-1.25, 2) {$\mathrm{PS}$};
	\end{pgfonlayer}
	\begin{pgfonlayer}{edgelayer}
		\draw (34.center) to (35.center);
		\draw (35.center) to (36.center);
		\draw (36.center) to (34.center);
		\draw [style=hyperbit] (42.center) to (40.center);
		\draw (43.center) to (44.center);
		\draw (44.center) to (45.center);
		\draw (45.center) to (43.center);
		\draw (48.center) to (47.center);
		\draw (49.center) to (50.center);
		\draw (50.center) to (51.center);
		\draw (51.center) to (49.center);
		\draw (54.center) to (53.center);
		\draw (57.center) to (58.center);
		\draw (57.center) to (59.center);
		\draw (59.center) to (58.center);
		\draw (64.center) to (63.center);
		\draw (63.center) to (62.center);
		\draw (62.center) to (61.center);
		\draw (61.center) to (64.center);
		\draw (60.center) to (66.center);
		\draw (67.center) to (65.center);
	\end{pgfonlayer}
\end{tikzpicture}}
\end{equation}
where $\mathrm{PS}$ is the postprocessing of the outcomes once the probabilistic discarding operation is applied. In the PW protocol $\mathrm{PS}=f_q$, which is the probabilistic flip
\begin{equation}
\begin{tikzpicture}
	\begin{pgfonlayer}{nodelayer}
		\node [style=none] (0) at (-0.75, 0.75) {};
		\node [style=none] (1) at (-0.75, -0.5) {};
		\node [style=none] (2) at (0.75, -0.5) {};
		\node [style=none] (3) at (0.75, 0.75) {};
		\node [style=none] (4) at (0, 1.5) {};
		\node [style=none] (5) at (0, -1.5) {};
		\node [style=none] (7) at (0, 0.75) {};
		\node [style=none] (8) at (-0.25, -0.5) {};
		\node [style=none] (9) at (-0.25, -0.5) {};
		\node [style=none] (10) at (-0.25, -0.5) {};
		\node [style=none] (11) at (0, -0.5) {};
		\node [style=none] (12) at (0.25, -0.5) {};
		\node [style=none] (13) at (0, 0) {$f_q$};
	\end{pgfonlayer}
	\begin{pgfonlayer}{edgelayer}
		\draw (0.center) to (1.center);
		\draw (0.center) to (3.center);
		\draw (3.center) to (2.center);
		\draw (2.center) to (1.center);
		\draw (4.center) to (7.center);
		\draw (11.center) to (5.center);
	\end{pgfonlayer}
\end{tikzpicture}}=q\, %
\begin{tikzpicture}
	\begin{pgfonlayer}{nodelayer}
		\node [style=none] (16) at (-1.25, 0) {};
		\node [style=none] (17) at (-2, 0.95) {};
		\node [style=none] (18) at (-0.5, 0.95) {};
		\node [style=none] (19) at (-2, -0.95) {};
		\node [style=none] (20) at (-0.5, -0.95) {};
		\node [style=none] (21) at (-1.25, 0.5) {$-$};
		\node [style=none] (22) at (-1.25, -0.525) {$+$};
		\node [style=none] (23) at (0, 0) {$+$};
		\node [style=none] (26) at (1.25, 0) {};
		\node [style=none] (27) at (0.5, 0.95) {};
		\node [style=none] (28) at (2, 0.95) {};
		\node [style=none] (29) at (0.5, -0.95) {};
		\node [style=none] (30) at (2, -0.95) {};
		\node [style=none] (31) at (1.25, 0.5) {$+$};
		\node [style=none] (32) at (1.25, -0.525) {$-$};
		\node [style=none] (33) at (-2.125, 1.25) {};
		\node [style=none] (34) at (-2.125, -1.25) {};
		\node [style=none] (35) at (2.175, 1.25) {};
		\node [style=none] (37) at (2.175, -1.25) {};
		\node [style=none] (38) at (0, 1.25) {};
		\node [style=none] (39) at (0, 1.75) {};
		\node [style=none] (40) at (0, -1.75) {};
		\node [style=none] (41) at (0, -1.25) {};
	\end{pgfonlayer}
	\begin{pgfonlayer}{edgelayer}
		\draw (17.center) to (18.center);
		\draw (18.center) to (16.center);
		\draw (17.center) to (16.center);
		\draw (16.center) to (19.center);
		\draw (19.center) to (20.center);
		\draw (16.center) to (20.center);
		\draw (27.center) to (28.center);
		\draw (28.center) to (26.center);
		\draw (27.center) to (26.center);
		\draw (26.center) to (29.center);
		\draw (29.center) to (30.center);
		\draw (26.center) to (30.center);
		\draw (33.center) to (34.center);
		\draw (33.center) to (35.center);
		\draw (35.center) to (37.center);
		\draw (37.center) to (34.center);
		\draw (39.center) to (38.center);
		\draw (41.center) to (40.center);
	\end{pgfonlayer}
\end{tikzpicture}}+(1-q)%
\begin{tikzpicture}
	\begin{pgfonlayer}{nodelayer}
		\node [style=none] (16) at (-1.25, 0) {};
		\node [style=none] (17) at (-2, 0.95) {};
		\node [style=none] (18) at (-0.5, 0.95) {};
		\node [style=none] (19) at (-2, -0.95) {};
		\node [style=none] (20) at (-0.5, -0.95) {};
		\node [style=none] (21) at (-1.25, 0.5) {$+$};
		\node [style=none] (22) at (-1.25, -0.55) {$+$};
		\node [style=none] (23) at (0, 0) {$+$};
		\node [style=none] (26) at (1.25, 0) {};
		\node [style=none] (27) at (0.5, 0.95) {};
		\node [style=none] (28) at (2, 0.95) {};
		\node [style=none] (29) at (0.5, -0.95) {};
		\node [style=none] (30) at (2, -0.95) {};
		\node [style=none] (31) at (1.25, 0.5) {$-$};
		\node [style=none] (32) at (1.25, -0.55) {$-$};
		\node [style=none] (33) at (-2.125, 1.25) {};
		\node [style=none] (34) at (-2.125, -1.25) {};
		\node [style=none] (35) at (2.175, 1.25) {};
		\node [style=none] (37) at (2.175, -1.25) {};
		\node [style=none] (38) at (0, 1.25) {};
		\node [style=none] (39) at (0, 1.75) {};
		\node [style=none] (40) at (0, -1.75) {};
		\node [style=none] (41) at (0, -1.25) {};
	\end{pgfonlayer}
	\begin{pgfonlayer}{edgelayer}
		\draw (17.center) to (18.center);
		\draw (18.center) to (16.center);
		\draw (17.center) to (16.center);
		\draw (16.center) to (19.center);
		\draw (19.center) to (20.center);
		\draw (16.center) to (20.center);
		\draw (27.center) to (28.center);
		\draw (28.center) to (26.center);
		\draw (27.center) to (26.center);
		\draw (26.center) to (29.center);
		\draw (29.center) to (30.center);
		\draw (26.center) to (30.center);
		\draw (33.center) to (34.center);
		\draw (33.center) to (35.center);
		\draw (35.center) to (37.center);
		\draw (37.center) to (34.center);
		\draw (39.center) to (38.center);
		\draw (41.center) to (40.center);
	\end{pgfonlayer}
\end{tikzpicture}}
\end{equation}
where the last term is the identity operation that can be equivalently represented by a solid vertical line.
\end{document}